\documentclass[12pt]{article}
\usepackage[left=1in,right =1in, top=1in, bottom=1in]{geometry}
\usepackage[linktocpage]{hyperref}
\usepackage[noadjust]{cite}
\usepackage{float}
\usepackage{amsfonts}
\usepackage{comment}
\usepackage{csquotes}
\usepackage{subcaption}
\usepackage{amsmath}
\usepackage{amssymb}
\usepackage{latexsym}
\usepackage{graphicx}
\usepackage[english]{babel}
\usepackage[font={small}]{caption}
\usepackage{extarrows}
\usepackage{fancybox}
\usepackage{mathtools}
\usepackage{physics}
\usepackage[none]{hyphenat}

\usepackage{url}
\usepackage{enumerate}
\usepackage[shortlabels]{enumitem}
\usepackage{ulem}
\usepackage{tikz}

\begin{document}
\input epsf

\def\p{\partial}
\def\h{{\frac12}}
\def\be{\begin{equation}}
\def\bea{\begin{eqnarray}}
\def\ee{\end{equation}}
\def\eea{\end{eqnarray}}
\def\d{\partial}
\def\la{\lambda}
\def\eps{\epsilon}
\def\b{\bigskip}
\def\m{\medskip}

\newcommand{\newsection}[1]{\section{#1} \setcounter{equation}{0}}

\def\at{\tilde{a}}
\def\bt{\tilde{b}}
\def\q{\quad}
\def\t{\tilde}
\def\r{\rightarrow}
\def\nn{\nonumber\\}
\def\Ad{\dot{A}}
\def\Bd{\dot{B}}
\def\Cd{\dot{C}}
\def\Dd{\dot{D}}
\def\wb{\bar{w}}
\def\wt{\tilde{w}}
\def\zt{\tilde{z}}
\def\tt{\tilde{t}}
\def\ep{\epsilon}

\newcommand{\Romk}[1]{|0^-\rangle_R^{(#1)}}
\newcommand{\Romb}[1]{{}^{\,(#1)}_{\ \,R}\langle0^-|}
\newcommand{\Ropk}[1]{|0^+\rangle_R^{(#1)}}
\newcommand{\Ropb}[1]{{}^{\,(#1)}_{\ \,R}\langle0^+|}
\newcommand{\Rok}[1]{|0\rangle_R^{(#1)}}
\newcommand{\Rob}[1]{{}^{\,(#1)}_{\ \,R}\langle0|}
\newcommand{\Rotk}[1]{|\tilde{0}\rangle_R^{(#1)}}
\newcommand{\Rotb}[1]{{}^{\,(#1)}_{\ \,R}\langle\tilde{0}|}

\newcommand{\Robmk}[1]{|\bar0^-\rangle_R^{(#1)}}
\newcommand{\Robmb}[1]{{}^{\,(#1)}_{\ \,R}\langle\bar0^-|}
\newcommand{\Robpk}[1]{|\bar0^+\rangle_R^{(#1)}}
\newcommand{\Robpb}[1]{{}^{\,(#1)}_{\ \,R}\langle\bar0^+|}
\newcommand{\Robk}[1]{|\bar0\rangle_R^{(#1)}}
\newcommand{\Robb}[1]{{}^{\,(#1)}_{\ \,R}\langle\bar0|}
\newcommand{\Rotbk}[1]{|\bar\tilde{0}\rangle_R^{(#1)}}
\newcommand{\Rotbb}[1]{{}^{\,(#1)}_{\ \,R}\langle\bar\tilde{0}|}

\newcommand{\Roobmk}[1]{|0^-,\bar0^-\rangle_R^{(#1)}}
\newcommand{\Roobmb}[1]{{}^{\,(#1)}_{\ \,R}\langle0^,\bar0^-|}

\def\NSo{0_{\scriptscriptstyle{N\!S}}}
\newcommand{\NSok}[1]{|0\rangle_{\scriptscriptstyle{N\!S}}^{(#1)}}
\newcommand{\NSob}[1]{{}^{\ \,(#1)}_{\ \;\scriptscriptstyle{N\!S}}\langle0|}

\def\phiR{\phi_R}
\def\phiNS{\phi_{\scriptscriptstyle{N\!S}}}
\newcommand{\phiRk}[1]{|\phi\rangle_R^{(#1)}}
\newcommand{\phiRb}[1]{{}^{\,(#1)}_{\ \,R}\langle\phi|}
\newcommand{\phiNSk}[1]{|\phi\rangle_{\scriptscriptstyle{N\!S}}^{(#1)}}
\newcommand{\phiNSb}[1]{{}^{\ \,(#1)}_{\ \;\scriptscriptstyle{N\!S}}\langle\phi|}

\newcommand{\MyRed}{\color [rgb]{0.8,0,0}}
\newcommand{\MyGreen}{\color [rgb]{0,0.7,0}}
\newcommand{\MyBlue}{\color [rgb]{0,0,0.8}}
\newcommand{\MyBrown}{\color [rgb]{0.8,0.4,0.1}}
\newcommand{\MyPurple}{\color [rgb]{0.6,0.0,0.6}}
\def\MH#1{{\MyRed [MH: #1]}}
\def\SM#1{{\MyPurple [SM: #1]}}   
\def\BG#1{{\MyBlue [BG: #1]}}
\def\MM#1{{\MyGreen [MM: #1]}}

\newcommand\blfootnote[1]{%
  \begingroup
  \renewcommand\thefootnote{}\footnote{#1}%
  \addtocounter{footnote}{-1}%
  \endgroup
}

\begin{flushright}
\end{flushright}
\hfill
\vspace{18pt}
\begin{center}
{\Large 
\textbf{Lifting of superconformal descendants in the\\ D1-D5 CFT
}}

\end{center}

\vspace{10mm}
\begin{center}
{\textsl{Marcel R. R. Hughes$^{a}$}\blfootnote{${}^{a}$hughes.2059@osu.edu}\textsl{, Samir D. Mathur${}^{b}$}\blfootnote{${}^b$mathur.16@osu.edu}\textsl{ and Madhur Mehta${}^{c}$}\blfootnote{${}^c$mehta.493@osu.edu}
\\}

\vspace{10mm}

\textit{\small ${}^{a,b,c}$ Department of Physics, The Ohio State University,\\ Columbus,
OH 43210, USA} \\  \vspace{6pt}

\end{center}

\vspace{12pt}
\begin{center}
\textbf{Abstract}
\end{center}

\vspace{4pt} {\small
\noindent
We consider D1-D5-P states in the untwisted sector of the D1-D5 orbifold CFT where we excite one copy of the seed CFT with a left-moving superconformal descendant. When the theory is deformed away from this region of moduli space these states can `lift', despite being BPS at the orbifold point. For descendants formed from the supersymmetry $G^{\alpha}_{\!\dot{A},-s}$ and R-symmetry $J^a_{-n}$ current modes we obtain explicit results for the expectation value of the lifts for various subfamilies of states at second order in the deformation parameter. A smooth $\sim\sqrt{h}$ behaviour is observed in the lifts of these subfamilies for large dimensions. Using covering space Ward identities we then find a compact expression for the lift of the above $J^a_{-n}$ descendant states valid for arbitrary dimensions. In the large-dimension limit this lift scales as $\sim\sqrt{h}\,$, strengthening the conjecture that this is a universal property of the lift of D1-D5-P states. We observe that the lift is not simply a function of the total dimension, but depends on how the descendant level is partitioned amongst modes.

\vspace{1cm}

\thispagestyle{empty}

\setcounter{footnote}{0}
\setcounter{page}{0}
\newpage

\tableofcontents

\setcounter{page}{1}

\numberwithin{equation}{section}

\section{Introduction}\label{secintro}
\normalem

One of the biggest ongoing problems in physics has to do with the study and understanding of black holes beyond the classical predictions of general relativity. A prototypical example black hole system that has lent itself well to study from various angles is the extremal 3-charge Strominger-Vafa black hole. The microscopic description of this classical black hole solution is that of type IIB string theory on $\mathbb{R}^{1,4}\times S^1_y\times \mathcal{M}^4$ (this compact manifold $\mathcal{M}^4$ can be either $T^4$ or $K3$). On this background are placed D1-branes, D5-branes and momentum (P) excitations wrapping the compact directions. This brane system preserves $1/8$th of the 32 real supercharges of the underlying string theory (referred to as $1/8$th BPS) and its many bound states are the microstates of the classical black hole each with the same mass and charges. The Bekenstein entropy of the Strominger-Vafa black hole $S_{\rm{bek}}=A/4G$ has been shown to be accounted for by the $e^{S_{\rm{bek}}}$ microstates of the brane system in \cite{Strominger:1996sh} and \cite{Maldacena:1999bp} for type IIB string theory compactified on $K3\times S^1_y$ and $T^4\times S^1_y$ respectively. These are lower bounds on the number of microstates due to the use of supersymmetric indices, however, their leading-order behavior matches exactly the Bekenstein entropy of the black hole.

Whilst the successful identification of the appropriate system in string theory that describes the Strominger-Vafa black hole, with the correct count of states, is a great achievement it is desirable to now ask more fine-grained questions. In particular: what do the individual microstates of a black hole look like, what are the dynamical properties of such states, and how does this microscopic description give rise to the relatively simple structure and properties of the classical solution in an appropriate coarse-grained description. The explicit construction and study of such black hole microstates is termed the fuzzball program. A long-standing effort to construct microstate geometries (a special class of fuzzballs for which suitably coherent behaviour leads to a semi-classical description) for the D1-D5-P system has led to the discovery of large classes of horizonless solutions to supergravity having the same asymptotic structure as the black hole, but with different infrared behaviours encoding microscopic details of the states (see \cite{Bena:2022ldq,Bena:2022rna} for reviews of the current state of these efforts and of the wider fuzzball proposal and also~\cite{Heidmann:2021cms,Bah:2022pdn,Heidmann:2022zyd,Bah:2022yji,Bah:2023ows} for recent interesting developments of non-BPS black hole microstate geometries).

Another reason for the richness of accessible information in the Strominger-Vafa black hole system is the existence of a supersymmetric conformal field theory (CFT) description that is holographically dual to the D1-D5-P system. A large program of work has been devoted to understanding the holographic dictionary between the black hole microstates and the spectrum of the dual `D1-D5 CFT'  \cite{Lunin:2001jy,Mathur:2005zp,Kanitscheider:2007wq,Bena:2007kg,Chowdhury:2010ct,Shigemori:2020yuo}. Despite the precise CFT dual to the weakly coupled brane system being at strong coupling, it is conjectured that there exists a locus of its moduli space at which the CFT has a description in terms of a $(1+1)$-dimensional free symmetric group orbifold theory \cite{Vafa:1995bm,Dijkgraaf:1998gf,Larsen:1999uk,Seiberg:1999xz,Arutyunov:1997gi,Arutyunov:1997gt,Jevicki:1998bm,David:2002wn} (this is analogous to free super Yang-Mills (SYM) in the D3-brane system \cite{Maldacena:1997re}). In this free theory we have much better control over the spectrum and its observables, in particular one can ask what form black hole microstates take in the dual field theory description. Since the black hole is BPS the dual states should also be equally BPS, however, while in the free orbifold theory any state with purely left-moving (or purely right-moving) excitations is BPS, this simple classification no longer holds at a general point in the D1-D5 CFT's moduli space. Upon deforming the free orbifold theory towards the strongly-coupled part of the moduli space, short multiplets of the free theory (which contain the BPS states) can join together into long multiplets of the theory with general moduli. This phenomena is termed `lifting' and is visible from an anomalous contribution to the conformal dimension of these states under deformation of the free theory. Therefore, only states that are globally BPS are seen by the index calculations of \cite{Strominger:1996sh,Maldacena:1999bp} and can correspond to microstates of extremal black holes. Whilst the precise form of the globally BPS states is moduli dependent, the number of such states (and some of their properties such as quantum numbers and 3-point functions) is not.

In \cite{Gava:2002xb}, the early work of Gava and Narain found that, as expected from the gravity theory, almost all low-lying states of the CFT were lifted; the only low-energy BPS states in the gravity picture describe supergravity quanta, while the remainder are string states whose energies are lifted to the string scale. However, the CFT spectrum should have a transition to a dense black hole phase at high enough energies which should contain the large number of expected black hole states seen by the index calculations\footnote{Some recent progress has been made for the case of $1/16$th BPS states in $\mathcal{N}=4$ SYM with gauge group $SU(2)$ above the black hole threshold~\cite{Chang:2022mjp,Chang:2023zqk,Budzik:2023vtr,Choi:2023znd}.}. Whilst there has been progress in computing the lift of various families of D1-D5-P states through the use of conformal perturbation theory, no simple answer has been found to the question of what dramatic change occurs in lifting calculations when above the black hole threshold.

Here we very briefly describe the families of states that have been considered so far\footnote{See \cite{Lima:2020boh,Lima:2020kek,Lima:2020nnx,Lima:2020urq,Lima:2021wrz,Lima:2021xqj,Lima:2022cnq,Benjamin:2022jin,Guo:2022and} for additional work about the lifting of states using conformal perturbation theory in the D1-D5 CFT and \cite{Guo:2022sos,Guo:2023czj} for some work related to the twist operators of the theory. Also of note are \cite{Burrington:2022dii,Burrington:2022rtr} in which for $S_N$ orbifold theories it is shown how to relate correlators of Virasoro descendants (using both integer and fractional modes) to correlators of a suitable set of ancestors using covering space methods.}. In \cite{Gava:2002xb}, states on highly-wound component strings (with winding $k\gg1$) were considered with purely left-moving excitations of level $n$ above the NS vacuum satisfying $n\gg k$. All of these states were found to be lifted bar those corresponding to the single-particle supergravity multiplet. In \cite{Gaberdiel:2015uca} the currents of the higher-spin symmetry algebra of the free orbifold theory were found to lift. This is expected since away from the free point in moduli space the higher-spin symmetry is broken to the $\mathcal{N}=4$ superconformal algebra. In \cite{Hampton:2018ygz}, a closed-form expression was found for the lift of states where a subset of copies of the orbifold theory were excited above the unique NS vacuum with an excitation formed from the modes of the left-moving part of the $SU(2)$ R-symmetry current. This excitation amounts to putting the maximal number of fermion excitations on top of the NS vacuum and was studied using the particular property that this excited state is the spectral flow of the vacuum. This 1-parameter family of states is labelled by the dimension of the excitation. For the case of the $N=2$ theory, \cite{Guo:2019ady,Guo:2020gxm} constructed and computed the lifts for all D1-D5-P primaries up to level 4 on a case-by-case basis. It was found that all such states were lifted for which suitable groups of short multiplets exist to join up into long multiplets of the deformed theory. The lift of untwisted-sector states in $(h,j)=(1,0)$ left-moving long multiplets and general $\bar{j}$ right-moving short multiplets was studied in \cite{Benjamin:2021zkn} where evidence was found for the undercounting of unlifted states by the index computations. In \cite{Guo:2022ifr}, untwisted-sector states were considered where all copies were in the NS vacuum bar one copy, which was excited with a superconformal primary of the $c=6$ seed theory. A closed-form expression was found for the lift of this family of states and it was shown that explicit knowledge of the superconformal primary is not necessary, only its conformal dimension. Most recently, \cite{Hughes:2023apl} studied the lift of untwisted-sector states where one copy was excited above the NS-NS vacuum by two left-moving modes of the bosons and fermions of the free theory. Patterns in the lift of this two-parameter family of states (labelled by the mode numbers $m$, $n$ of the two modes) were studied. Since the computation method used increased in complexity with the level of the state ($m+n$), this was done on a state-by-state basis and thus no closed-form expressions were found. The advantage of this method, however, was that lifts at much higher levels were obtainable as compared with the low-level searches of~\cite{Guo:2019ady,Guo:2020gxm}.

In these searches of the CFT spectrum, various patterns have emerged and it is hoped that a general answer to the question of which states do not lift will become clear given an understanding of the lift in enough special cases. One of these patterns in the lift of states is that for large dimensions, the lift appears to follow a $\sim\sqrt{h}$ growth independently of the particular state chosen. Due to the majority of previous lifting calculations being done on a state-by-state basis, this asymptotic growth has mostly been observed by fitting the lift of 1-parameter families of states. In this paper we aim to address this seemingly universal asymptotic behaviour of lifts analytically.

\subsection{Summary of results}

In this paper we find the following:
\begin{itemize}
    \item We study the second order lifting of states where one copy of the $c=6$ CFT is excited above the NS vacuum by a descendent of a superconformal primary $\phi$. Specifically, the single-copy descendants we discuss are
    \begin{equation} \label{eq.introstates}
        G^\alpha_{\!\Ad,-s}|\phi\rangle \quad \text{ and } \quad J^+_{-n}|\phi\rangle \ .
    \end{equation}
    The expectation value of lifts of these states is computed in Section~\ref{sec.main} using a new method that is well suited to having states where we do not specify their form, \textit{i.e.} the superconformal primary $|\phi\rangle$. This method, that we dub the ``series method" to distinguish it from the ``fields method" of \cite{Hughes:2023apl} that relied on explicit free-field representations of the states being considered. The lifts of the states \eqref{eq.introstates} are found in terms of contour integrals of various series, the complexity of which scales with the descendant level ($s$ and $n$). Therefore, similarly to \cite{Hughes:2023apl}, we evaluate these lifts on a state-by-state basis for various 1-parameter families of states to much higher levels than has been previously possible. In this lifting data we observe the same $\sim\sqrt{h}$ asymptotic growth that has been previously observed for other families of states in \cite{Guo:2022ifr,Hughes:2023apl}.

    \item The lifting of the descendant state $J^+_{-n}|\phi\rangle_h$ on a single copy is also studied using a method of Ward identities. In this method we use Ward identities on the covering space of the amplitude, required for the computation of the lifts, in order to remove the R-symmetry current modes. This approach turns out to allow for an analytic expression for the expectation value of the lift of these states given by
    \begin{align} \label{eq.EJphiIntro}
        E^{(2)}\big(|J\phi\rangle^+_{(h,n)}\big) = E^{(2)}_{h}\!\big(\phi\big) \Bigg[1 + a_n \frac{P_{n}(h)}{(h+1)_{2n-2}}\Bigg] \ , 
    \end{align}
    where
    \begin{equation} \label{eq.EphiIntro}
        E^{(2)}_{h}\!\big(\phi\big) = \pi^{\tfrac32}\lambda^2\frac{\Gamma(h+\tfrac12)}{\Gamma(h)} \ ,
    \end{equation}
    is the dimension-$h$ single-copy primary lift found in \cite{Guo:2022ifr}. The $a_n$ are rational coefficients listed in \eqref{eq.andata} and the $P_n(h)$ are polynomials in $h$ of degree $2n-4$, explicit forms of the first few are given in \eqref{eq.Pn}. In general these coefficients and polynomials are defined implicitly from the sum of \eqref{eq.E1def} and \eqref{eq.E2def}. Of note here is that the result \eqref{eq.EphiIntro} does not depend on the form of the primary $\phi$, only its dimension $h$.

    \item Having such a closed-form expression for the lift of this 2-parameter family of states allows us to partially answer previously posed conjectures on the universal behaviour of lifts. Namely, we discuss the following questions:
    \begin{enumerate}
        \item[(1)] \label{conject1} Is the leading $\sim\sqrt{h}$ behaviour for large conformal dimension, derived in \cite{Hughes:2023apl} for single-copy primaries, a universal property of second order lifts?
        \item[(2)] \label{conject2} Are second-order lifts in general purely a function of the total dimension, as in the case of \eqref{eq.EphiIntro}, or do they depend on the details of the state.
    \end{enumerate}
    With regards to point (1), we derive the asymptotic behaviour of the lift \eqref{eq.EJphiIntro} for large $h$ and fixed $n$ to the first subleading order to be
    \begin{equation} \label{eq.JphilargehIntro}
        \frac{E^{(2)}\!\big(|J\phi\rangle^+_{(h,n)}\big)}{E^{(2)}_{h}\!(\phi)} \approx 1 + \frac{n(n^2-1)}{8h^2} \ ,
    \end{equation}
    and so we see that the leading behaviour is $\sim\sqrt{h}$ as this is the leading behaviour of \eqref{eq.EphiIntro}. However, with regards to point (2) above we argue in Section~\ref{sssec:largeh} that the lift is not purely a function of the total dimension $h+n$. A follow up to point (2) is then:
    \begin{enumerate}
        \item[(2a)] Do the second order lifts of level-$n$ descendants of a superconformal primary of dimension $h$ depend on how $n$ is partitioned amongst descendant modes?
    \end{enumerate}
    By comparing \eqref{eq.JphilargehIntro} the lift of states with the single-copy excitation $J^+_{-(2m-1)}\cdots J^+_{-3}J^+_{-1}|\phi\rangle_h$ (whose descendant level is $m^2$) which we compute in closed-form in Appendix~\ref{app.3}, we find that the question of point (2a) is true.
        
\end{itemize}

\subsection*{Paper layout}

A summary of the present paper is as follows: Section~\ref{secD1D5} briefly reviews the D1-D5 CFT, its symmetries, spectral flow and deformations away from the free point in its moduli space. In particular, Section~\ref{ssec:SCPs} defines superconformal primaries in the D1-D5 CFT and gives a mechanism for the explicit construction of an infinite family using the fields of the free orbifold theory. Section~\ref{sec:LiftForm} briefly reviews a general method for, in principle, finding the expectation value of the second order lift in energy. In Sections~\ref{ssec:GphiLift} and \ref{ssec:Jphilift} we compute the lift of untwisted-sector states in which one copy is excited with a superconformal descendent of a primary $\phi$ to high levels using a new ``series method". Some patterns in the lifting data of these descendent states are discussed in Section~\ref{ssec:symmChecks} including the observation of asymptotic $\sim\sqrt{h}$ behaviour for large dimensions, inline with observations made in \cite{Guo:2022ifr,Hughes:2023apl}. Section~\ref{ssec:WardIdJphi} derives a closed-form for the lift of single-copy R-symmetry current descendants $\sim J^+_{-n}|\phi\rangle$ by a method of covering space Ward identities. This analytic expression then allows for an analysis of the large dimension behaviour of lifts in Sections~\ref{sec:AsympBehaviour}. The lift of states of the form $\sim J^+_{-(2m-1)}\cdots J^+_{-3}J^+_{-1}|\phi\rangle$ is derived analytically in Section~\ref{app.3} in order to strengthen conclusions about universal features of lifts in the D1-D5 CFT. We conclude with a discussion of the results in Section~\ref{sec.conc}.

\section{The D1-D5 CFT and its free-orbifold point} \label{secD1D5}

The setting is type IIB string theory on a background of
\begin{equation} \label{eq.background}
    M^{1,4}\times S^1\times T^4 \ ,
\end{equation}
containing $n_1$ D1-branes and $n_5$ D5-branes that both wrap the $S^1$ and with the D5-branes also wrapping the $T^4$. The bound states of this brane system generate the D1-D5 CFT -- a $(1+1)$-dimensional conformal field theory living on a cylinder made from the time direction and the spatial $S^1$. This theory is thought to have a point in its moduli space, dubbed the free-orbifold point, where there is a description of the theory in terms of
\begin{equation}
    N = n_1 n_5 \ ,
\end{equation}
copies of a seed $c=6$ free CFT. This free CFT contains $4$ free bosons and $8$ free fermions, 4 each in the left- and right-moving sectors. The Hilbert space of these free fields are then subject to an orbifolding by the permutation group $S_N$, leading to the Hilbert space factoring into twisted sectors labelled by integers $1\leq k\leq N$. The different twisted sectors in essence describe a CFT on a $k$-wound circle (sometimes called a component string). This orbifold point of the D1-D5 CFT has been shown~\cite{Gaberdiel:2018rqv,Eberhardt:2018ouy,Eberhardt:2019ywk,Eberhardt:2020akk,Eberhardt:2019qcl,Dei:2019osr} to be dual to the tensionless limit of a string in an AdS${}_3\times S^3\times T^4$ background with one unit of NS-NS flux.

\subsection{Symmetries of the CFT} \label{ssec:symms}

Generically in the moduli space of the D1-D5 CFT there is $\mathcal{N}=(4,4)$ supersymmetry~\cite{Schwimmer:1986mf,Sevrin:1988ew}, \textit{i.e.} an ${\cal N}=4$ superconformal symmetry algebra in both the left- and right-moving sectors. The chiral algebra generators are
\begin{equation} \label{eq.lcurr}
    \Big\{\, L_{n}\ ,\ \ G^{\alpha}_{\!\Ad,r}\ ,\ \ J^a_n \,\Big\} \ ,
\end{equation}
for the left movers associated with the stress-energy tensor, supercurrents and $\mathfrak{su}(2)$ R-currents respectively. The right-moving sector has analogous generators given by the set
\begin{equation} \label{eq.rcurr}
    \Big\{\,\bar L_{n}\ ,\ \ \bar G^{\bar \alpha}_{\!\Ad,r} \ ,\ \ \bar J^a_n\,\Big\} \ .
\end{equation}
The various indices used are associated with fundamental representations of the various $SU(2)$ factors of the symmetry algebra: $\alpha$ and $\bar{\alpha}$ are indices of the $SU(2)_L$ and $SU(2)_R$ factors of the
\begin{equation} \label{eq.SO4E}
    SO(4)_E\cong \big(SU(2)_L\times SU(2)_R\big)/\mathbb{Z}_2 \ ,
\end{equation}
`external' R-symmetry, originating from rotations in the noncompact spatial directions of the background \eqref{eq.background}. The broken `internal' $SO(4)_I$ global symmetry, coming from the $T^4$ factor of the background, provides a useful organising principle for the spectrum of the orbifold theory. The doublet indices $A, \dot A$ are from the $SU(2)_1$ and $SU(2)_2$ factors of
\begin{equation} \label{eq.SO4I}
    SO(4)_I\cong \big(SU(2)_1\times SU(2)_2\big)/\mathbb{Z}_2 \ .
\end{equation}
In \eqref{eq.lcurr} and \eqref{eq.rcurr} the index $a$ is a vector index of the $SO(4)_E$. The small $\mathcal{N}=(4,4)$ superconformal algebra spanned by the currents \eqref{eq.lcurr} and \eqref{eq.rcurr} and our conventions are outlined in Appendix~\ref{app_cft}. At the free orbifold point, where the theory has a realisation in terms of the free bosons and fermions
\begin{equation} \label{eq.freeFields}
    \Big\{\,\partial X_{\!A\Ad} \ ,\ \ \psi^{\,\alpha A} \ ,\ \ \bar{\psi}^{\,\bar{\alpha}\Ad}\,\Big\} \ ,
\end{equation}
with their corresponding modes (see the mode expansions \eqref{BosFerModes})
\begin{equation} \label{eq.freeFields2}
    \Big\{\, \alpha_{A\Ad,n}\ ,\ \ d^{\,\alpha A}_{s} \ ,\ \ \bar{d}^{\,\bar{\alpha}\Ad}_{s} \,\Big\} \ ,
\end{equation}
this symmetry algebra is in fact enlarged to the contracted large ${\cal N}=(4,4)$ superconformal symmetry \cite{Maldacena:1999bp,Sevrin:1988ew} (in fact exactly at the free point, this symmetry is boosted to include also a $\mathcal{W}_{\infty}$ algebra studied, for instance, in~\cite{Gaberdiel:2015mra,Gaberdiel:2015uca}).

\subsection{Superconformal primaries} \label{ssec:SCPs}

A class of states that will play an important role in this paper are superconformal primaries. Consider such a state $|\phi\rangle$ with conformal dimension $h$ and $J^3_0$ charge $m$ can be defined by the mode conditions (given in the NS sector)
\begin{equation} \label{def primary 2}
    L_{n}|\phi\rangle = G^{\alpha}_{\!\Ad,s}|\phi\rangle = J^{a}_{n}|\phi\rangle =0 \quad,\quad \text{ where } n>0\ ,\ s\geq \frac12 \ .
\end{equation}
If $\phi$ is to be a primary of the full $c=6N$ theory then these current algebra modes should be global modes. For instance, the global Virasoro modes $L^{(g)}_{n}$ in the untwisted sector are given by the diagonal sum
\begin{equation} \label{globalL}
    L^{(g)}_{n} = \sum_{i=1}^N L^{(i)}_{n} \ ,
\end{equation}
where $L^{(i)}_{n}$ are the modes on the $i$th copy (in the rest of the paper we will mostly drop the copy index since only single-copy modes are used). Likewise, a superconformal primary on a single copy would satisfy \eqref{def primary 2} with the modes on that particular copy of the $c=6$ seed theory. The superconformal primary states $|\phi\rangle^{(1)}$ used in this paper are primaries of one copy of the $c=6$ seed theory (written here as copy 1 for clarity), with all other copies being in the unique NS vacuum (the right-moving part of the state is in the NS vacuum on all copies). This type of state can be seen to be constrained to have $J^3_0$ charge $m$ of\footnote{Whilst the constraint \eqref{eq.mcond} on single-copy superconformal primaries has been written in terms of the eigenvalue of $J^3_0$, the action of the raising and lowering operators $J^{\pm}_0$ for $SU(2)_L$ multiplets commutes with the definition \eqref{def primary 2}. Therefore, the constraint \eqref{eq.mcond} must be satisfied by all members of a given $SU(2)_L$ multiplet. By labelling these multiplets by the eigenvalue $j$ of the quadratic Casimir of $SU(2)_L$ and since the range of $m$ in a multiplet is $m=-j,-j+1,\dots, j-1, j$, we see that there are only two possibilities. Single-copy superconformal primaries can either be in $j=\tfrac12$ or $j=0$ multiplets.}
\begin{equation} \label{eq.mcond}
    |m|\leq \frac12 \ ,
\end{equation}
as seen by considering the non-negativity of the norm of the $J^{+(1)}_{-1}$ descendent of this state. It was shown in \cite{Guo:2022ifr} that the states with $m=\frac12$ are forced to be chiral and as such do not lift. This leaves single-copy superconformal primaries with $m=0$. While it may seem possible that no such states exist, given the highly constraining conditions \eqref{def primary 2} and \eqref{eq.mcond}, we now give a prescription for the explicit construction of an infinite family in the free-field realisation of this theory. It should be noted that by moving away from the untwisted sector or by exciting more than one copy, these arguments are not as restrictive.

The key observation for constructing superconformal primaries of this theory is that none of the generators \eqref{eq.lcurr} of the $\mathcal{N}=4$ superconformal algebra \eqref{app com currents} are charged under the $SU(2)_1$ of the `internal' $SO(4)_I$ \eqref{eq.SO4I}. By selecting states with the \emph{minimal} conformal dimension for a given $SU(2)_1$ charge\footnote{More precisely, this charge $q_A$ is the eigenvalue of the zero mode of the third component of the generators of $SU(2)_1$. This is not to be confused with the R-symmetry charges discussed throughout for the $SU(2)_L$ of \eqref{eq.SO4E}. Since the $SU(2)_L$ is really a symmetry of the theory we can use it to label states, however, the $SU(2)_1$ is a broken symmetry and serves just to organise our discussion.} $q_A$, it is guaranteed that the positive current modes in the primary definition \eqref{def primary 2} (which all lower the dimension without changing the value of $q_A$) will annihilate the state. Trivially, the unique NS vacuum $|\NSo\rangle$ is the sole minimal dimension ($h=0$) state with charge $q_A=0$ and clearly satisfies the conditions \eqref{def primary 2}. In this section, for a superconformal primary of dimension $h$ we use the notation $|\phi^h_{j_A,q_A}\rangle$ with $q_A$ and $j_A$ being the eigenvalues of the third component and quadratic Casimir of $SU(2)_1$ respectively. The NS vacuum can then be labelled in this notation as
\begin{equation}
    |\NSo\rangle = |\phi^0_{0,0}\rangle \ .
\end{equation}
For $q_A=\frac12$ the options are to add either one bosonic or one fermionic mode of the free field realisation \eqref{eq.freeFields2} to the NS vacuum. Both options can yield the same $q_A$ charge, however, the minimal conformal dimension for that charge\footnote{A bosonic zero mode $\alpha_0$ would add charge without adding dimension, however, it would always commute through any other modes to annihilate the NS vacuum so we do not consider it here.} is from the use of a $d_{-\frac12}$. The $h=q_A=\frac12$ superconformal primary states are then
\begin{equation}
    |\phi^{\frac12}_{\frac12,\frac12}\rangle^{\alpha} = d^{\alpha+}_{-\frac12} |\NSo\rangle \ ,
\end{equation}
which are precisely the chiral/anti-chiral primaries with $h=|m|=\frac12$ discussed below \eqref{eq.mcond}. This state is the top member of an $SU(2)_1$ doublet and so the bottom member
\begin{equation}
    |\phi^{\frac12}_{\frac12,-\frac12}\rangle^{\alpha} = d^{\alpha-}_{-\frac12} |\NSo\rangle \ ,
\end{equation}
will also be a primary. In this paper we are interested in the primaries that are not chiral (since they trivially do not lift), so we will restrict ourselves to R-symmetry neutral states here. For $q_A=1$, the minimal dimension state is
\begin{equation} \label{eq.phi111}
    |\phi^{1}_{1,1}\rangle = d^{++}_{-\frac12} d^{-+}_{-\frac12}|\NSo\rangle \ ,
\end{equation}
which is R-symmetry neutral\footnote{Since $J^{\pm}_0$ annihilate all states in the triplet composed of the states in \eqref{eq.phi111} and \eqref{eq.phi110}, they have R-symmetry charges $m=j=0$ as expected.} and the top member of an $SU(2)_1$ triplet that is filled out by the states
\begin{align} \label{eq.phi110}
    |\phi^1_{1,0}\rangle &= \frac1{\sqrt{2}}\Big(d^{-+}_{-\frac12} d^{+-}_{-\frac12} + d^{--}_{-\frac12} d^{++}_{-\frac12}\Big) |\NSo\rangle \ ,\\
    |\phi^1_{1,-1}\rangle &= d^{--}_{-\frac12} d^{+-}_{-\frac12}|\NSo\rangle \ .
\end{align}
It should be noted that the singlet of the $\mathbf{2}\otimes \mathbf{2}= \mathbf{1} \oplus \mathbf{3}$ decomposition given by the state
\begin{equation}
    |\phi^1_{0,0}\rangle = \frac1{\sqrt{2}}\Big(d^{-+}_{-\frac12} d^{+-}_{-\frac12} - d^{--}_{-\frac12} d^{++}_{-\frac12}\Big) |\NSo\rangle \ ,
\end{equation}
is not a superconformal primary since $L_1|\phi^1_{0,0}\rangle\neq0$ for instance. For $q_A=\frac32$, the minimal dimension states are obtained by adding an $\alpha_{-1}$ mode on top of the $q_A=1$ states \eqref{eq.phi111} to obtain\footnote{A small subtlety here is that an $SU(2)_1$ charge of $q_A=+\frac12$ read off from an upper $A$ index is equivalent to a charge of $q_A=-\frac12$ from a lower $A$ index due to contractions being made by Levi-Civita symbols.}
\begin{equation} \label{eq.phi23/23/2}
    |\phi^2_{\frac32,\frac32}\rangle_{\Ad} = \alpha_{-\Ad,-1} d^{++}_{-\frac12} d^{-+}_{-\frac12}|\NSo\rangle \ ,
\end{equation}
which is the first state in our family of superconformal primaries that is charged under $SU(2)_2$. Both states of the $SU(2)_2$ doublet indexed by $\Ad$ are primaries. The $j_A=\frac32$ multiplet can then be filled out by the application of lowering operators of $SU(2)_1$ on the highest-weight state \eqref{eq.phi23/23/2}. It turns out that by continuing this construction for higher values of $q_A$ a pattern emerges: for $q_A\geq\frac32$ the minimal dimension states have 
\begin{equation}
    h= 2j_A-1 \quad,\quad \text{ for } j_A \geq1 \ ,
\end{equation}
and the top members of the $j_A=q_A$ representations of $SU(2)_1$ are of the schematic form
\begin{equation} \label{eq.phihqq}
    |\phi^{h}_{q_A,q_A}\rangle = \big(\alpha_{-,\Ad,-1}\big)^{2q_A-2} d^{++}_{-\frac12} d^{-+}_{-\frac12}|\NSo\rangle \ ,
\end{equation}
where these states are in the $(\mathbf{2q_A-1})\subset\mathbf{2}^{\otimes 2(q_A-1)}$ representation of $SU(2)_2$ due to the requirement of symmetry of the identical $\alpha$ modes. The total number of R-symmetry neutral superconformal primaries $N_{q_A}$ in this family for a given value of $q_A$ is then
\begin{equation} \label{eq.NqA}
    N_{q_A} = 4q_A^2-1 \quad\,\quad \text{ for } q_A=\tfrac12, 1, \tfrac32,\,\dots \ ,
\end{equation}
made by taking the state of the form \eqref{eq.phihqq} and filling out the suitable $SU(2)_1$ and $SU(2)_2$ representations. If this family of states represents the complete set of single-copy superconformal primaries, \eqref{eq.NqA} can be recast as the number of such states at a given level $h$
\begin{equation}
    N_h = h(h+2) \quad\,\quad \text{ for } h\geq1 \ .
\end{equation}
Clearly these states form an infinite family of superconformal primaries.

\subsection{Spectral flow} \label{sec.sf}

A $1$-parameter family of global automorphisms of the $\mathcal{N}=4$ superconformal algebra in $(1+1)$ dimensions labelled by an angle $\pi\eta$ is referred to as spectral flow. This map between equivalent algebras has the effect of changing the periodicity of the fermionic currents. Here we very briefly review the rules for spectral flow transformations \cite{Schwimmer:1986mf} that will be used in the main computations of Sections~\ref{sec.main} and \ref{ssec:WardIdJphi}. Under spectral flow by $\eta$ units%
\footnote{Here we have only mentioned spectral flow in the left-moving sector of the $\mathcal{N}=(4,4)$ superconformal algebra, however, both the left- and right-moving sectors can be spectral flowed independently by $\eta$ and $\bar{\eta}$ units respectively. See~\cite{Guo:2021uiu} for a smaller group of spectral flows, dubbed `partial spectral flow', which is related to the spectral flow of an $\mathcal{N}=2$ algebra contained within the $\mathcal{N}=4$ algebra.}%
, the quantum numbers of a state with dimension $h$ and $J^3_0$ charge $m$ transforms as
\begin{equation} \label{sfDims}
    h \rightarrow h' = h +m\eta +\frac{c \eta^2}{24} \quad\ ,\qquad m\rightarrow m' = m + \frac{c\eta}{12}\ ,
\end{equation}
where $c$ is the central charge of the theory. Note that if $\eta$ is an odd integer, spectral flow maps between the NS- and R-sector boundary conditions for the fermionic degrees of freedom of the theory. Whilst the transformations of the currents $T(z)$ and $J^3(z)$ are more involved due to mixing in the algebra and the central charge anomaly, the transformation properties of currents under spectral flow required in this paper are straightforward: under a spectral flow by $\eta$ units around the point $z_0$
\begin{align}
    G^{\pm}_{\!\Ad}(z) &\to (z-z_0)^{\mp\frac{\eta}{2}} G^{\pm}_{\!\Ad}(z) \ , \label{eq.GsfRule}\\
    J^{\pm}(z) &\to (z-z_0)^{\mp\eta} J^{\pm}(z) \ . \label{eq.JsfRule}
\end{align}
In Appendix C of \cite{Guo:2022ifr} it was shown that a superconformal primary field $\phi(z)$ on the plane also transforms simply via
\begin{equation} \label{sf_01}
    \phi(z)\to (z-z_0)^{-\eta m}\phi(z)\ .
\end{equation}

\subsection{Deformation away from the orbifold point} \label{ssec:deform}

It is thought that the free-orbifold theory can be deformed in the moduli space of the D1-D5 CFT towards a strong coupling regime, at which the theory would have a dual semiclassical gravity description. This deformation can be done by adding a deformation operator $D$ to the action such that
\begin{equation} \label{defor S}
    S\r S+\lambda \int d^2 z\, D(z, \bar z) \ ,
\end{equation}
where $\lambda$ is the deformation parameter and $D$ exactly marginal and so has conformal dimensions $(h, \bar h)=(1,1)$. In the theory there are $20$ exactly marginal operators to choose from; $16$ of these correspond to $T^4$ moduli, whilst the remaining $4$ are superdescendants of the order-2 twist chiral/anti-chiral primaries $\sigma^{\alpha\bar\alpha}$. The set of $T^4$ moduli are in a sense `trivial deformations', whereas the latter four break the higher-spin symmetry found at the orbifold point and move the theory towards the region with a semi-classical gravity description. The non-trivial deformation operator that is a singlet under all of the $SU(2)$ factors of the symmetry algebra discussed in \ref{ssec:symms} is given by
\begin{equation} \label{D 1/4}
    D = \frac{1}{4}\epsilon^{\dot A\dot B}\epsilon_{\alpha\beta}\epsilon_{\bar\alpha \bar\beta} \,G^{\alpha}_{\dot A, -\h} \bar G^{\bar \alpha}_{\dot B, -\h} \sigma^{\beta \bar\beta} \ .
\end{equation}
Here $G$ and $\bar{G}$ are the supercharge modes at the \emph{orbifold point}, \textit{i.e.} at $\lambda=0$. The remaining three non-trivial deformation operators form the triplet projection of $SU(2)_2\otimes SU(2)_2$. The $SO(4)_I\,$-invariant operator \eqref{D 1/4} corresponds to turning on the dual string tension. I should be noted that the chiral/anti-chiral operators $\sigma^{\alpha \bar\alpha}$ are formed by adding R-charge to the bare twist operator $\sigma_2$ using a mixture of R-symmetry currents and spin fields. In the particular case of $k=2$, this is done using only spin fields~\cite{Lunin:2001fv} to give
\begin{equation} \label{eq.sig2}
    \sigma^{\alpha\bar{\alpha}} = S_2^{\alpha} \bar{S}_2^{\bar{\alpha}} \sigma_2 \ .
\end{equation}
Spin fields change the boundary conditions of fermions around their insertion points. The operator $\sigma_2$ twists together two copies, $i$ and $j$, of the $c=6$ seed CFT, which in an $S_N$-invariant form is written as
\begin{equation}
    \sigma_2 = \sum_{\substack{i,j=1\\j<\:\!i}}^N \sigma_{(ij)} \ .
\end{equation}
We will be considering the deformation of the orbifold theory at second order in the parameter $\lambda$ since, as will be discussed below, this is the first non-trivial order when it comes to the lifting of states.

\subsection{Lift formulae} \label{sec:LiftForm}

The underlying method for the computation of second order lifts of states under the deformation discussed in Section \ref{ssec:deform} was developed in~\cite{Hampton:2018ygz} by the use of conformal perturbation theory. Here we do not give details of the derivation of this method and instead use it as a tool to learn about the behaviour of states under this deformation. At its core, this method requires the computation of the second order lift from the integrated correlator
\begin{equation}
    A^{(2)}(T) = \frac12 \Big<\phi\big(\tfrac{T}{2}\big)\Big| \int d^2w_2 D(w_2,\bar{w}_2) \int d^2w_1 D(w_1,\bar{w}_1) \Big|\phi\big(\!-\!\tfrac{T}{2}\big)\Big> \ ,
\end{equation}
where the factor of $\tfrac12$ comes from the second order perturbation of the action and the initial and final states are placed at finite Euclidean time $\tau=\frac{T}{2}$ in order to regularise the calculation. We find it convenient to break down these two insertions of the deformation operator $D(w,\wb)$ into different equivalent forms
\begin{equation} \label{eq.deformationDef}
    D(w,\bar{w}) = \epsilon^{\dot A \dot B} G^-_{\dot A,-\frac12} \bar{G}^-_{\dot B,-\frac12} \sigma^{++}_2(w,\bar w) = \epsilon^{\dot A \dot B} G^+_{\dot A,-\frac12} \bar{G}^+_{\dot B,-\frac12} \sigma^{--}_2(w,\bar w) \ ,
\end{equation}
both of which are equal to \eqref{D 1/4} but with the $SU(2)$ singlet structure obscured.
\begin{figure}[h]
    \centering
    \includegraphics[scale =0.55]{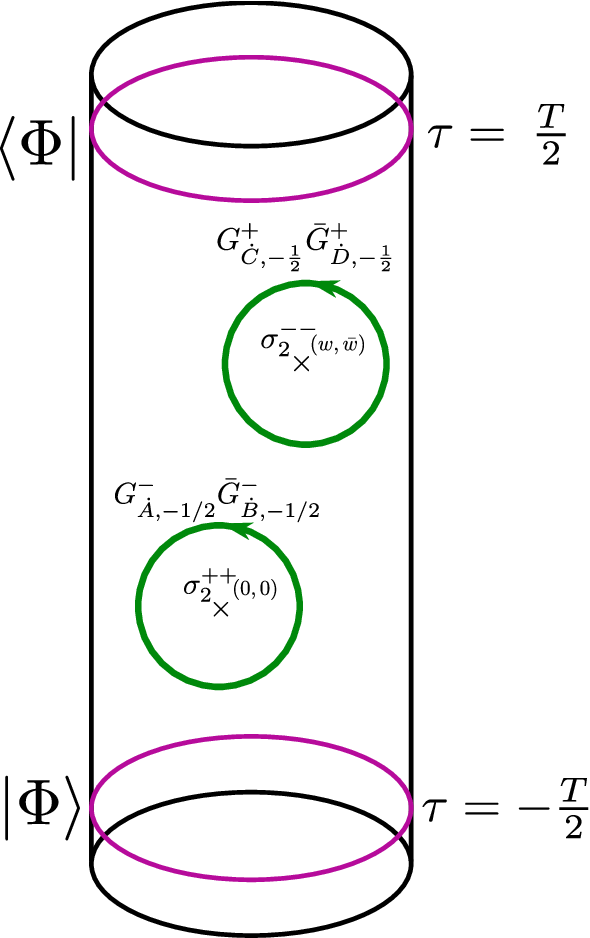}
    \caption{This figure depicts the correlator\eqref{eq.Xdef}. In purple, we see the initial $\ket{\Phi}$ and final states $\bra{\Phi}$ located at $\tau =-T/2$ and $\tau = T/2$ respectively. The green contours are the supersymmetry current modes from the two deformation operators placed at $(0,0)$ and $(w,\bar{w})$.}
    \label{fig:cylinderstate}
\end{figure}
Fixing the insertion of the first deformation operator to the origin and considering the large $T$ limit (since $T\to\infty$ at the end of the calculation anyway) we can write
\begin{equation}\label{amplitude}
    A^{(2)}(T) \approx \frac12 (2\pi T) e^{-E^{(0)} T} \big<\Phi\big| \int d^2w_2 D(w_2,\bar{w}_2) \,D(0,0) \big|\Phi\big> \ .
\end{equation}
From conformal perturbation theory it was shown in \cite{Hampton:2018ygz} that the second order lift in energy (the first order lift in energy is always vanishing by symmetry arguments) can be obtained from the coefficient of $-Te^{-E^{(0)}T}$ in the correlator $A^{(2)}(T\to\infty)$ and so we write
\begin{equation} \label{eq.lift}
    E^{(2)}\big(|\Phi\rangle\big) = -\pi \lambda^2 \lim_{T\to\infty} \frac{X(T)}{\langle\Phi|\Phi\rangle} \ ,
\end{equation}
where the integrated amplitude $X(T)$ is given by
\begin{equation}\label{eq.Xdef}
    X(T) \equiv \epsilon^{\dot C\dot D}\epsilon^{\dot A \dot B} \int d^2w\, \big<\Phi\big| \big(G^+_{\dot C,-\frac12}\bar{G}^+_{\dot D, -\frac12}\sigma^{--}_2\big)(w,\bar w)\, \big(G^-_{\dot A,-\frac12}\bar{G}^-_{\dot B, -\frac12}\sigma^{++}_2\big)(0,0)  \big|\Phi\big> \ ,
\end{equation}
with the initial and final states are located at $\tau =-\frac{T}{2}$ and $\tau = \frac{T}{2}$ respectively. This cylinder amplitude is depicted in Figure~\ref{fig:cylinderstate}. Since the right-moving part of the external states we consider is always the unique NS vacuum on all copies, the right-moving part of the correlator in \eqref{eq.Xdef} will be determined simply by the universal $D\times D$ OPE, thus contributing a factor independent of the initial and final states of the form
\begin{equation} \label{eq.RmovingCorr}
    \langle\NSo|\big(\bar{G}^{+}_{\Dd,-\frac12}\bar{\sigma}^-_2\big)(\wb)\, \big(\bar{G}^{-}_{\Bd,-\frac12}\bar{\sigma}^{+}_2\big)(0) |\NSo\rangle  = \frac{-\epsilon_{\dot D\dot B}}{4\sinh^2(\frac{\bar w}{2})} \ .
\end{equation}
This right-moving amplitude was derived in \cite{Hampton:2018ygz}. It is then convenient to write the integrated amplitude $X(T)$ as
\begin{align} \label{eq.XT}
    X(T) &= \epsilon^{\dot C\dot D}\epsilon^{\dot A \dot B} \int d^2w\, A(w,0)\epsilon_{\Ad\Cd}  \bigg(\frac{-\epsilon_{\dot D\dot B}}{4\sinh^2(\frac{\bar w}{2})}\bigg) \nonumber\\
    &= -\frac12\epsilon^{\dot C\dot D}\epsilon^{\dot A \dot B}\epsilon_{\dot D\dot B}\epsilon_{\Cd\Ad} \int d^2w\, A(w,0)\, \partial_{\wb}\Big(\coth(\tfrac{\bar w}{2})\Big) \nonumber\\
    &= \frac{i}{2} \int_C dw\, A(w,0) \coth(\tfrac{\bar w}{2}) \nonumber\\
    &\equiv I_{C_1} + I_{C_2} + I_{C_3} \ ,
\end{align}
where $A(w_2,w_1)$ is the non-integrated left-moving amplitude
\begin{equation}
    A(w_2,w_1) \equiv \big<\Phi\big| \Big(G^+_{-,-\frac12}\sigma^-_2\Big)(w_2)\Big(G^-_{+,-\frac12}\sigma^+_2\Big)(w_1)\big|\Phi\big> \ .
\end{equation}
The contour integrals $I_{C_1}$, $I_{C_2}$, $I_{C_3}$ in \eqref{eq.XT} are given by
\begin{subequations} \label{eq.IC123defs}
    \begin{align}
        I_{C_1}(T) &\equiv - \frac12\int^{2\pi}_0 \!d\sigma\, A(w,0) \coth(\tfrac{\bar w}{2}) \ ,\label{eq.IC1} \\
        I_{C_2}(T) &\equiv \frac12\int^{2\pi}_0 \!d\sigma\, A(w,0) \coth(\tfrac{\bar w}{2}) \ ,\label{eq.IC2} \\
        I_{C_3} &\equiv -\frac{i}{2} \oint_{|w|=\epsilon} \!dw\, A(w,0) \coth(\tfrac{\bar w}{2}) \ ,\label{eq.IC3}
    \end{align}
\end{subequations}
where the contours $C_1$, $C_2$, $C_3$ on the cylinder are at $\tau=\frac{T}{2},-\frac{T}{2}$ and around $|w|=\epsilon$ respectively and the complex coordinates on the cylinder have been written as $w=\tau+i\sigma$ and $\wb=\tau-i\sigma$. These contours are depicted in Figure~\ref{fig:ICs}.
\begin{figure}[H]
    \centering
    \includegraphics[scale =0.53]{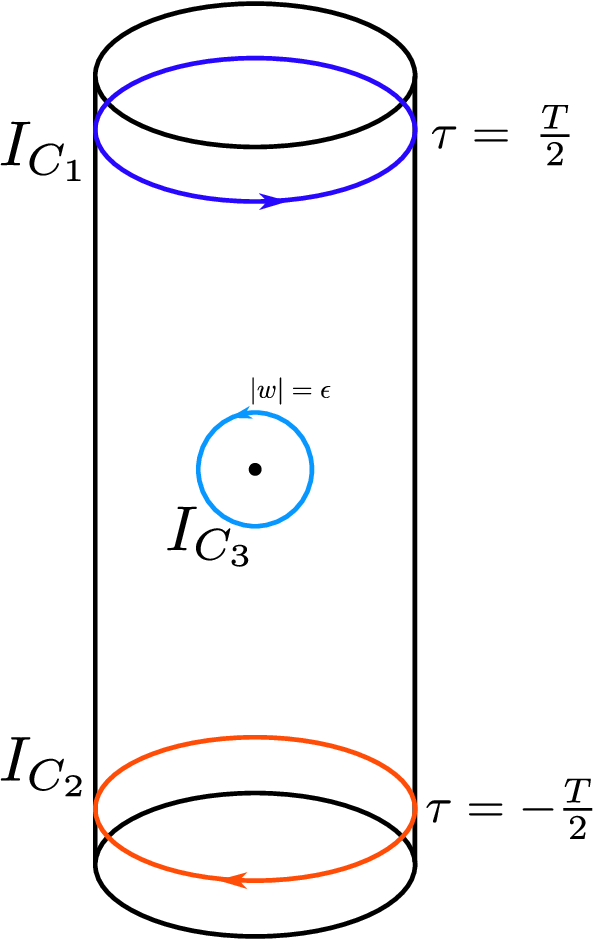}
    \caption{Here the three boundary contour integrals $I_{C_1}, I_{C_2}$ and $I_{C_3}$ as defined in equations \eqref{eq.IC123defs} are shown on the $w$-cylinder. The sum of these contour integrals gives the integrated amplitude in \eqref{eq.XT} required in the lift \eqref{eq.lift}.}\label{fig:ICs}
\end{figure}
In \eqref{eq.IC3} the cutoff $\epsilon$ should be taken to $0$ at the end of the calculation. The computation thus boils down to evaluating the non-integrated left-moving amplitude $A(w,0)$. We note that, as found in \cite{Hampton:2018ygz}, the integral $I_{C_3}$ for the states we consider will contribute only a universal divergent piece $\sim \frac1{\epsilon}$ which can be removed by a counterterm. Thus the integral $I_{C_3}$ does not contribute to the lift \eqref{eq.lift} for us.

The computation of the lift via the method described above only sees a single pair of copies at any given time\footnote{This is because of the order-2 twist fields $\sigma_2$ in the deformation operators $D$ act on a pair of copies of the seed theory and the lift being related to an expectation value. This forces the second twist operator to act on exactly the same pair of copies as the first.} and as such throughout the calculation it is sufficient to consider only a specific ordered pair of copies, with one copy being excited and one copy in the NS-NS vacuum. As derived in Section (3.1) of \cite{Guo:2022ifr} by combinatoric arguments, the final $N$-dependent lifting result is then
\begin{equation} \label{eq.GeneralN}
    E^{(2)}_N\!\big(|\Phi\rangle\big) = 2(N-1) E^{(2)}\!\big(|\Phi\rangle\big) \ ,
\end{equation}
where $E^{(2)}(|\Phi\rangle)$ is the lift that we work with and that will be quoted in this paper.

\section{Lifting of descendent states via series method} \label{sec.main}

In this section we will consider the lifting of descendants of superconformal primaries on a single copy. More precisely, let the state $|\phi\rangle$ be a superconformal primary on one copy of the seed $c=6$ CFT, then $|\phi\rangle$ satisfies the conditions \eqref{def primary 2}. Descendants of this primary are then found by acting with negative modes of the $\mathcal{N}=4$ superconformal algebra given in Appendix~\ref{app.comms}; namely acting with the current modes
\begin{equation}
    \Big\{ L_{-n}\ ,\ \ J^a_{-n} \ ,\ \ G^{\alpha}_{\!\Ad,-s}\Big\} \ ,
\end{equation}
for $n>0$ and $s\geq \frac12$. In actual fact, it is enough to study the lifting of descendants formed from acting with $J$ and $G$ modes. Specifically we consider the following states
\begin{align}
    |G\phi\rangle^{\alpha}_{\Ad(h,s)} &\equiv \frac1{\sqrt{2}} \Big( G^{\alpha(1)}_{\Ad,-s}|\phi\rangle^{(1)}|\NSo\rangle^{(2)} + G^{\alpha(2)}_{\Ad,-s}|\NSo\rangle^{(1)} |\phi\rangle^{(2)} \Big) \ ,\label{eq.Gphistatedef}\\
    |J\phi\rangle^{+}_{(h,n)} &\equiv \frac1{\sqrt{2}} \Big( J^{+(1)}_{-n}|\phi\rangle^{(1)}|\NSo\rangle^{(2)} + J^{+(2)}_{-n} |\NSo\rangle^{(1)} |\phi\rangle^{(2)}\Big) \label{eq.Jphistatedef}\ ,
\end{align}
where $\phi$ has conformal dimension $h$ and $J^3_0$ eigenvalue $j=0$ and the factor of $\frac{1}{\sqrt{2}}$ is a copy normalisation. The respective bra states are defined as
\begin{align}
    {}^{\quad \ \ \,\beta}_{(h,s)\Bd}\langle G\phi| &\equiv \frac1{\sqrt{2}}\bigg( \!\NSob{2}\,{}^{\,(1)}\langle\phi|G^{\beta(1)}_{\Bd,s} + {}^{(2)}\langle\phi|\NSob{1} G^{\beta(2)}_{\Bd,s} \bigg) \ , \label{eq.phiGstatedef}\\
    {}^{\quad\,-}_{(h,n)}\langle J\phi| &\equiv \frac1{\sqrt{2}}\bigg( \!\NSob{2}\,{}^{\,(1)}\langle\phi|J^{-(1)}_{n} + {}^{(2)}\langle\phi|\NSob{1}J^{-(2)}_{n} \bigg) \ , \label{eq.phiJstatedef}
\end{align}
where the Hermitian conjugate state of $|G\phi\rangle^{\alpha}_{\Ad(h,s)}$ should have opposite quantum numbers (see \ref{app.Herm} for our conventions). Whilst these states are not currently normalised on an individual copy, this is taken into account in the lift formula \eqref{eq.lift}. Since the method of lifting used here is for D1-D5-P states, the right-moving part of the states is necessarily the NS vacuum on all copies. The lift of single-copy superconformal primaries was previously computed in closed form in \cite{Guo:2022ifr}.

\subsection{Lifting of $|G\phi\rangle^{\alpha}_{\!\Ad(h,s)}$} \label{ssec:GphiLift}

In this section we compute the lift of the superdescendant $|G\phi\rangle^{\alpha}_{\Ad(h,s)}$. From Section~\ref{sec:LiftForm} the key quantity we need to compute is the left-moving amplitude
\begin{equation} \label{eq.A11Gphi}
    A^{(1)(1)}_{h,s}(w_2,w_1) \equiv \langle\phi| G^{\beta}_{\Bd,s}\,\big(G^+_{-,-\frac12}\sigma^-_2\big)(w_2) \big(G^-_{+,-\frac12}\sigma^+_2\big)(w_1)\,G^{\alpha}_{\!\Ad,-s}|\phi\rangle \ ,
\end{equation}
where we have taken the descendant excitation to be on a definite copy, namely copy 1, for both the initial and final state. The general method of computing this amplitude will be similar to that employed in \cite{Hughes:2023apl} with some modifications to account for the fact that we do not specify any explicit form of the state $|\phi\rangle$. The steps are as follows:
\begin{enumerate}[start=1,
    labelindent=\parindent,
    leftmargin =1.1\parindent,
    label=(\arabic*)]
    
    \item \label{Series1}\vspace{5pt}
    The amplitude $A^{(1)(1)}_{h,s}(w_2,w_1)$ is on a doubly-covered cylinder due to the presence of twist operators $\sigma_2$ joining and then splitting the two copies of the untwisted initial and final states. This effect of the twist operators can be resolved by mapping the amplitude to the covering space. As shown in~\cite{Lunin:2000yv} and used in the context of lifting of states in~\cite{Hampton:2018ygz,Guo:2022ifr,Hughes:2023apl}, the covering space map $z\to t$ for the case of two order-2 twist operators is given by 
    \begin{equation} \label{eq.ztotmap}
        z(t)=\frac{(t+a)(t+b)}{t} \ ,
    \end{equation}
    where $z=e^w$ is the doubly-covered plane. The twist operator insertions on the plane at $z_1=e^{w_1}$ and $z_2=e^{w_2}$ are mapped to $t_1=-\sqrt{ab}$ and $t_2=\sqrt{ab}$ on the covering space. In our conventions the points $z=0$ and $z=\infty$ on the first sheet of the doubly-covered plane map to $t=-a$ and $t=\infty$ on the covering space. Under this map the initial state transforms as
    \begin{align} \label{eq.GinitialStateTrans}
        G^{\alpha}_{\!\Ad,-s}|\phi\rangle &= \oint_{\tau\to-\infty} \frac{dw_3}{2\pi i}\, e^{-sw_3} \,G^{\alpha}_{\!\Ad}(w_3) |\phi\rangle \nonumber\\
        &= \lim_{\tt\to-a} \zt^h \bigg(\frac{d\tt}{d\zt}\bigg)^{\!h} \oint_{\tt}\,\frac{dt_3}{2\pi i}\, z_3^{-s+\frac12}\bigg(\frac{dt_3}{dz_3}\bigg)^{\!\frac12} G^{\alpha}_{\!\Ad}(t_3)\phi(\tt\,)|\NSo\rangle \ ,
    \end{align}
    and likewise the final state transforms as
    \begin{align} \label{eq.GfinalStateTrans}
        \langle\phi|G^{\beta}_{\Bd,s} &= -\oint_{\tau\to\infty} \frac{dw_4}{2\pi i}\, e^{sw_4} \langle\phi|G^{\beta}_{\Bd}(w_4) \nonumber\\
        &= -\lim_{t'\to\infty} z'^h \bigg(\frac{dt'}{dz'}\bigg)^{\!h} \oint_{t'}\,\frac{dt_4}{2\pi i}\, z_4^{s+\frac12}\bigg(\frac{dt_4}{dz_4}\bigg)^{\!\frac12} \langle\NSo|\phi(t')G^{\beta}_{\Bd}(t_4) \ ,
    \end{align}
    where the negative sign accounts for the direction of the contour at $\infty$. Since the covering space map \eqref{eq.ztotmap} is not conformal there will be an associated Liouville factor; this along with various conformal factors from the twist operators and spin fields are wrapped up into a ``base amplitude" \cite{Lunin:2000yv,Lunin:2001fv,Hampton:2018ygz}
    \begin{equation} \label{eq.baseamp}
        f_1 = \frac{a-b}{4\sqrt{ab}} \ ,
    \end{equation}
    that is independent of the external states.
    \item \label{Series2}
    While this map resolves the geometric effect of the two twist operators, we are left with insertions of the spin fields $S^-(t_2)$ and $S^+(t_1)$. These can in turn be removed by performing a spectral flow of the amplitude by $\eta=-1$ and $\eta=+1$ units around $t_1$ and $t_2$ respectively. Under these spectral flows, the state at $t=-a$ transforms as
    \begin{equation}
        G^{\alpha}_{\Ad}(t_3)\phi(\tt\,)|\NSo\rangle \rightarrow \bigg(\frac{t_3-t_1}{t_3-t_2}\bigg)^{\!q_\alpha} G^{\alpha}_{\Ad}(t_3)\phi(\tt\,)|\NSo\rangle \ ,
    \end{equation}
    where $q_\alpha$ is the $J^3_0$ charge of $G^\alpha_{\Ad,-s}$ and the state at $t=\infty$ transforms as
    \begin{equation}
        \langle\NSo|\phi(t')G^{\beta}_{\Bd}(t_4) \rightarrow \bigg(\frac{t_4-t_1}{t_4-t_2}\bigg)^{\!q_\beta} \langle\NSo|\phi(\tt\,)G^{\beta}_{\Bd}(t_4) \ .
    \end{equation}
    While the $G$ modes from the deformation operators do transform under the above maps, their transformations do not depend on the external states under consideration and so they contribute a universal factor of (see Section (4.3) of \cite{Hampton:2018ygz})
    \begin{equation} \label{eq.f2}
        f_2 = ab(a-b) \ .
    \end{equation}
    \item \label{Series3}
    Following steps \ref{Series1} and \ref{Series2} the amplitude \eqref{eq.A11Gphi} is now given by
    \begin{align} \label{eq.A11hsintegrals}
        A^{(1)(1)}_{h,s} = -f_1f_2f_3 \oint_{\infty}\frac{dt_4}{2\pi i}\oint_{-a}\frac{dt_3}{2\pi i}\, \frac{z_4^{s+\frac12}}{z_3^{s-\frac12}} \bigg(\frac{dt_3}{dz_3}\bigg)^{\!\frac12}\bigg(\frac{dt_4}{dz_4}\bigg)^{\!\frac12} \bigg(\frac{t_3-t_1}{t_3-t_2}\bigg)^{\!q_\alpha} \bigg(\frac{t_4-t_1}{t_4-t_2}\bigg)^{\!q_\beta} \widetilde{C}^{\beta\alpha}_{\!\Bd\Ad}(t_4,t_3) \ ,
    \end{align}
    where
    \begin{equation} \label{eq.f3}
        f_3 \equiv \lim_{\substack{\tt\to-a\\t'\to\infty}} \zt^h z'^h \bigg(\frac{d\tt}{d\zt}\bigg)^{\!h} \bigg(\frac{dt'}{dz'}\bigg)^{\!h} = \bigg(\frac{a}{a-b}\bigg)^{\!h} \ ,
    \end{equation}
    and $\widetilde{C}^{\beta\alpha}_{\Bd\Ad}$ is the correlator
    \begin{equation}
        \widetilde{C}^{\beta\alpha}_{\!\Bd\Ad}(t_4,t_3) \equiv \langle\phi|G^{\beta}_{\Bd}(t_4) G^+_-(t_2) G^-_+(t_1) G^{\alpha}_{\Ad}(t_3)|\phi\rangle \ .
    \end{equation}
    \item \label{Series4}
    In \cite{Hughes:2023apl}, the equivalent correlator on the $t$-plane was computed by writing all field insertions in terms of the bosons and fermions of the free-field realisation of the orbifold CFT and then using Wick contractions. Here, however, it is not convenient to utilise Wick contractions of free fields or OPEs of currents since the  $G\times \phi$ OPE will contain the superpartner of $\phi$, which is not known generally. Instead, we expand each of the fields in terms of modes around $t=-a$ using
    \begin{equation} \label{eq.GmodeExp}
        G^{\alpha}_{\Ad}(t) = \sum_{s\in \mathbb{Z}+\frac12}\, (t+a)^{-s-\frac32} G^{\alpha}_{\Ad,s} \ ,
    \end{equation}
    for the fields inserted at $t_1$ and $t_2$. The remaining $G$ fields are on contours around $-a$ and $\infty$ respectively and their mode expansions are more complicated. For these, the integrands of the contour integrals in \eqref{eq.A11hsintegrals} have to be expanded in powers of $t_3+a$ and $t_4+a$ yielding
    \begin{align} \label{eq.A11hsmodes}
        A^{(1)(1)}_{h,s} = f_1f_2f_3 \sum_{k_4=0}^{s-1/2}\sum_{k_3=0}^{s-1/2}\sum_{s_1,s_2} (t_2+a)^{-s_2-\frac32}(t_1+a)^{-s_1-\frac32} \,\mathcal{K}^{(G)}_{k_3}\widetilde{\mathcal{K}}^{(G)}_{k_4} \,C^{\beta\alpha}_{\!\Bd\Ad} \ ,
    \end{align}
    where $C^{\beta\alpha}_{\!\Bd\Ad}$ is the mode correlator
    \begin{equation} \label{eq.phiGGGGphidef}
        C^{\beta\alpha}_{\!\Bd\Ad} \equiv \langle\phi|G^{\beta}_{\Bd,s-k_4} G^+_{-,s_2} G^-_{+,s_1} G^{\alpha}_{\Ad,-s+k_3}|\phi\rangle \ .
    \end{equation}
    The expansion coefficients in \eqref{eq.A11hsmodes} can be written as
    \begin{subequations}
    \begin{align}
        \mathcal{K}^{(G)}_{k_3} &= \sum_{k_1=0}^{k_3}\sum_{k_2=0}^{k_3-k_1} \frac{(-1)^{k_1}a^{n-k_1}(a-b)^{1-n-k_2}}{\big(a+\sqrt{ab}\big)^{1+k_3-k_1-k_2}} \,{}^nC_{k_1}\, {}^{k_2+n-2}C_{k_2} \label{eq.Ck3}\ ,\\
        \widetilde{\mathcal{K}}^{(G)}_{k_4} &= \sum_{\tilde{k}_1=0}^{n}\sum_{\tilde{k}_2=0}^{k_4-\tilde{k}_1} (-1)^{\tilde{k}_1} a^{\tilde{k}_2} (a-b)^{\tilde{k}_1} \Big(a-\sqrt{ab}\Big)^{k_4-\tilde{k}_1-\tilde{k}_2}\; {}^{n}C_{\tilde{k}_1}\, {}^{n-2+\tilde{k}_2}C_{\tilde{k}_2} \label{eq.Ck4} \ ,
    \end{align}
    \end{subequations}
    where we use the integer $n=s+\frac12$.
    \item \label{Series5}
    The mode correlator \eqref{eq.phiGGGGphi2} can be calculated using the mode algebra of the $\mathcal{N}=4$ superconformal algebra \eqref{app com currents} and the definition of primaries \eqref{def primary 2}. We give the expression for $C^{\beta\alpha}_{\!\Bd\Ad}$ in Appendix~\ref{app.2}. 
    \item \label{Series6}
    Given the result of the mode correlator $C^{\beta\alpha}_{\!\Bd\Ad}$ the first step is to resum the $s_1$ and $s_2$ series in \eqref{eq.A11hsmodes} that came from the expansion of the $G$ fields of the deformation operators. Each term in the mode correlator contains either one or two Kronecker delta functions which localise one or both of these series. These series only have to be performed once for all mode numbers $s$ of the initial state.
    \item \label{Series7}
    The next step is to resum the remaining finite sums in \eqref{eq.A11hsmodes} which depend on the choice of initial state. The number of terms in each of these sums scales linearly with the level of the initial state $G$ mode so has to be done per data point.
    \item \label{Series8}
    The explicit dependence of the amplitude on $w_1$ and $w_2$ can then be recovered by using the relations
    \begin{equation} \label{abDef}
        a = e^S \cosh^2\!\Big(\frac{\Delta{w}}4\Big) \quad,\quad b = e^S \sinh^2\!\Big(\frac{\Delta{w}}4\Big) \ ,        
    \end{equation}
    where we define
    \begin{equation} \label{swDef}
        S \equiv \h(w_1+w_2) \quad,\quad \Delta w\equiv w_2-w_1 \ .
    \end{equation}
    This yields the amplitude $A^{(1)(1)}_{h,s}(w_2,w_1)$ where both the initial and final state have descendent excitations placed on copy 1. The left-moving copy symmetric amplitude required for the lift of the state \eqref{eq.Gphistatedef} is then given by
    \begin{align} \label{eq.GphiSymmAmp}
        A_{h,s}(w,0) &\equiv {}^{\quad\ \ \,\beta}_{(h,s)\Bd}\langle G\phi| \big(G^+_{-,-\frac12}\sigma^-_2\big)(w) \big(G^-_{+,-\frac12}\sigma^+_2\big)(0)|G\phi\rangle^{\alpha}_{\Ad(h,s)} \nonumber\\
        &= \frac12 \sum_{i,j=1}^2 A^{(j)(i)}_{h,s}(w,0) \nonumber\\
        &= A^{(1)(1)}_{h,s}(w,0) + A^{(2)(1)}_{h,s}(w,0) \ ,
    \end{align}
    where we have used that $A^{(2)(2)}=A^{(1)(1)}$ and $A^{(1)(2)}=A^{(2)(1)}$ on general symmetry grounds. The amplitude $A^{(2)(1)}$ can be obtained from $A^{(1)(1)}$ via the continuation
    \begin{equation} \label{eq.A21def}
        A^{(2)(1)}_{h,s}(w,0) = A^{(1)(1)}_{h,s}(w+2\pi i,0) \ ,
    \end{equation}
    which has the effect of interchanging the role of the two copies that the upper $\sigma_2$ twists.
    \item \label{Series9}
    The norm of our descendant state can be found from the correlator
    \begin{equation} \label{eq.GphiNorm}
        {}^{\quad \ \ \,\beta}_{(h,s)\Bd}\langle G\phi|G\phi\rangle^{\alpha}_{\Ad(h,s)} = \ep_{\Bd\Ad}\ep^{\beta\alpha}\Big(s^2-\frac14+h\Big) \ ,
    \end{equation}
    where the quantum numbers of the conjugate state need to be opposite to the ket state and appropriate negative signs added as per the conjugation conventions of Appendix~\ref{app.Herm}. This definition of the Hermitian conjugate state should also be enforced on the final state of the amplitude \eqref{eq.GphiSymmAmp} in order for it to be related to the lift of a particular state. The lift is then given by
    \begin{equation} \label{eq.GphiLiftform}
        E^{(2)}\Big(|G\phi\rangle^{\alpha}_{\Ad(h,s)}\Big) = \frac{\pi\lambda^2}{\big||G\phi\rangle^{\alpha}_{\Ad(h,s)}\big|^2} \lim_{T\to\infty} \int_{0}^{2\pi}\!d\sigma \, A_{h,s}(w,0) \coth\!\big(\tfrac{\wb}{2}\big) \ ,
    \end{equation}
    where $w= \frac{T}{2} + i \sigma$ and $\wb = \frac{T}{2} - i\sigma$ with $T\to \infty$ on this contour as given in \eqref{eq.IC1}. Note that of the three integrals in \eqref{eq.IC123defs} it was argued in Section~\ref{sec:LiftForm} that $I_{C_3}$ gives a vanishing contribution to the lift and for the states considered here we find that $\lim_{T\to\infty}I_{C_1}(T)=\lim_{T\to\infty}I_{C_2}(T)$ and so \eqref{eq.GphiLiftform} contains only one integral.
\end{enumerate}
A shallow matrix of lifts for the particular states $|G\phi\rangle^{+}_{+(h,s)}$ is given by:\\
\begin{align} \label{eq.Gphimat}
\tikz[remember picture, baseline=(mat.center)]{\node[inner sep=0](mat){$
\begin{pmatrix}
 \\[-5pt]
 \frac{\pi ^2}{2} & \frac{13 \pi ^2}{24} & \frac{1045 \pi ^2}{1792} & \frac{2047 \pi ^2}{3328} & \frac{880799 \pi ^2}{1376256} & \frac{167659 \pi ^2}{253952} & \frac{61061485 \pi ^2}{90177536} \\\\
 \frac{3 \pi ^2}{4} & \frac{201 \pi ^2}{256} & \frac{837 \pi ^2}{1024} & \frac{48471 \pi ^2}{57344} & \frac{1252047 \pi ^2}{1441792} & \frac{14893593 \pi ^2}{16777216} & \frac{10429173 \pi ^2}{11534336} \\\\
 \frac{15 \pi ^2}{16} & \frac{309 \pi ^2}{320} & \frac{2035 \pi ^2}{2048} & \frac{41721 \pi ^2}{40960} & \frac{12543015 \pi ^2}{12058624} & \frac{6105615 \pi ^2}{5767168} & \frac{90151127 \pi ^2}{83886080} \\\\
 \frac{35 \pi ^2}{32} & \frac{1715 \pi ^2}{1536} & \frac{9357 \pi ^2}{8192} & \frac{305375 \pi ^2}{262144} & \frac{3728135 \pi ^2}{3145728} & \frac{5045215 \pi ^2}{4194304} & \frac{3761238495 \pi ^2}{3087007744} \\\\
 \frac{315 \pi ^2}{256} & \frac{8955 \pi ^2}{7168} & \frac{917865 \pi ^2}{720896} & \frac{1441965 \pi ^2}{1114112} & \frac{11016927 \pi ^2}{8388608} & \frac{19529115 \pi ^2}{14680064} & \frac{67896576405 \pi ^2}{50465865728} \\\\
 \frac{693 \pi ^2}{512} & \frac{179487 \pi ^2}{131072} & \frac{22803 \pi ^2}{16384} & \frac{1480157 \pi ^2}{1048576} & \frac{311805369 \pi ^2}{218103808} & \frac{1552560779 \pi ^2}{1073741824} & \frac{12546253545 \pi ^2}{8589934592} \\\\
 \frac{3003 \pi ^2}{2048} & \frac{109109 \pi ^2}{73728} & \frac{786905 \pi ^2}{524288} & \frac{30277807 \pi ^2}{19922944} & \frac{928262027 \pi ^2}{603979776} & \frac{15420618983 \pi ^2}{9932111872} & \frac{13458915613 \pi ^2}{8589934592} \\\\
 \frac{6435 \pi ^2}{4096} & \frac{518661 \pi ^2}{327680} & \frac{5881395 \pi ^2}{3670016} & \frac{33986913 \pi ^2}{20971520} & \frac{3076088625 \pi ^2}{1879048192} & \frac{33706109085 \pi ^2}{20401094656} & \frac{357769837347 \pi ^2}{214748364800} \\\\
 \frac{109395 \pi ^2}{65536} & \frac{4843215 \pi ^2}{2883584} & \frac{14244759 \pi ^2}{8388608} & \frac{100735965 \pi ^2}{58720256} & \frac{215642435175 \pi ^2}{124554051584} & \frac{6092689185 \pi ^2}{3489660928} & \frac{4110711998445 \pi ^2}{2336462209024} \\\\
\end{pmatrix}
$};}
\begin{tikzpicture}[overlay,remember picture]
\draw[blue,thick,->] node[anchor=south west] (nn1) at (mat.north west)
{increasing $s$} (nn1.east) -- (nn1-|mat.north east) ;
\draw[red,thick,->] node[anchor=north west,align=center, inner xsep=0pt] (nn2) at 
(mat.north east)
{\rotatebox{-90}{increasing $h$}} (nn2.south) -- (nn2.south|-mat.south) 
;
\end{tikzpicture}
\end{align}

\begin{table}
\centering
\small
\begin{tabular}{|c||c|c||c|c|}
\hline
&&&&\\[-1em]
$s+\frac{1}{2}$ & $E^{(2)}\big(|G\phi\rangle^{+}_{+(1,s)}\big)/\lambda^2$ & $h$ & $E^{(2)}\big(|G\phi\rangle^{+}_{+(h,\frac{1}{2})}\big)/\lambda^2$  & $E^{(2)}\big(|G\phi\rangle^{+}_{+(h,h-\frac12)}\big)/\lambda^2$ \\
\hline&&&&\\[-1em]
1 & $\frac{\pi^2}{2}$ & 1 & $\frac{\pi^2}{2}$ & $\frac{\pi^2}{2}$ \\[3pt]
\hline
&&&&\\[-1em]
2 & $\frac{13\pi^2}{24}$ & 2 & $\frac{3\pi^2}{4}$  & $\frac{201\pi^2}{256}$ \\[3pt]
\hline
&&&&\\[-1em]
3 & $\frac{1045\pi^2}{1792}$ & 3 & $\frac{15\pi^2}{16}$ & $\frac{2035\pi^2}{2048}$ \\[3pt]
\hline
&&&&\\[-1em]
4 & $\frac{2047\pi^2}{3328}$ & 4 & $\frac{35\pi^2}{32}$  & $\frac{305375\pi^2}{262144}$ \\[3pt]
\hline
&&&&\\[-1em]
5 & $\frac{880799\pi^2}{1376256}$ & 5 & $\frac{315\pi^2}{256}$  & $\frac{11016927\pi^2}{8388608}$ \\[3pt]
\hline
&&&&\\[-1em]
6 & $\frac{167659\pi^2}{253952}$ & 6 & $\frac{693\pi^2}{512}$ &  $\frac{1552560779\pi^2}{1073741824}$ \\[3pt]
\hline
&&&&\\[-1em]
7 & $\frac{61061485\pi^2}{90177536}$ & 7 & $\frac{3003\pi^2}{2048}$ & $\frac{13458915613\pi^2}{8589934592}$ \\[3pt]
\hline
&&&&\\[-1em]
8 & $\frac{82679905\pi^2}{119537664}$ & 8 & $\frac{6435\pi^2}{4096}$ & $\frac{1845612417915\pi^2}{1099511627776}$ \\[3pt]
\hline
&&&&\\[-1em]
9 & $\frac{220850992003\pi^2}{313532612608}$ & 9 & $\frac{109395\pi^2}{65536}$  & $\frac{250928035737055\pi^2}{140737488355328}$ \\[3pt]
\hline
&&&&\\[-1em]
10 & $\frac{139865669375\pi^2}{195421011968}$ & 10 & $\frac{230945\pi^2}{131072}$ & $\frac{8472249907603625\pi^2}{4503599627370496}$ \\[3pt]
\hline
&&&&\\[-1em]
11 & $\frac{11074140097459\pi^2}{15255723835392}$ & 11 & $\frac{969969\pi^2}{524288}$ & $\frac{71132256003286443\pi^2}{36028797018963968}$ \\[3pt]
\hline
&&&&\\[-1em]
12 & $\frac{13438244584645\pi^2}{18279380811776}$ & 12 & $\frac{2028117\pi^2}{1048576}$ & $\frac{38055140340356147281\pi^2}{18446744073709551616}$ \\[3pt]
\hline
&&&&\\[-1em]
13 & $\frac{4107842603142835\pi^2}{5523946417946624}$ & 13 & $\frac{16900975\pi^2}{8388608}$ &  $\frac{1267812416041182611425\pi^2}{590295810358705651712}$ \\[3pt]
\hline
&&&&\\[-1em]
14 & $\frac{1209633084998329\pi^2}{1609685023064064}$ & 14 & $\frac{35102025\pi^2}{16777216}$ &  $\frac{84214113560913762070065\pi^2}{37778931862957161709568}$ \\[3pt]
\hline
&&&&\\[-1em]
15 & $\frac{180249923867363485\pi^2}{237564880343793664}$ & 15 & $\frac{145422675\pi^2}{67108864}$ & $\frac{697387735039463877232775\pi^2}{302231454903657293676544}$ \\[3pt]
\hline
&&&&\\[-1em]
16 & $\frac{207717974301142693\pi^2}{271341877549072384}$ & 16 & $\frac{300540195\pi^2}{134217728}$ &  $\frac{184381092183290428038099635\pi^2}{77371252455336267181195264}$ \\[3pt]
\hline
&&&&\\[-1em]
17 & $\frac{3887143893143114780555\pi^2}{5035961132122707591168}$ & 17 & $\frac{9917826435\pi^2}{4294967296}$ & $\frac{97300623587685344239590415359\pi^2}{39614081257132168796771975168}$ \\[3pt]
\hline
&&&&\\[-1em]
18 & $\frac{2202563353701091080685\pi^2}{2831575215314416173056}$ & 18 & $\frac{20419054425\pi^2}{8589934592}$ &  $\frac{3203499112728575370103597789355\pi^2}{1267650600228229401496703205376}$ \\[3pt]
\hline
&&&&\\[-1em]
19 & $\frac{158637441680129531942191\pi^2}{202471462953036038537216}$ & 19 & $\frac{83945001525\pi^2}{34359738368}$ &  $\frac{26326239008093207523832307852065\pi^2}{10141204801825835211973625643008}$ \\[3pt]
\hline
&&&&\\[-1em]
20 & $\frac{177415422817791031035205\pi^2}{224902703746666853302272}$ & 20 & $\frac{172308161025\pi^2}{68719476736}$ &  $\frac{3456688488715760869871367416311599\pi^2}{1298074214633706907132624082305024}$ \\[3pt]
\hline
\end{tabular}\caption{We give the values of lifts of the states $E^{(2)}\big(|G\phi\rangle^{+}_{+(h,s)}\big)/\lambda^2$ for three 1-parameter families: fixed small primary dimension but varying descendant mode number $s$, fixed small descendant mode number but varying primary dimension $h$, and a diagonal family with $s\sim h$. We choose to present the first twenty lifts in each of these families.\label{Gtab}}%
\end{table}

\begin{figure}[H]
    \centering
\fbox{\includegraphics[scale=0.41]{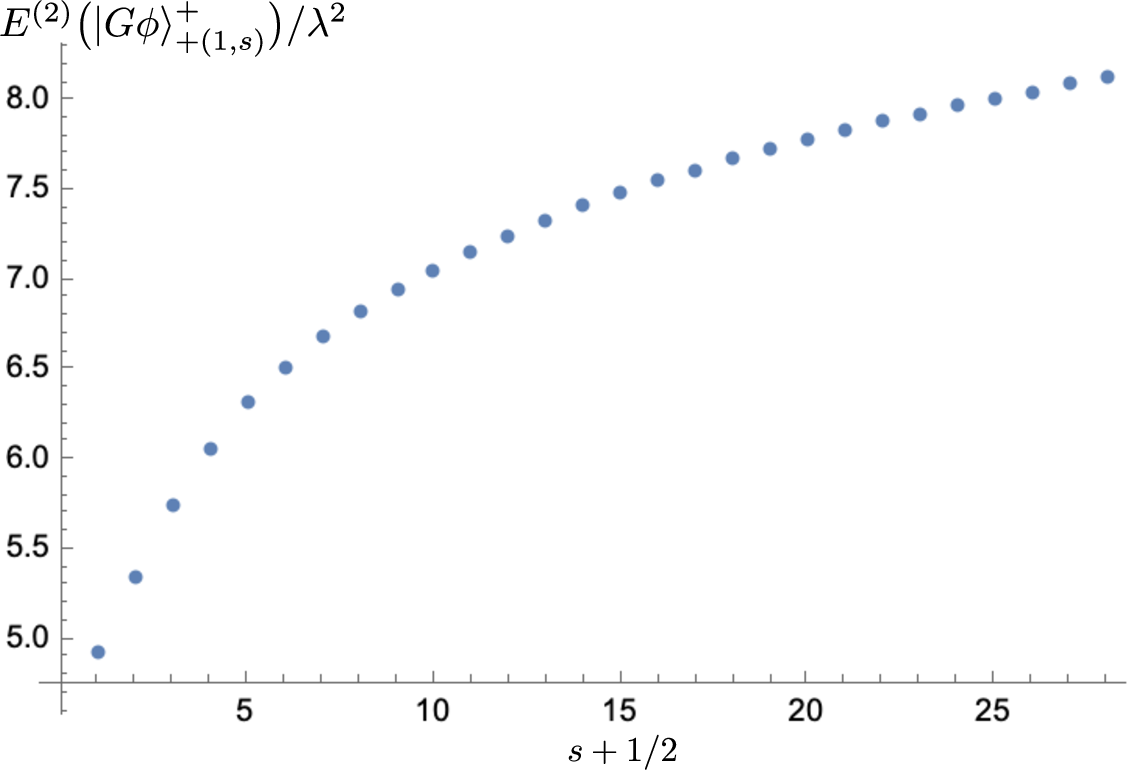}}
    \caption{Plot of the lifts $E^{(2)}\big(|G\phi\rangle^{+}_{+(1,s)}\big)/\lambda^2$ for varying $s$. The plot fits to the curve: $3.5943 + 1.45554 \sqrt{s+1/2} - 0.114676 (s+1/2)$.}
    \label{figGhn}
\end{figure}
\begin{figure}[H]
    \centering
\fbox{\includegraphics[scale=0.41]{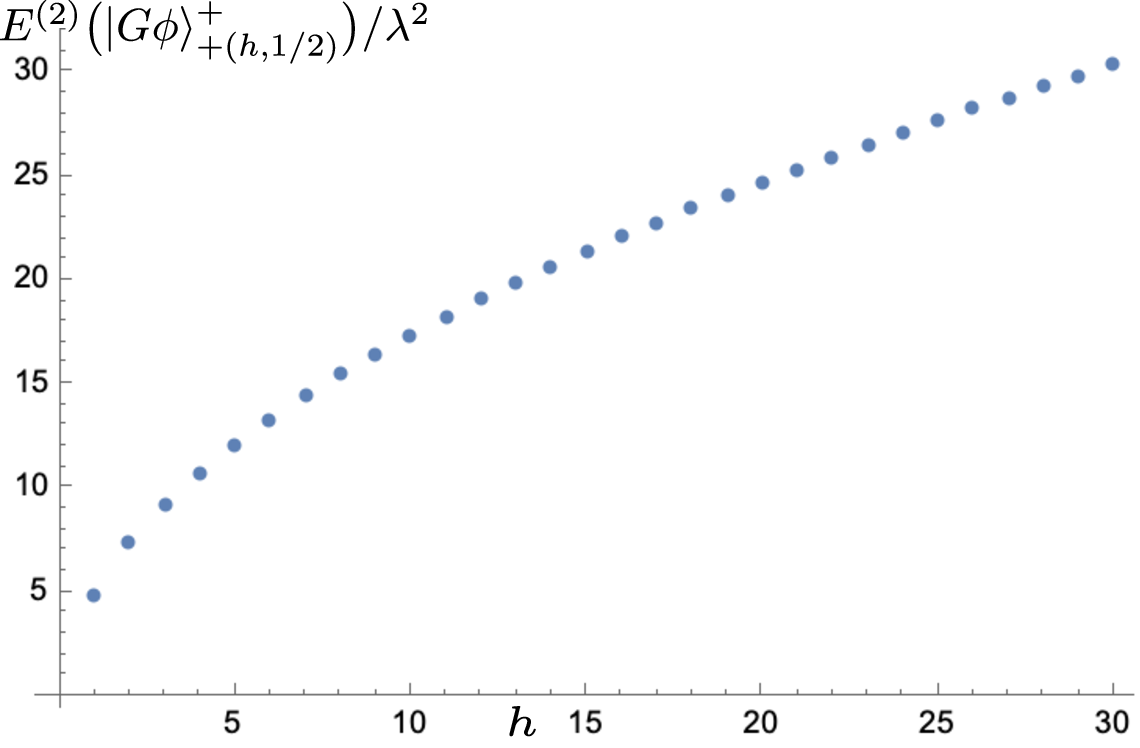}}
    \caption{Plot of the lifts $E^{(2)}\big(|G\phi\rangle^{+}_{+(h,1/2)}\big)/\lambda^2$ for varying $h$. The plot fits to the curve: $-0.707422 + 5.7688 \sqrt{h} - 0.0167021 h$.}
    \label{figGn1}
\end{figure}
\begin{figure}[H]
    \centering
\fbox{\includegraphics[scale=0.41]{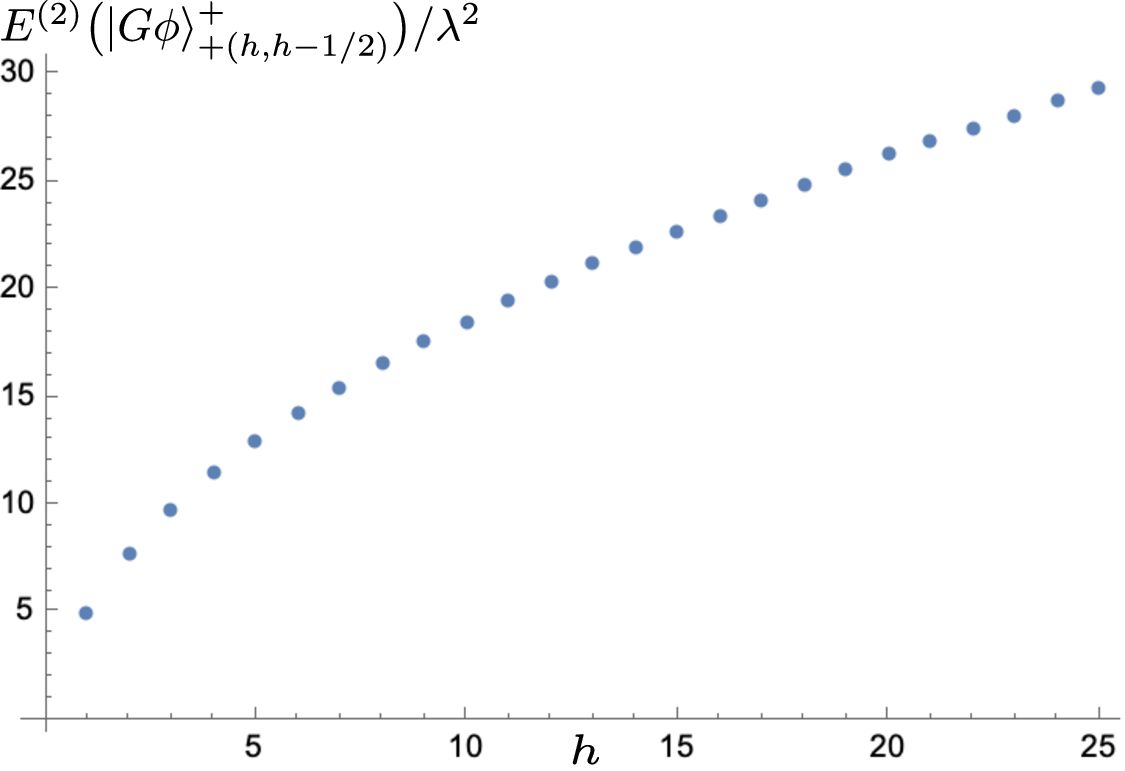}}
    \caption{Plot of the lifts $E^{(2)}\big(|G\phi\rangle^{+}_{+(h,h-1/2)}\big)/\lambda^2$ for varying $h$. The plot fits to the curve: $ -1.52527 + 6.68922\sqrt{h}  - 0.104878 h.$}
    \label{figGh1}
\end{figure}

\subsection{Lifting of $|J\phi\rangle^{+}_{(h,n)}$} \label{ssec:Jphilift}

In this section we consider the lifting of $J^+$ descendants of a superconformal primary state $|\phi\rangle$; the copy-symmetric state is defined in \eqref{eq.Jphistatedef}. The lift of these descendant states is amenable to the same method used in Section~\ref{ssec:GphiLift} and thus we will follow the same steps \ref{Series1}-\ref{Series9} described there. Since this computation is simpler than that of $G$ descendants, as will be described presently, and that in Section~\ref{ssec:WardIdJphi} it will turn out to be possible to obtain an exact analytic result for the lift using a method of Ward identities, we give fewer details in this section.

From Section~\ref{sec:LiftForm} the key quantity to be computed is the left-moving amplitude
\begin{equation} \label{eq.A11Jphi}
    A^{(1)(1)}_{h,n} \equiv \langle\phi| J^-_{n}\,\big(G^+_{-,-\frac12}\sigma^-_2\big)(w_2) \big(G^-_{+,-\frac12}\sigma^+_2\big)(w_1)\,J^+_{-n}|\phi\rangle \ ,
\end{equation}
with the initial and final state excitations placed on copy 1 for concreteness. Using step \ref{Series1} of the above-described method, we map this amplitude to the covering space $t$ under which, analogously to \eqref{eq.GinitialStateTrans} and \eqref{eq.GfinalStateTrans}, the initial state transforms as
\begin{align} \label{eq.JinitialStateTrans}
    J^+_{-n}|\phi\rangle &= \oint_{\tau\to-\infty} \frac{dw_3}{2\pi i}\, e^{-nw_3} \,J^+(w_3) |\phi\rangle \nonumber\\
    &= \lim_{\tt\to-a} \zt^h \bigg(\frac{d\tt}{d\zt}\bigg)^{\!h} \oint_{\tt}\,\frac{dt_3}{2\pi i}\, z_3^{-n} J^+(t_3)\phi(\tt\,)|\NSo\rangle \ ,
\end{align}
and the final state as
\begin{align} \label{eq.JfinalStateTrans}
    \langle\phi|J^-_{n} &= -\oint_{\tau\to\infty} \frac{dw_4}{2\pi i}\, e^{nw_4} \langle\phi|J^-(w_4) \nonumber\\
    &= -\lim_{t'\to\infty} z'^h \bigg(\frac{dt'}{dz'}\bigg)^{\!h} \oint_{t'}\,\frac{dt_4}{2\pi i}\, z_4^{n} \langle\NSo|\phi(t')J^-(t_4) \ ,
\end{align}
where the negative sign accounts for the direction of the contour at $\infty$. The universal factor \eqref{eq.baseamp} will still be present since it is independent of the external state under consideration. From step \ref{Series2} the spectral flows that remove spin field insertions on the covering space transform the state at $t=-a$ as
\begin{equation}
    J^+(t_3)\phi(\tt\,)|\NSo\rangle \rightarrow \frac{t_3-t_1}{t_3-t_2} J^+(t_3)\phi(\tt\,)|\NSo\rangle \ ,
\end{equation}
and the state at $t=\infty$ transforms as
\begin{equation}
    \langle\NSo| \phi(t')J^-(t_4) \rightarrow \frac{t_4-t_2}{t_4-t_1} \langle\NSo| \phi(t')J^-(t_4) \ .
\end{equation}
Once again, the transformations of the remaining $G$ field insertions at $t_1$ and $t_2$ contribute the universal factor \eqref{eq.f2}. The amplitude \eqref{eq.A11Jphi} can then be written as
\begin{align} \label{eq.A11hnintegrals}
    A^{(1)(1)}_{h,n} = -f_1f_2f_3 \oint_{\infty}\frac{dt_4}{2\pi i}\oint_{-a}\frac{dt_3}{2\pi i}\, \frac{z_4^{n}}{z_3^{n}} \bigg(\frac{t_3-t_1}{t_3-t_2}\bigg) \bigg(\frac{t_4-t_2}{t_4-t_1}\bigg) \widetilde{C}^{+}(t_4,t_3) \ ,
\end{align}
where the factor $f_3$ is given in \eqref{eq.f3} and $\widetilde{C}^{+}(t_4,t_3)$ is the correlator
\begin{equation} \label{eq.phiJGGJphiFields}
    \widetilde{C}^{+}(t_4,t_3) \equiv \langle\phi|J^-(t_4) G^+_-(t_2) G^-_+(t_1) J^+(t_3)|\phi\rangle \ .
\end{equation}
As explained in step \ref{Series4}, this correlator will be computed by expanding all currents in terms of their modes around the point $t=-a$ using \eqref{eq.GmodeExp} for the $G$ fields at $t_1$ and $t_2$ and expansions of the integrand in \eqref{eq.A11hnintegrals} for the remaining $J$ fields. This yields the form
\begin{align} \label{eq.A11hnmodes}
    A^{(1)(1)}_{h,n} = f_1f_2f_3 \sum_{k_4=0}^{n-1}\sum_{k_3=0}^{n-1}\sum_{s_1,s_2} (t_2+a)^{-s_2-\frac32}(t_1+a)^{-s_1-\frac32} \,\mathcal{K}^{(J)}_{k_3}\widetilde{\mathcal{K}}^{(J)}_{k_4} \,C^{+} \ ,
\end{align}
where $C^{+}$ is the mode correlator
\begin{equation} \label{eq.phiJGGJphidef}
    C^{+} \equiv \langle\phi|J^-_{n-k_4} G^+_{-,s_2} G^-_{+,s_1} J^+_{-n+k_3}|\phi\rangle \ .
\end{equation}
The expansion coefficients in \eqref{eq.A11hnmodes} can be written as
\begin{subequations}
\begin{align}
    \mathcal{K}^{(J)}_{k_3} &= \sum_{k=0}^1\sum_{k_1=0}^{k_3-k}\sum_{k_2=0}^{k_1}\frac{(-1)^{1+k_1-k_2} \;{}^{n+k_2-1}C_{k_2}\; ^{n}C_{k_1-k_2} (1-\delta_{k,0}(1+a-\sqrt{ab}))}{(a-b)^{k_2+n} a^{k_1-k_2-n}(a+\sqrt{ab})^{k_3-k-k_1+1}} \ ,\\
    \widetilde{\mathcal{K}}^{(J)}_{k_4} &= \sum_{\tilde{k}=0}^1\sum_{\tilde{k}_1=0}^{k_4+\tilde{k}-1}\sum_{\tilde{k}_2=0}^{\tilde{k}_1}\frac{(-1)^{\tilde{k}_2} \;{}^{n}C_{\tilde{k}_2}\; ^{n+\tilde{k}_1-\tilde{k}_2-1}C_{\tilde{k}_1-\tilde{k}_2} (1-\delta_{\tilde{k},0}(1+a+\sqrt{ab}))}{(a-b)^{-\tilde{k}_2} a^{\tilde{k}_2-\tilde{k}_1}(a-\sqrt{ab})^{1+\tilde{k}_1-\tilde{k}-k_4}} \ .
\end{align}
\end{subequations}
Computing the mode correlator \eqref{eq.phiJGGJphidef} using the $\mathcal{N}=4$ algebra \eqref{app com currents} and the primary definition \eqref{def primary 2} gives
\begin{align}
    C^+ &= \Big(s_2^2-\frac14 + h\Big) H\big[s_2-\tfrac12\big] \bigg(H[k_3-k_4-1] - H\big[n - s_1 - k_3-\tfrac12\big]\bigg)\,\delta_{s_1+s_2+k_3-k_4,0} \nonumber\\
    &\quad + (n-k_3)\Big(s_2^2-\frac14 + h\Big) H\big[\!-\!s_1-\tfrac12\big]  \delta_{s_1+s_2,0}\, \delta_{k_3,k_4} \nonumber\\
    &\quad - \Big(s_1^2-\frac14 + h\Big) H\big[\!-\!s_1-\tfrac12\big] H[k_3-k_4-1]\,\delta_{s_1+s_2+k_3-k_4,0} \nonumber\\
    &\quad + \Big((s_2+n-k_4)^2 - \frac14 +h\Big) H\big[n-s_1-k_3-\tfrac12\big] \,\delta_{s_1+s_2+k_3-k_4,0} \ ,
\end{align}
where we use the discrete step function
\begin{equation} \label{eq.Hfn}
    H[n] \equiv \begin{cases}
        \ 1 \quad\text{if } n\geq 0\vspace{4pt}\\
        \ 0 \quad \text{otherwise}
    \end{cases} \ .
\end{equation}
Resumming the series in \eqref{eq.A11hnmodes} as described in steps \ref{Series6} and \ref{Series7} can again be done per choice of $h,n$ and an amplitude as a function of $w_1$, $w_2$ obtained by using the relations \eqref{abDef}. As in \eqref{eq.GphiSymmAmp} the copy symmetric amplitude is given by
\begin{align} \label{eq.JphiSymmAmp}
    A_{h,n}(w,0) &\equiv {}^{\quad\,-}_{(h,n)}\langle J\phi| \big(G^+_{-,-\frac12}\sigma^-_2\big)(w) \big(G^-_{+,-\frac12}\sigma^+_2\big)(0)|J\phi\rangle^{+}_{(h,n)} \nonumber\\
    &= \frac12 \sum_{i,j=1}^2 A^{(j)(i)}_{h,n}(w,0) \nonumber\\
    &= A^{(1)(1)}_{h,n}(w,0) + A^{(2)(1)}_{h,n}(w,0) \ ,
\end{align}
with $A^{(2)(1)}$ given in terms of $A^{(1)(1)}$ by \eqref{eq.A21def}. Using the norm
\begin{equation}
    {}^{\quad\,-}_{(h,n)}\langle J\phi|J\phi\rangle^{+}_{(h,n)} = n \ ,
\end{equation}
the lift is then given by
\begin{equation} \label{eq.JphiLiftform}
    E^{(2)}\Big(|J\phi\rangle^{+}_{(h,n)}\Big) = \frac{\pi\lambda^2}{\big||J\phi\rangle^{+}_{(h,n)}\big|^2} \lim_{T\to\infty} \int_{0}^{2\pi}\!d\sigma \, A_{h,n}(w,0) \coth\!\big(\tfrac{\wb}{2}\big) \ ,
\end{equation}
where $w= \frac{T}{2} + i \sigma$ and $\wb = \frac{T}{2} - i\sigma$ with $T\to \infty$ on this contour as given in \eqref{eq.IC1} (noting again the comments below equation~\eqref{eq.GphiLiftform}).\\
\\
A shallow matrix of lifts for the states $|J\phi\rangle^{+}_{(h,n)}$ is given by:\\
\begin{align} \label{eq.Jphimat}
\tikz[remember picture, baseline=(mat.center)]{\node[inner sep=0](mat){$
\begin{pmatrix}
\\[-5pt]
 \frac{\pi ^2}{2} & \frac{9 \pi ^2}{16} & \frac{77 \pi ^2}{128} & \frac{645 \pi ^2}{1024} & \frac{21367 \pi ^2}{32768} & \frac{87857 \pi ^2}{131072} & \frac{719069 \pi ^2}{1048576} & \frac{11730309 \pi ^2}{16777216} \\\\
 \frac{3 \pi ^2}{4} & \frac{51 \pi ^2}{64} & \frac{213 \pi ^2}{256} & \frac{7035 \pi ^2}{8192} & \frac{57681 \pi ^2}{65536} & \frac{470757 \pi ^2}{524288} & \frac{1914825 \pi ^2}{2097152} & \frac{248646267 \pi ^2}{268435456} \\\\
 \frac{15 \pi ^2}{16} & \frac{249 \pi ^2}{256} & \frac{1029 \pi ^2}{1024} & \frac{33765 \pi ^2}{32768} & \frac{275565 \pi ^2}{262144} & \frac{2241015 \pi ^2}{2097152} & \frac{9089361 \pi ^2}{8388608} & \frac{1177469037 \pi ^2}{1073741824} \\\\
 \frac{35 \pi ^2}{32} & \frac{287 \pi ^2}{256} & \frac{4713 \pi ^2}{4096} & \frac{38505 \pi ^2}{32768} & \frac{626675 \pi ^2}{524288} & \frac{2542505 \pi ^2}{2097152} & \frac{82348469 \pi ^2}{67108864} & \frac{1331452941 \pi ^2}{1073741824} \\\\
 \frac{315 \pi ^2}{256} & \frac{2565 \pi ^2}{2048} & \frac{41925 \pi ^2}{32768} & \frac{341535 \pi ^2}{262144} & \frac{5547075 \pi ^2}{4194304} & \frac{22469475 \pi ^2}{16777216} & \frac{726821865 \pi ^2}{536870912} & \frac{11738994435 \pi ^2}{8589934592} \\\\
 \frac{693 \pi ^2}{512} & \frac{22473 \pi ^2}{16384} & \frac{91509 \pi ^2}{65536} & \frac{1487637 \pi ^2}{1048576} & \frac{12061809 \pi ^2}{8388608} & \frac{390395523 \pi ^2}{268435456} & \frac{1576971333 \pi ^2}{1073741824} & \frac{50897919303 \pi
   ^2}{34359738368} \\\\
 \frac{3003 \pi ^2}{2048} & \frac{97097 \pi ^2}{65536} & \frac{394303 \pi ^2}{262144} & \frac{6398777 \pi ^2}{4194304} & \frac{51816359 \pi ^2}{33554432} & \frac{1675479323 \pi ^2}{1073741824} & \frac{6762666295 \pi ^2}{4294967296} & \frac{218127013419 \pi^2}{137438953472} \\\\
 \frac{6435 \pi ^2}{4096} & \frac{51909 \pi ^2}{32768} & \frac{420693 \pi ^2}{262144} & \frac{3408555 \pi ^2}{2097152} & \frac{220587135 \pi ^2}{134217728} & \frac{890873295 \pi ^2}{536870912} & \frac{7186947963 \pi ^2}{4294967296} & \frac{115843402197 \pi
   ^2}{68719476736} \\\\
\end{pmatrix}
$};}
\begin{tikzpicture}[overlay,remember picture]
\draw[blue,thick,->] node[anchor=south west] (nn1) at (mat.north west)
{increasing $n$} (nn1.east) -- (nn1-|mat.north east) ;
\draw[red,thick,->] node[anchor=north west,align=center, inner xsep=0pt] (nn2) at 
(mat.north east)
{\rotatebox{-90}{increasing $h$}} (nn2.south) -- (nn2.south|-mat.south) 
;
\end{tikzpicture}
\end{align}

\begin{table}
\centering
\begin{tabular}{|c||c|c||c|c|}
\hline
&&&&\\[-1em]
$n$ & $E^{(2)}\big(|J\phi\rangle^{+}_{(1,n)}\big)/\lambda^2$ &$ h$ & $E^{(2)}\big(|J\phi\rangle^{+}_{(h,1)}\big)/\lambda^2$  & $E^{(2)}\big(|J\phi\rangle^{+}_{(h,h)}\big)/\lambda^2$ \\
\hline&&&&\\[-1em]
1 & $\frac{\pi^2}{2}$ & 1 & $\frac{\pi^2}{2}$  & $\frac{\pi^2}{2}$ \\[3pt]
\hline
&&&&\\[-1em]
2 & $\frac{9\pi^2}{16}$ & 2 & $\frac{3\pi^2}{4}$  & $\frac{51\pi^2}{64}$ \\[3pt]
\hline
&&&&\\[-1em]
3 & $\frac{77\pi^2}{128}$ & 3 & $\frac{15\pi^2}{16}$  & $\frac{1029\pi^2}{1024}$ \\[3pt]
\hline
&&&&\\[-1em]
4 & $\frac{645\pi^2}{1024}$ & 4 & $\frac{35\pi^2}{32}$  & $\frac{38505\pi^2}{32768}$ \\[3pt]
\hline
&&&&\\[-1em]
5 & $\frac{21367\pi^2}{32768}$ & 5 & $\frac{315\pi^2}{256}$  & $\frac{5547075\pi^2}{4194304}$ \\[3pt]
\hline
&&&&\\[-1em]
6 & $\frac{87857\pi^2}{131072}$ & 6 & $\frac{693\pi^2}{512}$  & $\frac{390395523\pi^2}{268435456}$ \\[3pt]
\hline
&&&&\\[-1em]
7 & $\frac{719069\pi^2}{1048576}$ & 7 & $\frac{3003\pi^2}{2048}$ & $\frac{6762666295\pi^2}{4294967296}$ \\[3pt]
\hline
&&&&\\[-1em]
8 & $\frac{11730309\pi^2}{16777216}$ & 8 & $\frac{6435\pi^2}{4096}$  & $\frac{115843402197\pi^2}{68719476736}$ \\[3pt]
\hline
&&&&\\[-1em]
9 & $\frac{1526945171\pi^2}{2147483648}$ & 9 & $\frac{109395\pi^2}{65536}$  & $\frac{125934725576295\pi^2}{70368744177664}$ \\[3pt]
\hline
&&&&\\[-1em]
10 & $\frac{6199001523\pi^2}{8589934592}$ & 10 & $\frac{230945\pi^2}{131072}$  & $\frac{2125128794403615\pi^2}{1125899906842624}$ \\[3pt]
\hline
&&&&\\[-1em]
11 & $\frac{50252720171\pi^2}{68719476736}$ & 11 & $\frac{969969\pi^2}{524288}$ &  $\frac{35672711189099061\pi^2}{18014398509481984}$ \\[3pt]
\hline
&&&&\\[-1em]
12 & $\frac{406850316977\pi^2}{549755813888}$ & 12 & $\frac{2028117\pi^2}{1048576}$ & $\frac{4769806285786438533\pi^2}{2305843009213693952}$ \\[3pt]
\hline
&&&&\\[-1em]
13 & $\frac{13161421594619\pi^2}{17592186044416}$ & 13 & $\frac{16900975\pi^2}{8388608}$ & $\frac{635476448618853061493\pi^2}{295147905179352825856}$ \\[3pt]
\hline
&&&&\\[-1em]
14 & $\frac{53172571377153\pi^2}{70368744177664}$ & 14 & $\frac{35102025\pi^2}{16777216}$ & $\frac{21101411540283429819765\pi^2}{9444732965739290427392}$ \\[3pt]
\hline
&&&&\\[-1em]
15 & $\frac{429306030983453\pi^2}{562949953421312}$ & 15 & $\frac{145422675\pi^2}{67108864}$ & $\frac{349425974935218889182525\pi^2}{151115727451828646838272}$ \\[3pt]
\hline
&&&&\\[-1em]
16 & $\frac{13855330294144389\pi^2}{18014398509481984}$ & 16 & $\frac{300540195\pi^2}{134217728}$ & $\frac{5773130365885258168314605\pi^2}{2417851639229258349412352}$ \\[3pt]
\hline
&&&&\\[-1em]
17 & $\frac{7150471815953612587\pi^2}{9223372036854775808}$ & 17 & $\frac{9917826435\pi^2}{4294967296}$ & $\frac{48738566651126418399719594511\pi^2}{19807040628566084398385987584}$ \\[3pt]
\hline
&&&&\\[-1em]
18 & $\frac{28815168723125441183\pi^2}{36893488147419103232}$ & 18 & $\frac{20419054425\pi^2}{8589934592}$ &$\frac{802233196347055292535813677217\pi^2}{316912650057057350374175801344}$ \\[3pt]
\hline
&&&&\\[-1em]
19 & $\frac{232135596668140476407\pi^2}{295147905179352825856}$ & 19 & $\frac{83945001525\pi^2}{34359738368}$ &  $\frac{13184066094015683210008550256055\pi^2}{5070602400912917605986812821504}$ \\[3pt]
\hline
&&&&\\[-1em]
20 & $\frac{1869337969360052852787\pi^2}{2361183241434822606848}$ & 20 & $\frac{172308161025\pi^2}{68719476736}$ &$\frac{432733110292413371373686195637405\pi^2}{162259276829213363391578010288}$ \\[3pt]
\hline
\end{tabular}
\caption{We give the values of lifts of the states $E^{(2)}\big(|J\phi\rangle^{+}_{(h,n)}\big)/\lambda^2$ for three 1-parameter families: fixed small primary dimension but varying descendant mode number $n$, fixed small descendant mode number but varying primary dimension $h$, and a diagonal family with $n=h$. We choose to present the first twenty lifts in each of these families.\label{Jtab}}%
\end{table}
\begin{figure}[ht]
    \centering
\fbox{\includegraphics[scale=0.41]{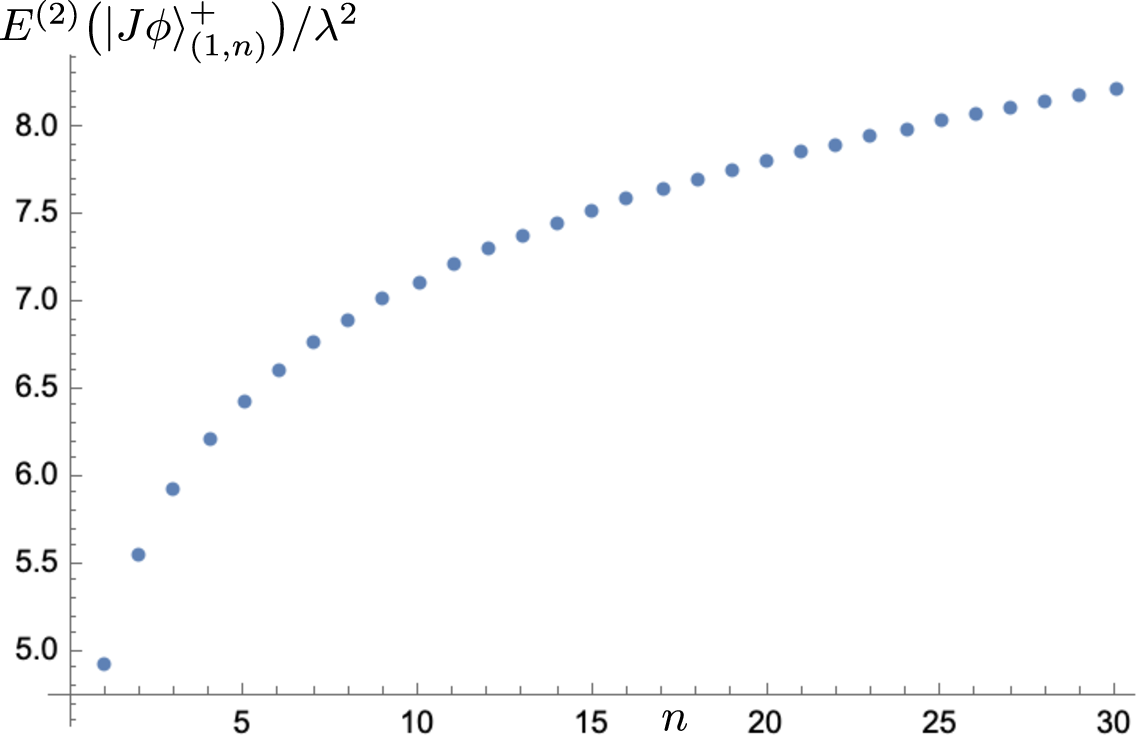}}
    \caption{Plot of the lifts $E^{(2)}\big(|J\phi\rangle^{+}_{(1,n)}\big)/\lambda^2$ for varying $n$. The plot fits to the curve: $3.79158 + 1.39824 \sqrt{n} - 0.109782 n$.}
    \label{figJh1}
\end{figure}

\begin{figure}[H]
    \centering
\fbox{\includegraphics[scale=0.41]{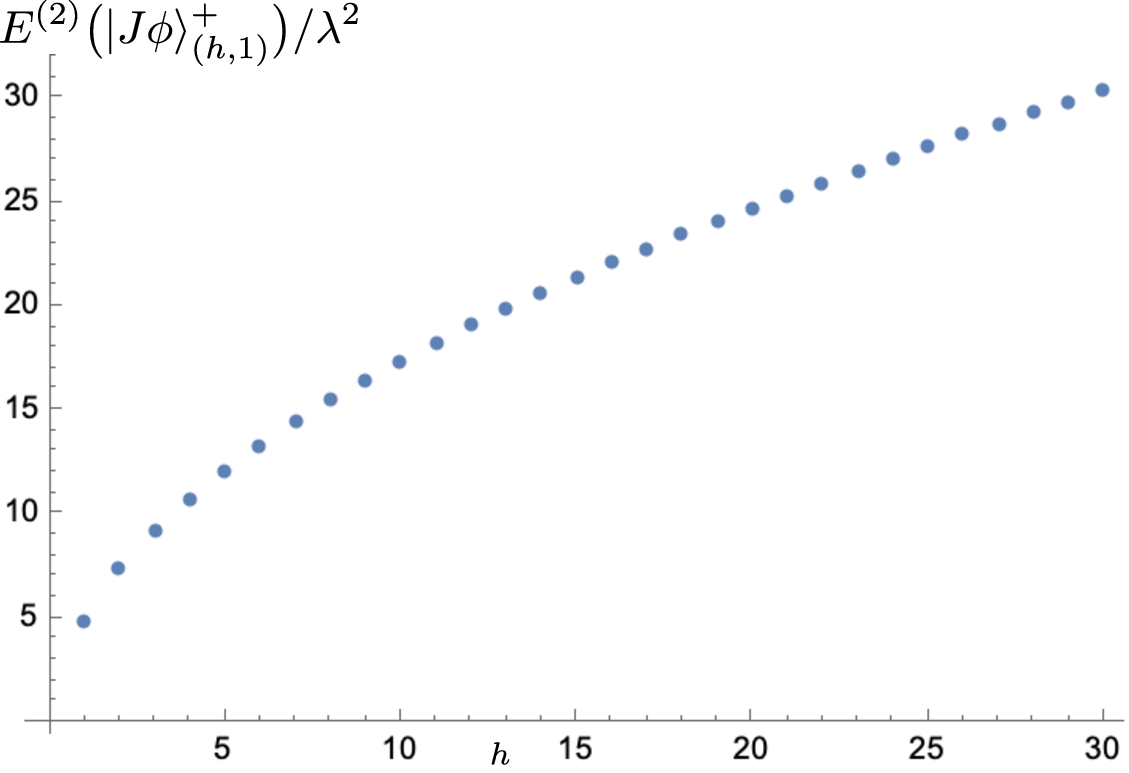}}
    \caption{Plot of the lifts $E^{(2)}\big(|J\phi\rangle^{+}_{(h,1)}\big)/\lambda^2$ for varying $h$. The plot fits to the curve: $-0.820134 + 5.85066 \sqrt{h} - 0.0292201 h$.}
    \label{figJn1}
\end{figure}
\begin{figure}[H]
    \centering
\fbox{\includegraphics[scale=0.41]{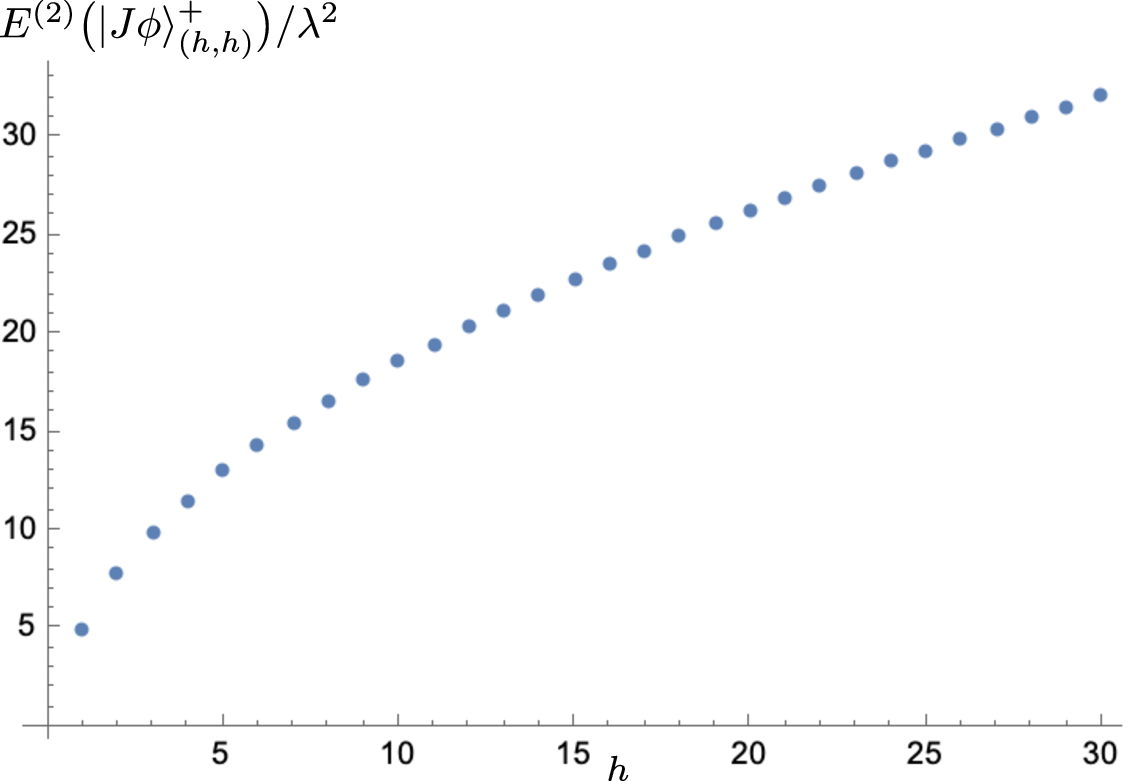}}
    \caption{Plot of the lifts $E^{(2)}\big(|J\phi\rangle^{+}_{(n,n)}\big)/\lambda^2$ for varying $n$. The plot fits to the curve: $-1.35675 + 6.61605 \sqrt{h} - 0.0935374 h$.}
    \label{figJhn}
\end{figure}

\subsection{Symmetries and checks} \label{ssec:symmChecks}

There are a few immediate checks of the results of Sections~\ref{ssec:GphiLift} and \ref{ssec:Jphilift} and therefore, on the method used. Firstly and most simply, the lift of states in the same $SU(2)$ multiplet -- be it the $SU(2)_L$ or $SU(2)_2$ under which the states $|G\phi\rangle$ and $|J\phi\rangle$ are charged -- should be the same. The calculation of Section~\ref{ssec:GphiLift} in steps~\ref{Series1}-\ref{Series9} is written for generic $\alpha$ and $\Ad$ charges on the $G$ mode and so this can be immediately checked. The cases of $J^-$ and $J^3$ work similarly.
\begin{figure}[h]
    \centering
    \includegraphics[width=\textwidth]{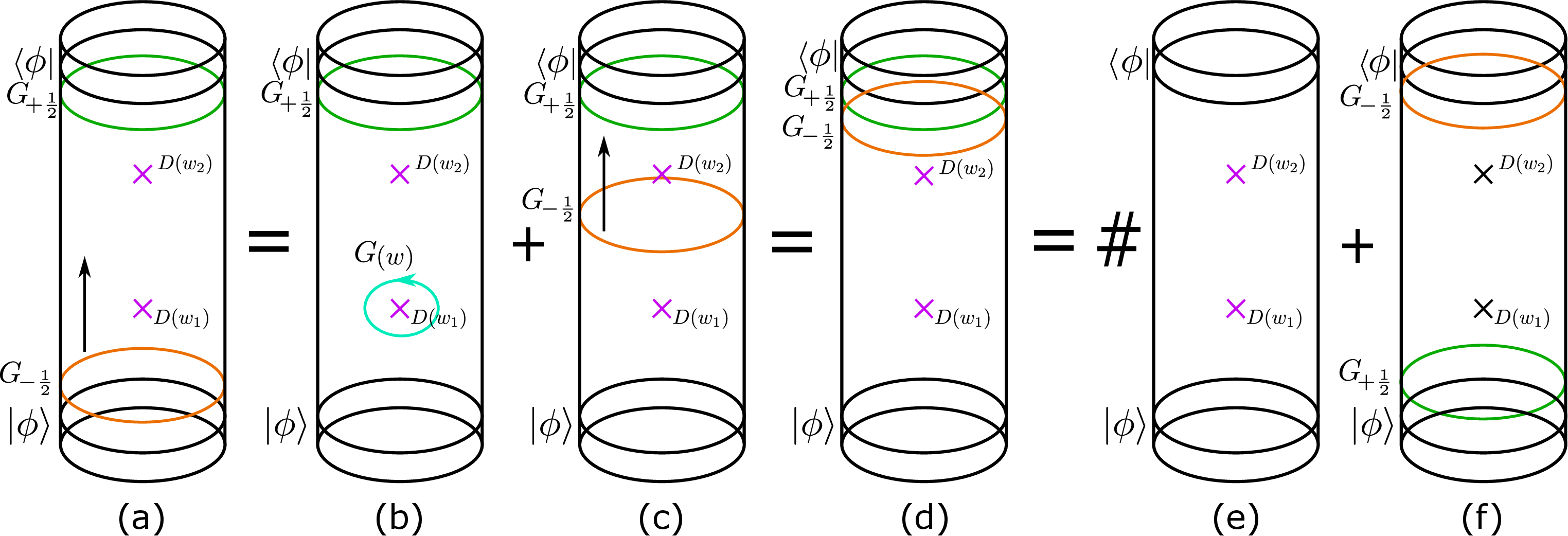}
    \caption{
    The general tactic in deriving these relations between lifts will be to start with the correlator \eqref{eq.AGphi1} containing descendants of $|\phi\rangle$ as initial and final states (a), deform the contour of the $G_{-1/2}$ mode through the insertions of the deformation operators (b)-(d) and then commute the $G_{-1/2}$ mode with the $G_{+/2}$ mode from the final state. This leaves (up to constant coefficients) a correlator that computes the lift of the state $|\phi\rangle$ (e) and one that computes the lift of the state $G_{1/2}|\phi\rangle$ (f) which vanishes due to $|\phi\rangle$ being a superconformal primary. We have suppressed the trivial right-moving part throughout. See \cite{Hughes:2023apl} for a complete derivation.}
    \label{fig:Gcheck}
\end{figure}
The second check against known results emerges by considering an explicit realisation of the state $|J\phi\rangle^+_{(1,1)}$ in terms of the free boson and fermion realisation of the theory. A prescription for finding explicit realisations of (at least an infinite family of) superconformal primaries at arbitrary levels was given in Section~\ref{ssec:SCPs}, however, here we simply need a level-1 example. Taking the $(h,m)=(1,0)$ superconformal primary with minimal charge under $SU(2)_1$
\begin{equation} \label{eq.phi1}
    |\phi^1_{1,-1}\rangle = d^{--}_{-\frac12}d^{+-}_{-\frac12}|\NSo\rangle \ ,
\end{equation}
the following is then a realisation of its level-1 $J$ descendant
\begin{equation}
    J^+_{-1}|\phi^1_{1,-1}\rangle = d^{+-}_{-\frac32}d^{+-}_{-\frac12}|\NSo\rangle \ .
\end{equation}
The lift of this two-mode state was found in \cite{Hughes:2023apl} to be (using the notation of that paper)
\begin{equation}
    E^{(2)}\big(|dd\rangle^{+-+-}_{(3/2,1/2)}\big) = \frac{\pi^2\lambda^2}{2} \ ,
\end{equation}
matching the first element in the lifting matrix \eqref{eq.Jphimat}. Similarly, considering an explicit realisation of the descendant state $|G\phi\rangle^{+}_{+(1,1/2)}$ using \eqref{eq.phi1}, giving the two-mode state
\begin{equation}
    G^+_{+,-\frac12}|\phi^1_{1,-1}\rangle = i\, \alpha_{++,-1}d^{+-}_{-\frac12}|\NSo\rangle \ ,
\end{equation}
whose lift was found in \cite{Hughes:2023apl} to be
\begin{equation}
    E^{(2)}\big(|\alpha d\rangle^{+-}_{++(1,1/2)}\big) = \frac{\pi^2\lambda^2}{2} \ ,
\end{equation}
in agreement with the first element of the lifting matrix \eqref{eq.Gphimat}.

The third check is for the families of states $|G\phi\rangle^{\alpha}_{\!\Ad(h,1/2)}$ and relating their lifts to that of the single-copy superconformal primary $|\phi\rangle$. Here we only briefly describe the key steps in deriving such relations, however, full details can be found in Section (4) of \cite{Hughes:2023apl}. We start with the initial state $G^{\alpha(1)}_{\!\Ad,-\frac12}|\phi\rangle^{(1)}|\NSo\rangle^{(2)}$ in the amplitude \eqref{eq.A11Gphi} on the $w$-cylinder. The single-copy mode $G^{\alpha(1)}_{\!\Ad,-\frac12}$ can be trivially made into a global mode $G^{\alpha}_{\!\Ad,-\frac12}$, as defined around Equation \eqref{globalL}, since it annihilates the NS vacuum state on all but one copy. This starting point is depicted in Figure~\ref{fig:Gcheck}(a). By writing this global mode as a contour of a $G$ field it can be passed through the two deformation operator insertions to the final state (see Figure~\ref{fig:Gcheck}(b)-(c)), yielding the equality between amplitudes
\begin{align} \label{eq.AGphi1}
     A^{\alpha}_{\!\Ad}\big(G\phi\big) &\equiv \langle\NSo|\langle\phi| \big(G^\alpha_{\!\Ad,-\frac12}\big)^{\dagger}\int\! d^2w_2\, D(w_2,\bar{w}_2)\int\! d^2w_1\, D(w_1,\bar{w}_1)\,G^\alpha_{\!\Ad,-\frac12}|\phi\rangle|\NSo\rangle \nonumber\\
     &= \langle\NSo|\langle\phi| \big(G^\alpha_{\!\Ad,-\frac12}\big)^{\dagger}G^\alpha_{\!\Ad,-\frac12}\int\! d^2w_2\, D(w_2,\bar{w}_2)\int\! d^2w_1\, D(w_1,\bar{w}_1)|\phi\rangle|\NSo\rangle \ .
\end{align}
The initial state $G$ mode can then be anti-commuted through the final state $G$ mode to annihilate both the superconformal primary $\langle\phi|$ (using the defining relations \eqref{def primary 2}) and the NS vacuum $\langle\NSo|$, leaving a term with an $L_0$ insertion from the anti-commutator (see Figure~\ref{fig:Gcheck}(d)-(f)). The amplitude \eqref{eq.AGphi1} is then given by
\begin{align}
    A^{\alpha}_{\!\Ad}\big(G\phi\big) = K^\alpha_{\!\Ad}\, \langle\NSo|\langle\phi| \int\! d^2w_2\, D(w_2,\bar{w}_2)\int\! d^2w_1\, D(w_1,\bar{w}_1)\,|\phi\rangle|\NSo\rangle = K^\alpha_{\!\Ad}\, A\big(\phi\big) \ ,
\end{align}
where the factor $K^\alpha_{\!\Ad}$ depends on the dimension of $\phi$ and the $G$ mode chosen. This relation between amplitudes can then be converted to a simple relation between lifts
\begin{equation}
    E^{(2)}\big(|G\phi\rangle^{\alpha}_{\Ad(h,1/2)}\big) = E^{(2)}\big(|\phi\rangle\big) \ ,
\end{equation}
for all $\alpha$ and $\Ad$ values. The data for $E^{(2)}(G^+_{+,-\frac12}|\phi\rangle)$ for the first $20$ values of $h$ is given in the first column of Table~\ref{Gtab} which can be seen to match the analytic expression found in \cite{Guo:2022ifr} for the lift of single-copy superconformal primaries
\begin{equation} \label{eq.phiLift}
    E^{(2)}_h\big(|\phi\rangle\big) = \pi^{\frac32}\lambda^2\frac{\Gamma(h+\tfrac12)}{\Gamma(h)} \ .
\end{equation}
Likewise, from the second column of Table~\ref{Jtab} we see that the lifts of the states $|J\phi\rangle^+_{(h,1)}$ also equal those of the single-copy superconformal primary $|\phi\rangle$.

\section{Lifting of descendants by method of Ward identities}

In the previous section the lift of descendent states was found using a method that required the computation of a 6-point function of fields on the covering space. This was done by expanding the four currents in modes, evaluating the mode correlator using the $\mathcal{N}=4$ algebra, and resumming all of the series. Due to the complexity of the functions involved, this resummation and thus the lifts were done on a state-by-state basis. This is unfortunately not well-suited for understanding the asymptotic behaviour of lifts for states of large dimension, comparing primaries and descendants, and testing the dependence of the second order lift on the partition of the dimension between current modes generating the descendant state. In order to study these aspects, analytic expressions for lifts are required. In this section we again study the case of the state $J^+_{-n}|\phi\rangle$ and find an analytic formula for its lift in terms of a handful of finite sums.

\subsection{Lifting of $|J\phi\rangle^+_{(h,n)}$ by Ward identity} \label{ssec:WardIdJphi}

The starting point here will be the covering space correlator \eqref{eq.phiJGGJphiFields} given by
\begin{equation} \label{eq.phiJGGJphi2}
    \widetilde{C}^+(t_4,t_3) = \langle\phi(\infty)J^-(t_4) G^+_-(t_2) G^-_+(t_1) J^+(t_3)\phi(-a)\rangle \ ,
\end{equation}
that is, we follow steps \ref{Series1}-\ref{Series3} for $J^+_{-n}|\phi\rangle$ as in Section~\ref{ssec:Jphilift}, however, instead of expanding fields in terms of modes we evaluate \eqref{eq.phiJGGJphi2} by a Ward identity. We do this by first trivially rewriting the field $J^+(t_3)$ as a contour integral centred around $t_5=t_3$
\begin{align}
    J^{+}(t_3) = \oint_{t_3}\frac{dt_5}{2\pi i} \frac{J^{+}(t_5)}{t_5-t_3} \ ,
\end{align}
followed by unwrapping this contour to get a sum of contours around each insertion on the $t$-plane
\begin{align}
    \widetilde{C}^+(t_4,t_3) = \widetilde{C}^+_{t_1} + \widetilde{C}^+_{t_2} + \widetilde{C}^+_{t_4} + \widetilde{C}^+_{-a} + \widetilde{C}^+_{\infty} \ .\label{contoursum}
\end{align}
\begin{figure}[t]
    \centering
    \includegraphics[scale=0.5]{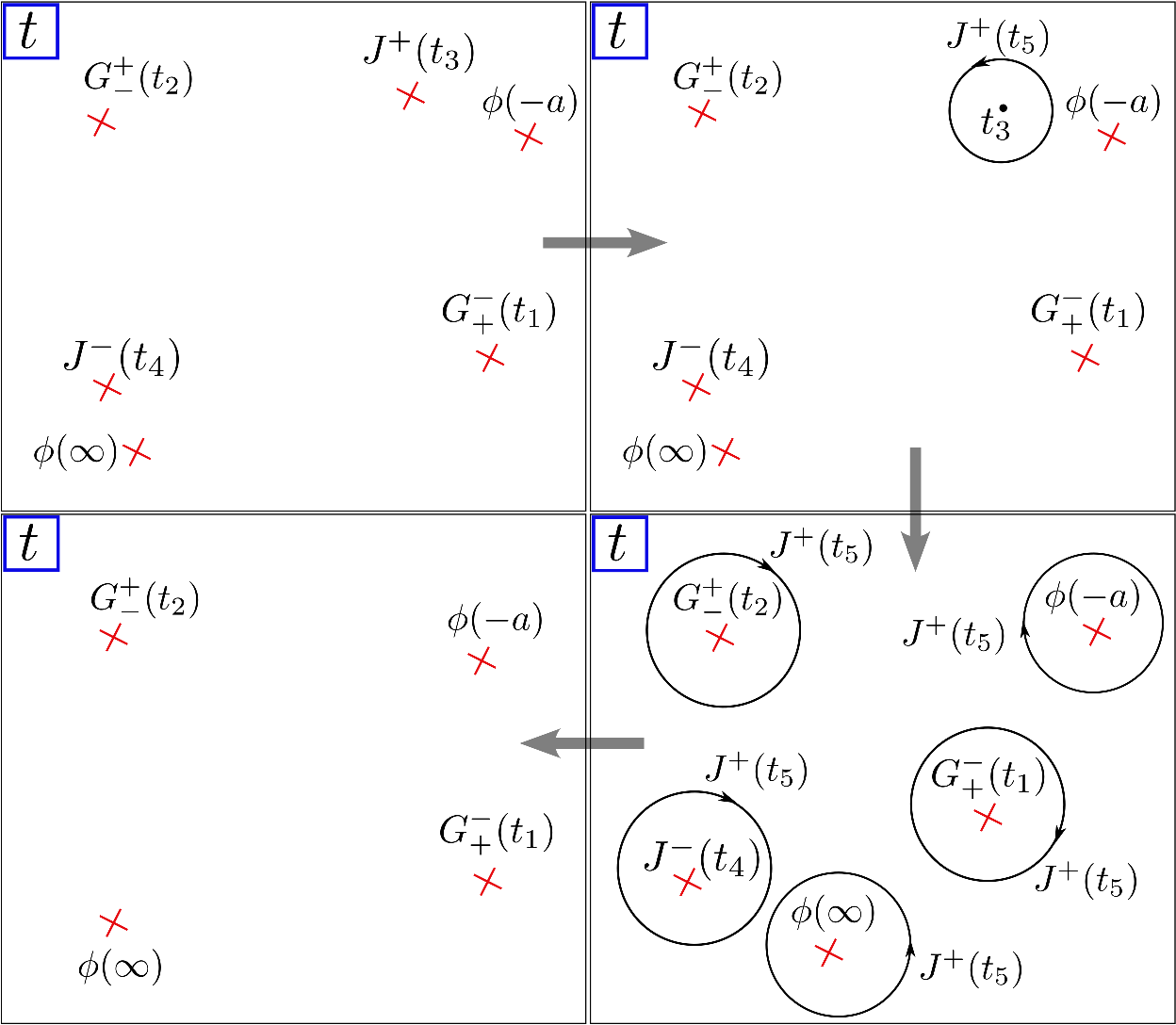}
    \caption{The general idea in deriving $|J\phi\rangle^+_{(h,n)}$ lifts using Ward identities is be to start with the correlator \eqref{eq.phiJGGJphi2} containing different insertions on the $t$-plane. Moving clockwise in the figure, trivially rewrite the field $J^+(t_3)$ as a contour integral centred around $t_5=t_3$. Unwrapping this contour yields a sum of contours around each insertion on the $t$-plane. Different terms in the sum \eqref{contoursum}evaluate to different correlators one of which is shown as a 4-point insertion. Hence the descendant $J^+$ can be removed using this method leading to a different method of lifting computation.}
    \label{figJward}
\end{figure}
The term $\widetilde{C}^+_{t_2}$ where the $t_5$ contour encircles the point $t_2$ will vanish due to the OPE $G^+_{\!\Ad}(t_2)\times J^+(t_5)$ being purely regular. Likewise the term $\widetilde{C}^+_{-a}$ where the $t_5$ contour wraps $\phi(-a)$ can also be seen to vanish due to
\begin{align}
    -\oint_{-a}\frac{dt_5}{2\pi i} \frac{J^+(t_5)\phi(-a)}{t_5-t_3} &= \sum_{m=0}^{\infty} \oint_{-a}\frac{dt_5}{2\pi i} \frac{(t_5+a)^{m}}{(t_3+a)^{m+1}} J^+(t_5)\phi(-a) \nonumber\\
    &= \sum_{m=0}^{\infty} (t_3+a)^{-m-1} J^+_{m}\phi(-a) = 0 \ ,
\end{align}
where the properties of a superconformal primary \eqref{def primary 2} and the fact that $\phi$ has zero $J^2$ charge were used. Thus $\widetilde{C}^+_{-a}=0$ and by similar arguments one sees that also $\widetilde{C}^+_{\infty}=0$. The contour around $t_1$ gives
\begin{align}
     -\oint_{t_1}\frac{dt_5}{2\pi i} \frac{J^+(t_5)G^-_+(t_1)}{t_5-t_3} = -\oint_{t_1}\frac{dt_5}{2\pi i} \frac{G^{+}_{+}{(t_1)}}{(t_5-t_3)(t_5-t_1)} = -\frac{G^{+}_{+}{(t_1)}}{(t_1-t_3)} \ ,
\end{align}
using the current OPEs \eqref{eq.currOPE} and so
\begin{equation}
    \widetilde{C}^+_{t_1} = -\frac1{t_1-t_3} \langle\phi(\infty)J^-(t_4) G^+_-(t_2) G^+_+(t_1) \phi(-a)\rangle \ .
\end{equation}
Lastly, the contour around $t_4$ yields
\begin{align}
    -\oint_{t_4}\frac{dt_5}{2\pi i} \frac{J^{+}(t_5) J^{-}{(t_4)}}{t_5-t_3} &=-\oint_{t_4}\frac{dt_5}{2\pi i}\bigg( \frac{1}{(t_5-t_3)(t_5-t_4)^2} +  \frac{2J^{3}(t_4)}{(t_5-t_3)(t_5-t_4)}\bigg) \nonumber\\
    &= \frac{1}{(t_4-t_3)^2} -  \frac{2J^{3}(t_4)}{(t_4-t_3)} \ ,
\end{align}
and so the correlator $\widetilde{C}^+_{t_4}$ becomes
\begin{equation}
    \widetilde{C}^+_{t_4} = \frac{A^{+-}_{-+}(t_2,t_1;h)}{(t_4-t_3)^2} - \frac{2}{t_4-t_3} \langle\phi(\infty)J^3(t_4) G^+_-(t_2) G^-_+(t_1) \phi(-a)\rangle \ ,
\end{equation}
where $A^{+-}_{-+}(t_2,t_1;h)$ is defined from
\begin{equation} \label{eq.phiGGphibase}
    A^{\alpha\beta}_{\!\Ad\Bd}(t_2,t_1;h) \equiv \langle\phi|G^{\alpha}_{\Ad}(t_2)G^{\beta}_{\Bd}(t_1) |\phi\rangle = \ep_{\Ad\Bd}\ep^{\alpha\beta} \frac{a+b(2h-1)}{4(a-b)(ab)^{3/2}} \ .
\end{equation}
Next we trivially write the field $J^-(t_4)$ in $\widetilde{C}^+_{t_1}$ as the contour
\begin{align}
    J^{-}(t_4) = \oint_{t_4}\frac{dt_6}{2\pi i} \frac{J^{-}(t_6)}{t_6-t_4} \ ,
\end{align}
and again unwrap this $t_6$ contour yielding non-zero terms around $t_2$ and $t_1$. Using
\begin{align}
    \frac{1}{(t_1-t_3)}\oint_{t_2}\frac{dt_6}{2\pi i}\frac{J^{-}(t_6)G^+_-(t_2)}{t_6-t_4} = \frac{1}{(t_1-t_3)}\oint_{t_2}\frac{dt_6}{2\pi i}\frac{G^-_-(t_2)}{(t_6-t_4)(t_6-t_2)} = \frac{G^-_-(t_2)}{(t_1-t_3)(t_2-t_4)} \ ,
\end{align}
and 
\begin{align}
    \frac{1}{(t_1-t_3)}\oint_{t_1}\frac{dt_6}{2\pi i}\frac{J^{-}(t_6)G^+_+(t_1)}{t_6-t_4} = \frac{1}{(t_1-t_3)}\oint_{t_1}\frac{dt_6}{2\pi i}\frac{G^-_+(t_1)}{(t_6-t_4)(t_6-t_1)} = \frac{G^-_+(t_1)}{(t_1-t_3)(t_1-t_4)} \ ,
\end{align}
we get
\begin{equation}
    \widetilde{C}^+_{t_1} = \frac{A^{-+}_{-+}(t_2,t_1;h)}{(t_1-t_3)(t_2-t_4)} + \frac{A^{+-}_{-+}(t_2,t_1;h)}{(t_1-t_3)(t_1-t_4)} \ .
\end{equation}
Likewise, the $J^3(t_4)$ field in $\widetilde{C}^+_{t_4}$ is treated similarly and one finds
\begin{equation}
    \widetilde{C}^+_{t_4} = \frac{A^{+-}_{-+}(t_2,t_1;h)}{(t_4-t_3)^2} + \frac{A^{+-}_{-+}(t_2,t_1;h)}{(t_2-t_4)(t_4-t_3)} -\frac{A^{+-}_{-+}(t_2,t_1;h) }{(t_1-t_4)(t_4-t_3)} \ .
\end{equation}
All in all the amplitude \eqref{eq.phiJGGJphi2} is given in terms of \eqref{eq.phiJGGJphi2} as
\begin{align} \label{eq.WardidResult}
    \widetilde{C}^+(t_4,t_3) &= \frac{A^{-+}_{-+}(t_2,t_1;h)}{(t_1-t_3)(t_2-t_4)} + A^{+-}_{-+}(t_2,t_1;h) \bigg(\frac{1}{(t_1-t_3)(t_1-t_4)} + \frac{1}{(t_4-t_3)^2} \nonumber\\
    &\hspace{6cm}+ \frac{1}{(t_2-t_4)(t_4-t_3)} -\frac{1}{(t_1-t_4)(t_4-t_3)} \bigg) \ .
\end{align}
From this the amplitude required for the lift is then
\begin{equation} \label{eq.A11Jdef}
    A^{(1)(1)}_{h,n}(w_2,w_1) \equiv \langle\phi|J^-_n\,\big(G^+_{-,-\frac12}\sigma_2^-\big)(w_2) \big(G^-_{+,-\frac12}\sigma_2^+\big)(w_1) \,J^+_{-n}|\phi\rangle = f_1f_2f_3\,A^{+-}_{-+}(t_2,t_1;h) \,\mathcal{I}_n\ ,
\end{equation}
in terms of factors $f_1$, $f_2$, $f_3$ given in \eqref{eq.baseamp}, \eqref{eq.f2} and \eqref{eq.f3} that do not depend on the presence of the $J$ modes and some integrals $\mathcal{I}_n$ originating from the $J$ modes. The $n$-dependent integrals are
\begin{align} \label{eq.Iintegral}
    \mathcal{I}_n &\equiv \oint_{\infty}\frac{dt_4}{2\pi i}\oint_{-a}\frac{dt_3}{2\pi i}\, z_4^n z_3^{-n}\, \frac{t_3-t_1}{t_3-t_2}\frac{t_4-t_2}{t_4-t_1}\; \widetilde{C}^+(t_4,t_3) \nonumber\\
    &= \oint_{\infty}\frac{dt_4}{2\pi i}\oint_{-a}\frac{dt_3}{2\pi i}\, z_4^n z_3^{-n} \frac{1}{(t_4-t_3)^2} \ ,
\end{align}
where the result of the Ward identity \eqref{eq.WardidResult} was used in the second line and $z_3$ and $z_4$ are functions of $t_3$ and $t_4$ respectively as seen from \eqref{eq.ztotmap}. The integrand has $n$th order poles in both $t_3$ and $t_4$; consider the $t_3$ integral first
\begin{equation} \label{eq.I3Integraldef}
    \mathcal{I}_n^{(3)} \equiv \oint_{-a}\frac{dt_3}{2\pi i}\,z_3^{-n} \frac{1}{(t_4-t_3)^2} = \oint_{-a}\frac{dt_3}{2\pi i}\,(t_3+a)^{-n} \frac{t_3^{\,n}}{(t_3+b)^n(t_4-t_3)^2} \ .
\end{equation}
By expanding the part of the integrand that is regular at $t_3=-a$ as a power series around $-a$ we find
\begin{align}\label{eq.I3Integral}
    \mathcal{I}_n^{(3)} &= \oint_{-a}\frac{dt_3}{2\pi i} \sum_{k_5=0}^{\infty} \sum_{k_3=0}^{\min(k_5,n)}\sum_{k_1=0}^{k_5-k_3} \frac{(-1)^{k_3}(k_1+1)a^{n-k_3}(t_3+a)^{k_5-n}}{(t_4+a)^{k_1+2}(a-b)^{n+k_5-k_1-k_3}}\, {}^{n+k_5-k_1-k_3-1}C_{k_5-k_1-k_3} \,{}^{n}C_{k_3} \nonumber\\
    &= \sum_{k_3=0}^{n-1}\sum_{k_1=0}^{n-k_3-1} \frac{(-1)^{k_3}(k_1+1)a^{n-k_3}}{(t_4+a)^{k_1+2}(a-b)^{2n-k_1-k_3-1}}\, {}^{2n-k_1-k_3-2}C_{n-1} \,{}^{n}C_{k_3} \ ,
\end{align}
where the integral picks out the term $k_5=n-1$. The $t_4$ integral of \eqref{eq.Iintegral} is then given by
\begin{equation} \label{eq.I4Integraldef}
    \mathcal{I}_n^{(4)} \equiv \oint_{\infty}\frac{dt_4}{2\pi i} \frac{z_4^n}{(t_4+a)^{k_1+2}} = \oint_{\infty}\frac{dt_4}{2\pi i} \,t_4^{n-k_1-2}\, \Big(1+\frac{a}{t_4}\Big)^{n-k_1-2}\Big(1+\frac{b}{t_4}\Big)^n \ .
\end{equation}
By converting this to a contour integral around $t_4=0$ and expanding the regular part of the integrand as a power series around $0$ we find
\begin{align} \label{eq.I4Integral}
    \mathcal{I}_n^{(4)} &= -\oint_{0}\frac{d\tt_4}{2\pi i} \,\tt_4^{\,k_1-n}\, \Big(1+ a\tt_4\Big)^{n-k_1-2}\Big(1+b\tt_4\Big)^n \nonumber\\
    &= -\oint_{0}\frac{d\tt_4}{2\pi i} \sum_{j_3=0}^{2n-k_1-2}\sum_{j_2=0}^{\min(n,j_3)} a^{j_3-j_2}\, b^{j_2}\, \tt_4^{\,k_1-n+j_3}\; {}^{n-k_1-2}C_{j_3-j_2} {}^{n}C_{j_2} \nonumber\\
    &= - \sum_{j_2=0}^{n-k_1-1} a^{n-k_1-1-j_2}\, b^{j_2} \; {}^{n-k_1-2}C_{j_2-1}\, {}^{n}C_{j_2} \ ,
\end{align}
where the integral picks out the term $j_3=n-k_1-1$. The total integral $\mathcal{I}_n$ is then
\begin{align} \label{eq.Iintegral2}
    \mathcal{I}_n = \Big(\frac{a}{a-b}\Big)^{\!2n-1}\! \sum_{k_3=0}^{n-1}\sum_{k_1=0}^{n-k_3-1}\sum_{j_2=0}^{n-k_1-1} \frac{(-1)^{k_3+1}(k_1+1)b^{j_2}}{(a-b)^{-k_1-k_3}a^{k_1+k_3+j_2}}\, {}^{2n-k_1-k_3-2}C_{n-1} {}^{n}C_{k_3} {}^nC_{j_2} {}^{n-k_1-2}C_{j_2-1} \ ,
\end{align}
and the amplitude \eqref{eq.A11Jdef} is then given by
\begin{align} \label{eq.A11J2}
    A^{(1)(1)}_{h,n} &= \frac{a+b(2h-1)}{16b} \bigg(\frac{a}{a-b}\bigg)^{\!h+2n-2} \sum_{k_1=0}^{n-1}\sum_{k_1=0}^{n-1-k_3}\sum_{j_2=0}^{n-k_1-1} (-1)^{k_3+1}(k_1+1) \bigg(\frac{b}{a}\bigg)^{\!j_2} \bigg(\frac{a}{a-b}\bigg)^{\!\!-k_1-k_3} \nonumber\\
    &\hspace{7cm} \times {}^{2n-k_1-k_3-2}C_{n-1} {}^{n}C_{k_3} {}^nC_{j_2} {}^{n-k_1-2}C_{j_2-1} \ .
\end{align}
After replacing $a$ and $b$ using \eqref{abDef} this is an analytic function of $w$ and the copy-symmetric amplitude
\begin{align}
    A_{h,n}(w,0) &\equiv {}^{\ \ \ \,-}_{(h,n)}\langle J\phi| \big(G^+_{-,-\frac12}\sigma^-_2\big)(w) \big(G^-_{+,-\frac12}\sigma^+_2\big)(0)|J\phi\rangle^{+}_{(h,n)} \nonumber\\
    &= \frac12 \sum_{i,j=1}^2 A^{(j)(i)}_{h,n}(w,0) \nonumber\\
    &= A^{(1)(1)}_{h,n}(w,0) + A^{(2)(1)}_{h,n}(w,0) \ ,
\end{align}
where we have used that $A^{(2)(2)}=A^{(1)(1)}$ and $A^{(1)(2)}=A^{(2)(1)}$ by copy symmetry. The amplitude $A^{(2)(1)}$ can be obtained from $A^{(1)(1)}$ via the continuation
\begin{equation}
    A^{(2)(1)}_{h,n}(w,0) = A^{(1)(1)}_{h,n}(w+2\pi i,0) \ .
\end{equation}
The lift is then obtained by integrating over $w$, for which we will consider the contour integral $I_{C_2}$ where $C_2$ is at $\Re w\to-\infty$ as defined in \eqref{eq.IC2}. The right-moving part of the computation is simple ends up contributing a factor of $-1$ and the integral  then picks out the order $(e^w)^0$ term from $A^{(1)(1)}$ and $A^{(2)(1)}$. It turns out that the contribution to this integral from $A^{(1)(1)}$ and $A^{(2)(1)}$ are equal in this case, with the contribution from \eqref{eq.A11J2} being
\begin{align} \label{eq.A11J3}
    A^{(1)(1)}_{h,n}(w,0)\Big|_{(e^w)^0} &= \sum_{k_3=0}^{n-1}\sum_{k_1=0}^{n-1-k_3}\sum_{j_2=0}^{n-k_1-1}  {}^{2n-k_1-k_3-2}C_{n-1}\, {}^{n}C_{k_3}\, {}^nC_{j_2}\, {}^{n-k_1-2}C_{j_2-1} \frac{(-1)^{k_3+1}(k_1+1)}{2^{2(h+2n-k_1-k_3)-1}} \nonumber\\
    &\quad\times \frac{\sum_{\nu=0}^{2}c_\nu x^\nu}{x^{h+2n-k_1-k_3-2}} \sum_{\ell_3=0}^{\infty}\sum_{\ell_2=0}^{\min(q,\ell_3)} (-1)^{\ell_3-\ell_2} \; {}^{2(j_2-1)}C_{\ell_3-\ell_2}\, {}^{q}C_{\ell_2}\,x^{\ell_3}\Bigg|_{x^0} \nonumber\\
    &= \sum_{k_3=0}^{n-1}\sum_{k_1=0}^{n-1-k_3}\sum_{j_2=0}^{n-k_1-1}\sum_{\nu=0}^{2}\sum_{\ell_2=0}^{\min(q,L_3)} \frac{c_\nu(-1)^{h+k_1+\ell_2+\nu}(k_1+1)}{2^{2(h+2n-k_1-k_3)-1}} \nonumber\\
    &\hspace{2cm}\times {}^{2n-k_1-k_3-2}C_{n-1}\, {}^{n}C_{k_3}\, {}^nC_{j_2}\, {}^{n-k_1-2}C_{j_2-1}\, {}^{2(j_2-1)}C_{L_3-\ell_2}\, {}^{q}C_{\ell_2} \ .
\end{align}
where the shorthand $x=e^{w/2}$ was used and the projection onto $x^0$ picked out the term $\ell_3=L_3$ where
\begin{equation} \label{eq.L3qdefs}
    L_3=h+2n-k_1-k_3-2-\nu \quad,\quad q=2(h+2n-k_1-k_3-j_2-2) \ ,
\end{equation}
and $c_\nu$ are coefficients such that
\begin{equation} \label{eq.cnudef}
    \sum_{\nu=0}^{2}c_\nu\, x^\nu = hx^2 -2(h-1)x + h \ .
\end{equation}
From \eqref{eq.lift} the lift is then given by
\begin{align} \label{eq.EJphihn}
    E^{(2)}\big(|J\phi\rangle^+_{(h,n)}\big) &= \frac{4\pi\lambda^2}{\big||J\phi\rangle^+_{(h,n)}\big|} \,A^{(1)(1)}_{h,n}(w,0)\Big|_{(e^w)^0} \nonumber\\
    &= \frac{8\pi^2\lambda^2}{n}\! \sum_{k_3=0}^{n-1}\sum_{k_1=0}^{n-1-k_3}\sum_{j_2=0}^{n-k_1-1}\sum_{\nu=0}^{2}\sum_{\ell_2=0}^{\min(q,L_3)} \frac{c_\nu(-1)^{h+k_1+\ell_2+\nu}}{2^{2(h+2n-k_1-k_3)}}(k_1+1) \nonumber\\
    &\hspace{2cm}\times {}^{2n-k_1-k_3-2}C_{n-1}\, {}^{n}C_{k_3}\, {}^nC_{j_2}\, {}^{n-k_1-2}C_{j_2-1}\, {}^{2(j_2-1)}C_{L_3-\ell_2}\, {}^{q}C_{\ell_2} \ .
\end{align}
It is useful to note that if $2(j_2-1)<L_3-\ell_2$ then the binomial ${}^{2(j_2-1)}C_{L_3-\ell_2}$ in \eqref{eq.EJphihn} vanishes and the ranges of $L_3$ and $q$ are
\begin{align} \label{eq.L3qRanges}
    h-2\leq L_3 \leq h+2(n-1) \quad,\quad 2h \leq q\leq 2h+4(n-1) \ .
\end{align}
While the expression \eqref{eq.EJphihn} for the lift of $|J\phi\rangle^+_{(h,n)}$ is not particularly compact, it is valid for all values of $h$ and $n$ and as will be seen in the following section some information about the large dimension behaviour will be extractable. One aspect of the form \eqref{eq.EJphihn} that makes the large dimension regime difficult to analyse is the dependence of the range of the $\ell_2$ sum on $h$. It turns out that we can resum the $\nu$ and $\ell_2$ sums by considering the $j_2=0$ and $j_2\geq1$ terms separately. We find that for $j_2=0$
\begin{align} \label{eq.F0def}
    F_{0} &\equiv \sum_{\nu=0}^{2} c_{\nu}(-1)^{\nu} \sum_{\ell_2=0}^{L_3} (-1)^{\ell_2}\: {}^{-2}C_{L_3-\ell_2} \,{}^{2L_3+2\nu}C_{\ell_2} \nonumber\\
    &= \sum_{\nu=0}^{2} c_{\nu}(-1)^{h-k_1-k_3} \sum_{\ell_2=0}^{L_3} (L_3-\ell_2+1)\: {}^{2L_3+2\nu}C_{\ell_2} \nonumber\\
    &= \frac{2(h+L_3)\,\Gamma(2L_3)}{\Gamma(L_3)\,\Gamma(L_3+1)}\bigg|_{\nu=0} \ ,
\end{align}
and for $J=j_2-1\geq0$
\begin{align} \label{eq.FJdef}
    F_{J} &\equiv \sum_{\nu=0}^{2} c_{\nu}(-1)^{\nu}\sum_{\ell_2=0}^{L_3} (-1)^{\ell_2}\: {}^{2J}C_{L_3-\ell_2} \,{}^{2L_3+2\nu-2J-2}C_{\ell_2} \nonumber\\
    &= \frac{(-1)^{L_3+J} 2^{2L_3-1} \big(L_3-h(2J+1)\big) \,\Gamma(J+\tfrac12)\,\Gamma(L_3-J-\tfrac12)}{\pi\,\Gamma(L_3+1)}\bigg|_{\nu=0} \ .
\end{align}
The term \eqref{eq.F0def} in the lift \eqref{eq.EJphihn} can be fully resummed to give
\begin{align} \label{eq.E1def}
    E_1 &= \frac{\pi^2\lambda^2}{n\, 2^{2h+4n-3}}\! \sum_{k_1,k_3=0}^{n-1} \frac{(-1)^{h+k_1}(k_1+1)}{2^{2(-k_1-k_3)}}\, {}^{2n-k_1-k_3-2}C_{n-1}\, {}^{n}C_{k_3}\, \delta_{k_1,n-1}\, \widetilde{F}_{0} \nonumber\\
    &= \frac12 \pi^{\frac32}\lambda^2\,\frac{(2h+n-1)\,\Gamma(h+n-\tfrac12)}{(h+n-1)\,\Gamma(h+n-1)} \ ,
\end{align}
where the terms $k_1=n-1$ and $k_3=0$ were picked out by the various binomials. We find that the lift \eqref{eq.EJphihn} can be written as
\begin{align} \label{eq.EJphiE12}
    E^{(2)}\big(|J\phi\rangle^+_{(h,n)}\big) &= E_1 + E_2 \ ,
\end{align}
with $E_2$ being given by
\begin{align} \label{eq.E2def}
    E_2 &= \frac{\pi\lambda^2}{4n} \sum_{K=1}^{n}\sum_{k_1=1}^{\min(K,n-1)}\sum_{J=0}^{n-k_1-1} \frac{(-1)^{K-k_1+J}k_1}{\Gamma(h+2n-K)} \,{}^{n}C_{K-k_1}\,{}^{2n-K-1}C_{n-1}\, {}^{n}C_{J+1}\, {}^{n-k_1-1}C_{J}\nonumber\\
    &\hspace{3.5cm}\times (2n-K-1-2hJ)\,\Gamma(J+\tfrac12)\,\Gamma(h+2n-K-J-\tfrac32) \ .
\end{align}
where $K=k_1+k_3$ was used. We note that the upper limit of the $k_1$ sum can be written as $K$ since the one term where $\min(K,n-1)\neq K$ is $K=k_1=n$ and then the binomial ${}^{n-k_1-1}C_J$ vanishes.

A simple check of this exact result for the lift is that, as seen in the series method data of Section~\ref{ssec:Jphilift}, the case of $n=1$ should reduce to the lift of a superconformal primary, for which an analytic form was found in \cite{Guo:2022ifr} and rewritten in \eqref{eq.phiLift}. Considering the case of $n=1$ in \eqref{eq.E2def} we see that $E_2=0$ and so 
\begin{align} \label{eq.EJphin1}
    E^{(2)}_{h,1}\big(J^+|\phi\rangle\big) = E_1\big|_{n=1} = \pi^{\frac32}\lambda^2\,\frac{\Gamma(h+\tfrac12)}{\Gamma(h)} = E^{(2)}_h\big(|\phi\rangle\big) \ .
\end{align}
Despite the complicated form of \eqref{eq.E2def} it turns out that the lift is polynomial in $h$ and we find that for a given value of $n$
\begin{equation} \label{eq.JphiLifthn}
    E^{(2)}\big(|J\phi\rangle^+_{(h,n>1)}\big) = E^{(2)}_{h}\!\big(\phi\big) \bigg[1 + a_n \frac{P_{n}(h)}{(h+1)_{2n-2}}\bigg] \ , 
\end{equation}
where $P_{n}(h)$ is a polynomial in $h$ of degree $2n-4$ and $(x)_n$ is the Pochhammer symbol
\begin{equation}
    (x)_n \equiv \frac{\Gamma(x+n)}{\Gamma(x)} \ .
\end{equation}
The rational coefficients $a_n$ in \eqref{eq.JphiLifthn} for the first few values of $n$ are
\begin{equation} \label{eq.andata}
    \big\{a_{n\geq2}\big\} = \bigg\{\,\frac{3}{4},\, \frac{3}{8},\, \frac{15}{32},\, \frac{15}{128},\, \frac{105}{256},\, \frac{21}{512},\, \dots \bigg\} \ ,
\end{equation}
and the respective polynomials are
\begin{align} \label{eq.Pn}
    P_2 &= 1 \ ,\nonumber\\
    P_3 &= 8h^2 + 16h + 41 \ ,\nonumber\\
    P_4 &= 16h^4 + 96h^3 + 578 h^2 + 798h + 1305 \ ,\nonumber\\
    P_5 &= 128 h^6+1536 h^5+16976 h^4+64608 h^3+237920 h^2+268512 h+352107 \ ,\nonumber\\
    P_6 &= 64 h^8+1280 h^7+22592 h^6+163520 h^5+1077776 h^4+3208160 h^3+9205198 h^2 \nonumber\\
    &\hspace{8.97cm}+9091050 h+10377045 \ .
\end{align}
Given the knowledge of the form \eqref{eq.JphiLifthn} for the lift of the $|J\phi\rangle^+_{(h,n)}$ states, we were able to find that the series method data of Section~\ref{ssec:GphiLift} for the lift of $|G\phi\rangle^{+}_{+(h,n)}$ follows a similar pattern of
\begin{equation} \label{eq.GphiLifthn}
     E^{(2)}\big(|G\phi\rangle^+_{+(h,n>1)}\big) = E^{(2)}_{h}\!\big(\phi\big) \bigg[1 + b_n \frac{\widetilde{P}_n(h)}{(h+1)_{2(n-1)}(h+n^2-n)} \bigg] \ ,
\end{equation}
where the rational coefficients are
\begin{equation} \label{eq.bndata}
    \big\{b_{n\geq2}\big\} = \bigg\{\, \frac{3}{4},\,\frac{15}{16},\,\frac{21}{16},\,\frac{45}{256},\,\frac{165}{256},\,\frac{273}{1024},\, \dots \bigg\} \ ,
\end{equation}
and $b_1=0$ and $\widetilde{P}_n(h)$ are degree $2n-3$ polynomials in $h$ of the form
\begin{align} \label{eq.Ptn}
    \widetilde{P}_2 &= h + 1 \ ,\nonumber\\
    \widetilde{P}_3 &= 4 h^3+16 h^2+51 h+78 \ ,\nonumber\\
    \widetilde{P}_4 &= 8 h^5+72 h^4+601 h^3+2193 h^2+3711 h+4905 \ ,\nonumber\\
    \widetilde{P}_5 &= 128 h^7+2048 h^6+31952 h^5+219392 h^4+979664 h^3+2908064 h^2+3700845 h+4296180 \ ,\nonumber\\
    \widetilde{P}_6 &= 64 h^9+1600 h^8+39952 h^7+446320 h^6+3690736 h^5+19756240 h^4+64904013 h^3 \ ,\nonumber\\
    &\hspace{6.77cm}+164470575 h^2+176337585 h+185479875 \ .
\end{align}
We do not, however, have a derivation of this form for the lift of $|G\phi\rangle^{+}_{+(h,n)}$.

\subsection{Asymptotic behaviour} \label{sec:AsympBehaviour}

In \cite{Guo:2022ifr} the lift of a single-copy superconformal primary was found in a closed form -- displayed in \eqref{eq.phiLift} -- and it was observed that in the limit of large dimension the leading behaviour of the lift was of the form
\begin{equation} \label{eq.philargeh}
    E^{(2)}_h\big(\phi\big) \approx \pi^{3/2}\lambda^2 \,\sqrt{h} \ .
\end{equation}
This $\sim\sqrt{h}$ behaviour has also been hinted at in computations of lifts for which closed form expressions are not known, such as in \cite{Hughes:2023apl} where the lift of states with excitations on one copy from two bosonic or fermionic modes was computed state by state. Looking at the lifts of states where one mode had a fixed and small mode number and the other was varied to, say, level 30 a square root behaviour was observed in the data. Likewise, in the state-by-state computations of Section~\ref{sec.main} for descendants, similar behaviour was observed. It is natural to then ask more nuanced questions: for a level-$n$ descendant of a primary of dimension $h$, is the asymptotic behaviour of lifts the same as that of the primary and is the important quantity the total dimension $h_{tot}=h+n$ or are the dimensions of the primary and descendant excitation treated differently? If the latter case is true, it is interesting to then ask whether the partitioning of the descendant excitation dimension amongst different modes is important when it comes to the lift of the state. In this section we hope to answer exactly these questions.

\subsubsection{Large dimension limits} \label{sssec:largeh}

Since the lifting formula of the descendant state $|J\phi\rangle^+_{(h,n)}$ is known exactly, albeit in terms of finite sums, we may derive a large dimension limit. Namely, consider the large $h$ regime
\begin{equation} \label{eq.largeh}
    h \gg n \sim O(1) \ .
\end{equation}
In this regime the dominant part of $E_2$ in \eqref{eq.E2def} comes from the function
\begin{align}
    \frac{h\,\Gamma(h+2n-K-J-\tfrac32)}{\Gamma(h+2n-K)} \sim h^{-\frac12-J} \ ,
\end{align}
which for all values of $J\in [0,n-2]$ is subleading compared with the leading order behaviour of $E_1$ in \eqref{eq.E1def} and so we find
\begin{align} \label{eq.Jphilargeh12}
    E^{(2)}\big(|J\phi\rangle^+_{(h,n)}\big) \approx E_1 \approx \pi^{\frac32}\lambda^2 \,\sqrt{h} \ ,
\end{align}
which exactly matches the leading asymptotic behaviour of the lift of superconformal primaries given in \eqref{eq.philargeh} and notably \eqref{eq.Jphilargeh12} is independent of the descendant excitation level $n$.

It turns out that more can be said about the asymptotic forms of the lifts of the descendant states $J^+_{-n}|\phi\rangle$ and $G^{\alpha}_{\Ad,-s}|\phi\rangle$ that will shine a light on the questions raised at the beginning of this section. Firstly, and most simply, we have already argued in Section~\ref{ssec:symmChecks} that the lifts of the lowest-level descendants ($n=1$ and $s=\frac12$ respectively) are exactly equal to the lift of the primary, \textit{i.e.}
\begin{equation} \label{eq.n1lift}
    E^{(2)}\big(|J\phi\rangle^+_{(h,1)}\big) = E^{(2)}\big(|G\phi\rangle^+_{+(h,1/2)}\big) = E^{(2)}_{h}\!\big(\phi\big) = \pi^{\frac32}\lambda^2 \frac{\Gamma(h+\tfrac12)}{\Gamma(h)}\ .
\end{equation}
In the case of $J^+_{-1}|\phi\rangle$ this statement was derived from the exact result in \eqref{eq.EJphin1}. The large $h$ behaviour of these families is thus trivially also equal to that of $\phi$, given in \eqref{eq.philargeh}.

Although the expressions \eqref{eq.JphiLifthn} and \eqref{eq.GphiLifthn} are not in closed form, it is possible to find a closed form for the subleading large $h$ behaviour
\begin{align}
    \frac{E^{(2)}\!\big(|J\phi\rangle^+_{(h,n)}\big)}{E^{(2)}_{h}\!(\phi)} &\approx 1 + \frac{n(n^2-1)}{8h^2} \ , \label{eq.Jphilargeh2}\\
    \frac{E^{(2)}\!\big(|G\phi\rangle^+_{+(h,n)}\big)}{E^{(2)}_{h}\!(\phi)} &\approx 1 + \frac{n(n-1)(2n-1)}{8h^2} \ , \label{eq.Gphilargeh2}
\end{align}
where again $n=s+\frac12$ for the $G$ descendant. The subleading term carries information about the descendant excitation, however, if the asymptotic behaviour for large total dimension $h_{tot}=h+n$ was $\sim\sqrt{h_{tot}}$ as might be expected, then the subleading corrections should appear at order $1/h$ and not $1/h^2$ since for $h\gg n$
\begin{equation}
    \sqrt{h+n} \approx \sqrt{h}\bigg(1 + \frac{n}{2h} + O\big(h^{-2}\big)\bigg) \ .
\end{equation}
It therefore appears that the dimension of the primary and that of the descendant excitation are treated differently in the lift. In Appendix~\ref{app.3} we present a special two parameter family of states for which we have complete control of the lifting calculation and can obtain a closed form expression for the lift. These untwisted sector states are in the NS vacuum on all copies bar one, on which there is the following excited state
\begin{equation} \label{eq.Jmphidef}
    \mathcal{J}^{+,(m)}|\phi\rangle \equiv J^+_{-(2m-1)}\cdots J^+_{-3}J^+_{-1}|\phi\rangle \ ,
\end{equation}
where $\phi$ is again a single-copy superconformal primary of dimension $h$. This state has a total dimension $h^{(m)}$ and $J^3_0$ charge $j^{(m)}$ of 
\begin{equation} \label{eq.JmphiCharges}
    h^{(m)} = h + m^2 \quad,\quad j^{(m)} = m \ .
\end{equation}
The excitation $\mathcal{J}^{+,(m)}$ is special because it is the spectral flow of the NS vacuum under a flow by $\eta=2m$ units, a fact that enables the simple calculation of the lift of \eqref{eq.Jmphidef} to be
\begin{align} \label{eq.JmphiLift}
    E^{(2)}\big(\mathcal{J}^{+,(m)}|\phi_h\rangle\big) = \frac{\pi^{3/2}\lambda^2}{2} \frac{(2h+m^2-1)\,\Gamma(h+m^2-\frac12)}{\Gamma(h+m^2)} \ .
\end{align}
This result is the first closed-form expression we have obtained (that does not contain sums as in \eqref{eq.E2def}) for a two-parameter family of states, albeit a very special one. We can then use \eqref{eq.JmphiLift} to understand the questions posed at the start of Section~\ref{sec:AsympBehaviour}. Firstly, in the regime $h\gg m^2\sim O(1)$ (analogous to the regime \eqref{eq.largeh}) we find the asymptotic growth
\begin{equation}
    E^{(2)}\big(\mathcal{J}^{+,(m)}|\phi_h\rangle\big) \approx \pi^{3/2}\sqrt{h}\bigg(1-\frac{1}{8h} + \frac{16m^4-16m^2+1}{128h^2}\bigg) \ ,
\end{equation}
which should be compared against the large $h$ expansion of the lift of $\phi$ \eqref{eq.phiLift}
\begin{equation}
    E^{(2)}_h\big(|\phi\rangle\big) \approx \pi^{3/2}\sqrt{h}\bigg(1-\frac{1}{8h} + \frac{1}{128h^2}\bigg) \ ,
\end{equation}
and so of interest is the asymptotic behaviour of the ratio
\begin{equation} \label{eq.Jmphilargeh}
    \frac{E^{(2)}\big(\mathcal{J}^{+,(m)}|\phi_h\rangle\big)}{E^{(2)}_h(|\phi\rangle)} \approx 1 + \frac{m^4-m^2}{8h^2} \ .
\end{equation}
This once again shows that information of the excitation on top of a superconformal primary comes only at order $h^{-2}$ in the large $h$ limit, as seen in \eqref{eq.Jphilargeh2} and \eqref{eq.Gphilargeh2}, so once again we see that even asymptotically the lift cannot be a function of the total dimension. However, in contrast to those asymptotic results, \eqref{eq.Jmphilargeh} displays a different dependence on the excitation dimension $h_{ex}$. This was $\sim h_{ex}^3=n^3$ in the cases of \eqref{eq.Jphilargeh2} and \eqref{eq.Gphilargeh2} and $\sim h_{ex}^2=m^4$ in the case of \eqref{eq.Jmphilargeh}, thus suggesting that the lift depends not only on $h_{ex}$ but on how this excitation dimension is partitioned amongst modes. Considering instead the regime of large excitation 
\begin{equation}
    m^2\gg h\sim O(1) \ ,
\end{equation}
then the lift \eqref{eq.JmphiLift} has the leading behaviour
\begin{equation}
    E^{(2)}\big(\mathcal{J}^{+,(m)}|\phi_h\rangle\big) \approx \frac{\pi^{3/2}m}{2}\bigg(1+ \frac{12h-5}{8m^2}\bigg) \ ,
\end{equation}
where information of the primary comes in already at the first subleading order, in contrast to \eqref{eq.Jmphilargeh}. From the exact result \eqref{eq.JmphiLift} we can also look at the regime of large total dimension
\begin{equation}
    h,m^2 \gg1 \ ,
\end{equation}
in which the lift behaves as
\begin{equation}
    E^{(2)}\big(\mathcal{J}^{+,(m)}|\phi_h\rangle\big) \approx \frac{\pi^{3/2}\lambda^2}{2} \frac{2h+m^2}{\sqrt{h+m^2}} \ ,
\end{equation}
which is the first time that we have had an expression valid in this part of parameter space. From this one can then see exactly why there is no $\sim h^{-1}$ term if $h\gg m^2$ since
\begin{equation}
   \frac{2h+m^2}{2\sqrt{h+m^2}} \approx \sqrt{h}\bigg(1+\frac{m^2}{2h}\bigg)\bigg(1-\frac{m^2}{2h}\bigg) \approx \pi^{3/2}\lambda^2 \sqrt{h}\Big(1+O(h^{-2})\Big) \ .
\end{equation}

\section{Discussion} \label{sec.conc}

Understanding the BPS spectrum of the D1-D5 CFT is relevant to identifying and studying the states dual to microstates of certain extremal black holes. BPS states are classified very simply at the locus of moduli space where the theory has a description in terms of a free orbifold theory, in terms of states with purely left- or right-moving excitations above the NS-NS vacuum. However, away from this special region their classification is unknown. This is because multiple short multiplets of the free theory can group together into a long multiplet of the general theory and so `lift'. Thus, members of those free-theory short multiplets are not in fact globally BPS. Progress towards identifying the globally BPS states that can be the duals of black hole microstates can be made by looking at the lifting of various families of states of the free orbifold theory in order to understand general patterns in which states lift and by how much. In this paper we have studied new families of states in the untwisted sector where one copy of the $c=6$ seed theory is excited above the NS-NS vacuum with a descendant of a superconformal primary. More precisely, we have considered the single-copy states
\begin{equation} \label{eq.states}
    \Big\{\, G^{\alpha}_{\Ad,-s}|\phi\rangle \ ,\ \ J^+_{-n}|\phi\rangle\,\Big\} \ ,
\end{equation}
where $|\phi\rangle$ is a primary of the $c=6$ theory with dimension $h$. In Section~\ref{sec.main} the expectation value of the lift of states in the 2-parameter families \eqref{eq.states} was found in terms of an integral of a fixed number of nested sums on a given summand; this summand depends on the descendant level of the state (\textit{i.e.} on $s$ and $n$ of the states \eqref{eq.states}) and the dimension of the primary $h$. Explicit lifting matrices for mode numbers and primary dimensions up to order $7$ were presented for examples of the two types of descendants in \eqref{eq.Gphimat} and \eqref{eq.Jphimat}. We evaluate these sums to obtain the explicit value of the lifts for various 1-parameter subfamilies of states: (i) with the descendant mode number held to be small and $h$ allowed to grow large, (ii) with $h$ held fixed and the descendant level allowed to grow large, and (iii) with both the descendant mode number and primary dimension taken to have the same value. In each case we presented the first $20$ lifts with this explicit data found in tables~\ref{Gtab} and \ref{Jtab}. The new ``series" method used to compute these lifts is different to the ``fields method" used in \cite{Hughes:2023apl} and is well suited to states of the form \eqref{eq.states} where we do not specify the form of the superconformal primary. We chose to present explicit lifts for these states purely for space and computational reasons; similarly to the method used in \cite{Hughes:2023apl}, in principle the method described in this paper can scale arbitrarily given the resources. By having values for the lift of these states up to a suitably high level allows us to observe what seems to be $\sim\sqrt{h}$ growth for large $h$ in the various subfamilies considered. This is similar to the behaviour observed in the two-modes states of \cite{Hughes:2023apl} and the asymptotic $\sqrt{h}$ behaviour seen in the closed-form expression obtained in \cite{Guo:2022ifr} for the lift of single-copy superconformal primaries.

We stress that the lifts found in this paper are in the theory for a general $N$ number of copies of the $c=6$ seed CFT. The method used to compute lifts sees only two copies of the $c=6$ theory at any given time and so it is sufficient to consider a particular pair of copies in intermediate steps. As written in \eqref{eq.GeneralN}, the existence of the $N$ total copies is easily reinstated using combinatorics.

In Section~\ref{ssec:WardIdJphi} we study the expectation value of lifts again for the $J^+_{-n}$ descendants of $|\phi\rangle$, however, this time using a method of Ward identities on the covering space. This method allows us to find a compact form for these lifts valid for arbitrary $h$ and $n$. Because of this exact expression for the lift we are able to derive a leading order $\sim\sqrt{h}$ growth from it in Section~\ref{sssec:largeh}. This leading asymptotic behaviour matches that seen in the closed-form expression of \cite{Guo:2022ifr} for the lift of single-copy primaries. In that paper it was conjectured that this $\sim\sqrt{h}$ behaviour may be a universal feature of the lift of D1-D5-P states and it has since been hinted at in explicit values of lifts for various families of states in \cite{Hughes:2023apl} and Section~\ref{sec.main}. However, what was not known is whether expectation values of lifts are functions purely of the total dimension of the state. The result of \cite{Guo:2022ifr} was that for single-copy superconformal primaries the lift did not depend of the form of the state under consideration, only its total conformal dimension. From the results of Section~\ref{sec:LiftForm} we find that for single-copy superconformal descendant states, the lift depends differently on the primary's dimension $h$ and the descendant excitation $h_{ex}$. Not only this, but by comparison of the lift of $J^+_{-n}|\phi\rangle_h$ from Section~\ref{ssec:WardIdJphi} and the lift of $J^+_{-(2m-1)}\cdots J^+_{-3}J^+_{-1}|\phi\rangle_h$ from Appendix~\ref{app.3} we find that the lift depend on the partitioning of $h_{ex}$ amongst descendant modes. This is seen even at the first subleading order in the large $h$ regime, with the $O(h^{-2})$ term scaling as $\sim h_{ex}^3=n^3$ and $\sim h_{ex}^2=m^4$ for $J^+_{-n}|\phi\rangle_h$ and $J^+_{-(2m-1)}\cdots J^+_{-3}J^+_{-1}|\phi\rangle_h$ respectively.

These results have strengthened the conjecture that the second order expectation value of lifts of D1-D5-P states universally grow as $\sim\sqrt{h}$ in the large $h$ limit. It would be interesting to understand whether this seemingly universal behaviour gives insight into the nature of these very stringy states in the dual string theory. It would also be interesting to see if any lessons can be brought over from the recent successes of identifying candidate black hole states \cite{Chang:2022mjp,Chang:2023zqk,Budzik:2023vtr,Choi:2023znd} in $\mathcal{N}=4$ SYM with gauge group $SU(2)$.

\section*{Acknowledgements}

We would like to thank Bin Guo for discussions. This work is supported in part by DOE grant DE-SC0011726.

\appendix

\section{The $\mathcal N=4$ superconformal algebra} \label{app_cft}

The conventions and notation that we follow is that of appendix A of \cite{Hampton:2018ygz}. In our conventions the indices $\alpha=(+,-)$ and $\bar{\alpha}=(+,-)$ correspond to the fundamental representations of the subgroups $SU(2)_L$ and $SU(2)_R$ arising from rotations on the $S^3$, whereas the indices $A=(+,-)$ and $\dot{A}=(+,-)$ correspond to the subgroups $SU(2)_1$ and $SU(2)_2$ arising from rotations in the internal $T^4$. We use the convention
\begin{equation}
    \epsilon_{+-}=1 \quad,\quad\epsilon^{+-}=-1 \ .
\end{equation}

\subsection{Commutation relations} \label{app.comms}

The current algebra of the small $\mathcal N=4$ superconformal algebra are
\begin{subequations} \label{app com currents}
\begin{align}
\big[L_m,L_n\big] &= \frac{c}{12}m(m^2-1)\delta_{m+n,0}+ (m-n)L_{m+n} \ , \label{LLcomm}\\
\big[J^a_{m},J^b_{n}\big] &= \frac{c}{12}m\,\delta^{ab}\delta_{m+n,0} +  i\epsilon^{ab}_{\,\,\,\,c}\,J^c_{m+n} \ ,\label{JJcomm}\\
\big\{ G^{\alpha}_{\dot{A},r} , G^{\beta}_{\dot{B},s} \big\} &=  \epsilon_{\dot{A}\dot{B}}\bigg[\epsilon^{\alpha\beta}\frac{c}{6}\Big(r^2-\frac14\Big)\delta_{r+s,0} + \big(\sigma^{aT}\big)^{\alpha}_{\gamma}\:\epsilon^{\gamma\beta}(r-s)J^a_{r+s} + \epsilon^{\alpha\beta}L_{r+s} \bigg] \ ,\label{GGcomm}\\
\big[J^a_{m},G^{\alpha}_{\dot{A},r}\big] &= \h\big(\sigma^{aT}\big)^{\alpha}_{\beta}\, G^{\beta}_{\dot{A},m+r} \ ,\label{JGcomm}\\
\big[L_{m},G^{\alpha}_{\dot{A},r}\big] &= \Big(\frac{m}{2}  -r\Big)G^{\alpha}_{\dot{A},m+r} \ ,\label{LGcomm}\\
\big[L_{m},J^a_n\big] &= -nJ^a_{m+n} \ , \label{LJcomm}
\end{align}
\end{subequations}
where $\sigma^{aT}$ are the transpose of the Pauli sigma matrices and with the right-moving currents satisfying an analogous algebra. A different basis for the $J$ currents is also often used; instead of $J^a$ we have $J^3,J^{\pm}$ where
\begin{equation} \label{eq.Jpmdef}
    J^{\pm} = J^1 \pm i J^2 \ .
\end{equation}
The full contracted large $\mathcal{N}=4$ superconformal algebra of the D1-D5 CFT also includes the modes $d^{\alpha A}_r$ and $\alpha_{A\dot{A},n}$ of the free fermions $\psi^{\alpha A}$ and bosons $\partial X_{\!A\dot{A}}$ respectively. The mode expansions of these free fields are
\begin{subequations} \label{BosFerModes}
    \begin{align}
        \partial X_{A\dot{A}}(z) &= -i \sum_n z^{-n-1}\,\alpha_{A\dot{A},n} \ ,\\
        \psi^{\alpha A}(z) &= \sum_r z^{-r-1/2}\, d^{\alpha A}_{r} \ ,
    \end{align}
\end{subequations}
and likewise the inverse relations are
\begin{subequations} \label{BosFerModes2}
    \begin{align}
        \alpha_{A\Ad,n} &= i \oint \frac{dz}{2\pi i} z^{n} \partial X_{A\Ad}(z) \ ,\\
        d^{\alpha A}_{s} &= \oint \frac{dz}{2\pi i} z^{n-\frac12} \psi^{\alpha A}(z) \ .
    \end{align}
\end{subequations}
The brackets of the $\alpha$ and $d$ modes are
\begin{subequations}\label{FundComms}
    \begin{align}
        \big[\alpha_{A\dot{A},n},\alpha_{B\dot{B},m}\big] &= - n\frac{c}{6}\epsilon_{AB}\,\epsilon_{\dot{A}\dot{B}}\, \delta_{n+m,0} \ ,\label{eq.alalcomm}\\
        \big\{d^{\alpha A}_{r},d^{\beta B}_{s}\big\} &= - \frac{c}{6}\epsilon^{\alpha\beta}\epsilon^{AB} \delta_{r+s,0} \ ,\label{eq.ddcomm}\\
        \big[\alpha_{A\dot{A},n}, d^{\beta B}_{s}\big] &= 0 \ ,
    \end{align}
\end{subequations}
and similarly for the right-moving modes. The remaining parts of the full contracted large $\mathcal{N}=4$ superconformal algebra are the current-free field brackets
\begin{subequations}
    \begin{align}
        \big[L_n, \alpha_{\!A\Ad,m}\big] &= -m\, \alpha_{\!A\Ad,n+m} \ ,\\
        \big[L_n,d^{\alpha A}_{s}\big] &= -\Big(\frac{n}{2} + s\Big) d^{\alpha A}_{n+s} \ , \\
        \big[J^a_n, d^{\alpha A}_{s}\big] &= \frac12 \big(\sigma^{Ta}\big)^{\alpha}_{\ \beta}\, d^{\beta A}_{s+n} \ ,\\
        \big[G^{\alpha}_{\Ad,s}, \alpha_{B\Bd,n}\big] &= -in\, \epsilon_{AB}\epsilon_{\Ad\Bd}\, d^{\alpha A}_{s+n} \ ,\\
        \big\{G^{\alpha}_{\Ad,r}, d^{\beta B}_{s}\big\} &= i \epsilon^{\alpha\beta}\epsilon^{AB} \alpha_{\!A\Ad,r+s} \ .
    \end{align}
\end{subequations}
In the computation of Section~\ref{eq.WardidResult} we will need the current OPEs which are given below
\begin{subequations}\label{eq.currOPE}
    \begin{align}
        T(z)T(w) &\sim \frac{c/2}{(z-w)^4} + \frac{2T(w)}{(z-w)^2} + \frac{\partial T(w)}{z-w} \ ,\\
        J^-(z) J^+(w) &\sim \frac{c/6}{(z-w)^2} - \frac{2J^3(w)}{z-w} \ ,\ \ J^3(z) J^-(w) \sim -\frac{J^-(w)}{z-w} \ ,\ \ J^3(z) J^+(w) \sim \frac{J^+(w)}{z-w} \ ,\\
        T(z) J^a(w) &\sim \frac{J^a(w)}{(z-w)^2} + \frac{\partial J^a(w)}{z-w} \ \ ,\quad T(z) G^{\alpha}_{\Ad}(w) \sim \frac32 \frac{G^{\alpha}_{\Ad}(w)}{(z-w)^2} + \frac{\partial G^{\alpha}_{\Ad}(w)}{z-w} \ ,\\
        G^{\alpha}_{\Ad}(z) G^{\beta}_{\Bd}(w) &\sim \ep_{\Ad\Bd}\bigg[\ep^{\alpha\beta} \frac{c/3}{(z-w)^3} + \big(\sigma^{aT}\big)^{\alpha}_{\,\gamma} \ep^{\gamma\beta} \bigg(\frac{2J^a(w)}{(z-w)^2} + \frac{\partial J^a(w)}{z-w} \bigg) + \epsilon^{\alpha\beta}\frac{T(w)}{z-w} \bigg] \ ,\\
        J^3(z) G^{\alpha}_{\Ad}(w) &\sim \frac12 \frac{(\sigma^3)^{\alpha}_{\,\beta} \,G^{\beta}_{\Ad}(w)}{z-w} \ \ ,\quad J^+(z) G^-_{\Ad}(w) \sim \frac{G^+_{\Ad}(w)}{z-w} \ \ ,\quad J^-(z) G^+_{\Ad}(w) \sim \frac{G^-_{\Ad}(w)}{z-w} \ ,
    \end{align}
\end{subequations}
where $\sim$ here is equality up to regular terms.

\subsection{Hermitian conjugation} \label{app.Herm}

Contractions between $su(2)$ indices are done using antisymmetric tensors such as $\epsilon_{\alpha\beta}$ and this gives rise to certain extra negative signs in the definitions of Hermitian conjugates. For the left-moving cylinder supercharge fields we use the following conjugation rules
\begin{equation} \label{Gconj}
    \Big( G^{+}_{+}(\tau,\sigma)\Big)^{\dagger} =-G^{-}_{-}(-\tau,\sigma)\quad,\quad \Big( G^{+}_{-}(\tau,\sigma)\Big)^{\dagger}=G^{-}_{+}(-\tau,\sigma) \ ,
\end{equation}
which leads to the conjugation rules for the modes
\begin{equation}
    \big(G^+_{+,s}\big)^{\dagger} = -G^-_{-,-s} \quad,\quad \big(G^+_{-,s}\big)^{\dagger} = G^-_{+,-s} \ ,
\end{equation}
with the right-moving supercharges obeying similar conjugation conventions. For the degree-2 twist operators our conjugation conventions are
\begin{equation} \label{TwistConj}
    \big(\sigma^{--}(\tau,\sigma)\big)^{\dagger}=-\sigma^{++}(-\tau,\sigma)\quad\ ,\quad \big(\sigma^{-+}(\tau,\sigma)\big)^{\dagger}=\sigma^{+-}(-\tau,\sigma) \ .
\end{equation}
Likewise, in terms of the free fields of the orbifold theory the conjugation conventions we use are
\begin{subequations}
\begin{align} \label{eq.adDagConv}
    \big(\alpha_{++,n}\big)^{\dagger} &= -\alpha_{--,-n} \quad,\quad \big(\alpha_{+-,n}\big)^{\dagger} = \alpha_{-+,-n} \ , \\
    \big(d^{++}_{s}\big)^{\dagger} &= - d^{--}_{-s} \quad,\quad \big(d^{+-}_{s}\big)^{\dagger} =  d^{-+}_{-s} \ .
\end{align}
\end{subequations}

\section{The mode correlator $C^{\beta\alpha}_{\!\Bd\Ad}$} \label{app.2}

In Section~\ref{ssec:GphiLift} the lift of the state $G^{\alpha}_{\Ad,-s}|\phi\rangle$ is found in terms of the $t$-plane mode correlator
\begin{equation} \label{eq.phiGGGGphidef2}
    C^{\beta\alpha}_{\!\Bd\Ad} \equiv \langle\phi|G^{\beta}_{\Bd,s-k_4} G^+_{-,s_2} G^-_{+,s_1} G^{\alpha}_{\Ad,-s+k_3}|\phi\rangle \ .
\end{equation}
Using the shorthand $\Delta\equiv\delta_{s_1+s_2+k_3-k_4,0}$ and the discrete step function
\begin{equation}
    H[n] \equiv \begin{cases}
        \ 1 \quad\text{if } n\geq 0\vspace{4pt}\\
        \ 0 \quad \text{otherwise}
    \end{cases} \ ,
\end{equation}
this correlator can be found straightforwardly using the $\mathcal{N}=4$ superconformal algebra \eqref{app com currents} and the definition of a superconformal primary in this theory \eqref{def primary 2}. We give the result here for reference
\begin{align} \label{eq.phiGGGGphi2}
    C^{\beta\alpha}_{\!\Bd\Ad} &= \ep_{+\Ad}\ep^{-\alpha}\ep_{\Bd-}\ep^{\beta+}\Big(s_1^2-\frac14\Big)\Big((s-k_4)^2-\frac14+h\Big) \delta_{s_1-s+k_3,0}\, \delta_{s_2+s-k_4,0} \,H\big[\!-\!s_2-\tfrac12\big] \nonumber\\
    &\ \ \ -\frac12 \ep_{+\Ad}\big(\sigma^{aT}\big)^{\!-}_{\gamma}\ep^{\gamma\alpha}\ep_{\Bd-}(s_1+s-k_3)\Delta \Bigg[\big(\sigma^{aT}\big)^{+}_{\delta} \ep^{\beta\delta} \Big((s-k_4)^2 -\frac14+h\Big) H\big[s-k_4-\tfrac12\big] \nonumber\\
    &\ \ \ +\big(\sigma^{aT}\big)^{\beta}_{\delta}\ep^{\delta+} \Big((s_1+k_3-k_4)^2-\frac14+h\Big) H\big[\!-\!s_2-\tfrac12\big] \Bigg] H[s-s_1-k_3-1] \nonumber\\
    &\ \ \ -\frac12 \ep_{+\Ad}\ep^{-\alpha}\ep_{\Bd-}\ep^{\beta+}\Delta \Bigg[(s_1-s+k_3-2s_2) \Big((s-k_4)^2-\frac14+h\Big) H\big[s-s_1-s_2-k_3-\tfrac12\big] \nonumber\\
    &\ \ \ +(s_1-3s+k_3+2k_4)\Big((s_1+k_3-k_4)^2-\frac14+h\Big) H\big[\!-\!s_2-\tfrac12\big] \Bigg]H[s-s_1-k_3] \nonumber\\
    &\ \ \ + \ep_{+\Ad}\ep^{-\alpha}\ep_{\Bd-}\ep^{\beta+}h\Big((s-k_4)^2-\frac14+h\Big) \delta_{s_2+s-k_4,0}\, \delta_{s_1-s+k_3,0}\, H[s-s_1-k_3]H\big[\!-\!s_2-\tfrac12\big] \nonumber\\
    &\ \ \ - \ep_{-\Ad}\ep^{+\alpha}\ep_{\Bd+}\ep^{\beta-}\Big(s_2^2-\frac14\Big)\Big((s-k_4)^2-\frac14+h\Big)\delta_{s_2-s+k_3,0}\, \delta_{s_1+s-k_4,0}\, H\big[\!-\!s_1-\tfrac12\big] \nonumber\\
    &\ \ \ -\frac12\ep_{-\Ad}\big(\sigma^{aT}\big)^{\!+}_{\gamma}\ep^{\gamma\alpha}\ep_{\Bd+}(s_2+s-k_3)\Delta \Bigg[\big(\sigma^{aT}\big)^{-}_{\delta}\ep^{\beta\delta}\Big((s-k_4)^2-\frac14+h\Big) H\big[s-k_4-\tfrac12\big] \nonumber\\
    &\ \ \ + \big(\sigma^{aT}\big)^{\beta}_{\delta} \ep^{\delta-}\Big((s-k_4)^2-s_1^2\Big) H[s-s_2-k_3-1]H[s_1+s-k_4-1] \Bigg] H\big[\!-\!s_1-\tfrac12\big] \nonumber\\
    &\ \ \ - \frac12\ep_{-\Ad}\ep^{+\alpha}\ep_{\Bd+}\ep^{\beta-}\Delta \Bigg[(s_2-s+k_3-2s_1)\Big((s-k_4)^2-\frac14+h\Big)H\big[s-k_4-\tfrac12\big] \nonumber\\
    &\ \ \ + (s_1+s-k_4)\Big((s_1+s-k_4)^2-1\Big)H[s-s_2-k_3]H[s+s_1-k_4] \nonumber\\
    &\ \ \ + 2h(s_1-s_2+2s-k_3-k_4) H[s-s_2-k_3]H[s+s_1-k_4]\Bigg] H\big[\!-\!s_1-\tfrac12\big] \nonumber\\
    &\ \ \ - \ep_{-\Ad}\ep^{+\alpha}\ep_{\Bd+}\ep^{\beta-} \bigg( h\Big((s-k_4)^2-\frac14\Big) + h^2 \bigg) \delta_{s_1+s-k_4,0}\,\delta_{s_2-s+k_3,0}\, H\big[\!-\!s_1-\tfrac12\big] \nonumber\\
    &\ \ \ + \ep_{\Bd\Ad}\ep^{\beta\alpha} \Big((s-k_4)^2-\frac14\Big)\Big(s_2^2-\frac14+h\Big) \delta_{k_3-k_4,0}\,\delta_{s_1+s_2,0}\,H\big[\!-\!s_1-\tfrac12\big] \nonumber\\
    &\ \ \ - \frac12\ep_{\Bd\Ad}\big(\sigma^{aT}\big)^{\!\beta}_{\gamma}\ep^{\gamma\alpha}(2s-k_3-k_4)\Delta \Bigg[ \big(\sigma^{aT}\big)^{+}_{\delta}\ep^{\delta-} (s_2-s_1)(k_3-k_4) H[-s_1-s_2-1] \nonumber\\
    &\ \ \ +\big(\sigma^{aT}\big)^{-}_{\delta}\ep^{\delta+}\Big((s_1+k_3-k_4)^2-\frac14+h\Big) H\big[\!-\!s_2-\tfrac12\big] \Bigg] H[k_3-k_4-1] H\big[\!-\!s_1-\tfrac12\big] \nonumber\\
    &\ \ \ + \ep_{\Bd\Ad}\ep^{\beta\alpha} H\big[\!-\!s_1-\tfrac12\big] \bigg(h \Big(s_2^2 -\frac14\Big) + h^2 \bigg) \delta_{s_1+s_2,0}\,\delta_{k_3-k_4,0} \nonumber\\
    &\ \ \ + \frac12\ep_{\Bd\Ad}\ep^{\beta\alpha}H\big[\!-\!s_1-\tfrac12\big]H[k_3-k_4]\,\Delta \Bigg[ (k_3-k_4)\Big((k_3-k_4)^2-1\Big) H[-s_1-s_2] \nonumber\\
    &\ \ \ + 2h(k_3-k_4-s_1-s_2) H[-s_1-s_2] - (k_3-k_4-2s_1)\Big(s_2^2-\frac14+h\Big) H\big[\!-\!s_2-\tfrac12\big] \Bigg] \, ,
\end{align}

\section{Lifting of $\mathcal{J}^{+,(m)}|\phi\rangle$} \label{app.3}

In this appendix we consider lift of the family of untwisted sector states with the NS vacuum on all copies bar one, on which there us the excited state
\begin{equation}
    \mathcal{J}^{+,(m)}|\phi\rangle \equiv J^+_{-(2m-1)}\cdots J^+_{-3}J^+_{-1}|\phi\rangle \ ,
\end{equation}
where $\phi$ is a single-copy superconformal primary of dimension $h$. The Hermitian conjugate of this state is given by
\begin{equation}
    \langle\phi|\mathcal{J}^{-,(m)} \equiv \langle\phi| J^-_{-1}J^-_{-3}\cdots J^-_{-(2m-1)} \ ,
\end{equation}
and its quantum numbers are 
\begin{equation}
    (h^{(m)},j^{(m)}) = (h+m^2,m) \ .
\end{equation}
This particular family of states has been chosen for the fact that it is the unique spectral flow of the single-copy superconformal primary $|\phi\rangle$ under a flow by $\eta=2m$ units. Since $m\in \mathbb{Z}^+$ here the spectral flow maps the NS sector theory back to the NS sector. The mapping of quantum numbers under spectral flow can be found in \eqref{sfDims}. This fact can then be leveraged in the lifting computation in a manner very similar to that used in \cite{Hampton:2018ygz}: once on the covering space of the key amplitude required for the lift, the inverse spectral flow can be used to effectively remove the $\mathcal{J}^{+,(m)}$ excitations. This roughly reduces the problem to the calculation of the lift of $|\phi\rangle$ which was done in \cite{Guo:2022ifr}, albeit with a different method than used here.

As explained in Section~\ref{sec:LiftForm}, we need to compute the $w$-cylinder left-moving amplitude
\begin{equation} 
    A^{(1)(1)}_{h,m}(w_2,w_1) \equiv \langle\phi|\mathcal{J}^{-,(m)}\,\big(G^+_{-,-\frac12}\sigma_2^-\big)(w_2) \big(G^-_{+,-\frac12}\sigma_2^+\big)(w_1) \,\mathcal{J}^{+,(m)}|\phi\rangle \ ,
\end{equation}
which upon mapping to the covering space as per the steps~\ref{Series1}-\ref{Series3} of Section~\ref{ssec:GphiLift} is given by
\begin{align} \label{eq.A11Jmphi1}
    A^{(1)(1)}_{h,m}(w_2,w_1) &= f_1 f_2 f_3 \bigg(\frac{dt}{dz}\bigg|_{-a}\bigg)^{\!m^2} \bigg(\frac{-a-t_1}{-a-t_2}\bigg)^{\!m} \,C^{(m)}_{h} \ ,
\end{align}
where the factors $f_1$, $f_2$ and $f_3$ are given in \eqref{eq.baseamp}, \eqref{eq.f2} and \eqref{eq.f3} respectively and $C^{(m)}_h$ is the $t$-plane correlator
\begin{equation} \label{eq.Cmh}
     C^{(m)}_h \equiv \big< \big(\mathcal{J}^{-,(m)}\phi\big)(\infty)\, G^+_-(t_2) \,G^-_+(t_1)\, \big(\mathcal{J}^{+,(m)}\phi\big)(-a) \big> \ .
\end{equation}
In \eqref{eq.A11Jmphi1} the factors of $\big(\frac{dt}{dz}\big|_{-a}\big)^{\!m^2}$ and $\big(\frac{-a-t_1}{-a-t_2}\big)^{\!m}$ come respectively from the transformation of the initial state $\mathcal{J}^{+,(m)}$ modes under the map to the covering space \eqref{eq.ztotmap} and the spectral flows used to remove spin fields (see step \ref{Series2}). As explained above, the $t$-plane correlator \eqref{eq.Cmh} can be evaluated by spectral flowing away the $\mathcal{J}^{+,(m)}$ by a spectral flow by $\eta=-2m$ units around $t=-a$, which also automatically removes the $\mathcal{J}^{-,(m)}$ excitation at $t=\infty$. Under this spectral flow, the $G$ field insertions transform as per \eqref{eq.GsfRule} (the $\phi$ do not since they are not charged under $J^3_0$) yielding
\begin{align}
    C^{(m)}_h = \bigg(\frac{t_2+a}{t_1+a}\bigg)^{\!m} A^{+-}_{-+}(w_2,w_1;h) \ , 
\end{align}
with $A^{+-}_{-+}(w_2,w_1;h)$ defined in \eqref{eq.phiGGphibase}. The amplitude $A^{(1)(1)}$ is then
\begin{align}
    A^{(1)(1)}_{h,m}(w_2,w_1) = \frac{a+b(2h-1)}{16\,b} \bigg(\frac{a}{a-b}\bigg)^{\!h+m^2-1} \ .
\end{align}
Replacing $a$ and $b$ as functions of $w$ as per \eqref{abDef} we get
\begin{align}
    A^{(1)(1)}_{h,m}(w_2,w_1) = \frac{hx^2-2(h-1)x+h}{2^{2h+2m^2+1} x^{h+m^2-1}} \frac{(x+1)^{2(h+m^2-1)}}{(x-1)^{2}} \ ,
\end{align}
where we use the shorthand $x=e^{w/2}$. The amplitude with the other independent copy structure is obtained by $x\to -x$ as
\begin{equation}
    A^{(2)(1)}_{h,m}(w_2,w_1) = (-1)^{h+m^2+1}\frac{hx^2+2(h-1)x+h}{2^{2h+2m^2+1} x^{h+m^2-1}} \frac{(x-1)^{2(h+m^2-1)}}{(x+1)^{2}} \ .
\end{equation}
Considering the contour integral $I_{C_2}$ as defined in \eqref{eq.IC2}, the right-moving part contributes a factor of $-1$ and we then project onto the $x^0$ part of the amplitude, giving
\begin{align}
    I^{(1)(1)}_{C_2} &\equiv - A^{(1)(1)}_{h,m}(w_2,w_1)\Big|_{x^0} \nonumber\\
    &= -\sum_{\nu=0}^{2}\frac{c_{\nu}}{2^{2h+2m^2+1} x^{h+m^2-1-\nu}} \sum_{\ell_3=0}^{\infty}\sum_{\ell_2=0}^{\min(\ell_3,2(h+m^2-1))} (\ell_3-\ell_2+1)\, {}^{2(h+m^2-1)}C_{\ell_2} \,x^{\ell_3} \bigg|_{x^0} \nonumber\\
    &= -\sum_{\nu=0}^{2}\sum_{\ell_2=0}^{h+m^2-1-\nu} \frac{c_{\nu}(h+m^2-\nu-\ell_2)}{2^{2h+2m^2+1}}\, {}^{2(h+m^2-1)}C_{\ell_2} \nonumber\\
    &= -\frac{(2h+m^2-1)\,\Gamma(h+m^2-\frac12)}{8\sqrt{\pi}\,\Gamma(h+m^2)} \ ,
\end{align}
where the term $\ell_3=h+m^2-1-\nu$ was picked out and $c_\nu$ are coefficients such that
\begin{equation} \label{eq.cnudefapp}
    \sum_{\nu=0}^{2}c_\nu\, x^\nu = hx^2 -2(h-1)x + h \ .
\end{equation}
For $A^{(2)(1)}$ the integral results in
\begin{align}
    I^{(2)(1)}_{C_2} &\equiv - A^{(2)(1)}_{h,m}(w_2,w_1)\Big|_{x^0} \nonumber\\
    &= \sum_{\nu=0}^{2} \frac{c_{\nu}(-1)^{h+m^2+\nu}}{2^{2h+2m^2+1}} \sum_{\ell_3=0}^{\infty}\sum_{\ell_2=0}^{\min(\ell_3,2(h+m^2-1))} (-1)^{\ell_3}\,(\ell_3-\ell_2+1)\, {}^{2(h+m^2-1)}C_{\ell_2} \,x^{\ell_3} \bigg|_{x^0} \nonumber\\
    &= \sum_{\nu=0}^{2} \sum_{\ell_2=0}^{h+m^2-1-\nu} \frac{(-1)^{2\nu+1} c_{\nu}}{2^{2h+2m^2+1}} (h+m^2-\nu-\ell_2)\, {}^{2(h+m^2-1)}C_{\ell_2} \nonumber\\
    &= I^{(1)(1)}_{C_2} \ .
\end{align}
From \eqref{eq.lift} the lift is then
\begin{align} \label{eq.JsphiLift}
    E^{(2)}\big(\mathcal{J}^{+,(m)}|\phi_h\rangle\big) &= \pi^{3/2}\lambda^2 \frac{(2h+m^2-1)\,\Gamma(h+m^2-\frac12)}{2\,\Gamma(h+m^2)} \ .
\end{align}
The asymptotic behaviour of this lift for large dimension is discussed in Section \ref{sssec:largeh}. We note that for $m=1$ the state is a single-copy descendant state of $\phi$ and, as expected from Section~\ref{ssec:symmChecks}, the lift obeys the relation
\begin{equation}
    E^{(2)}\big(\mathcal{J}^{+,(1)}|\phi_h\rangle\big) = \pi^{3/2}\lambda^2\frac{\Gamma(h+\frac12)}{\Gamma(h)} = E^{(2)}\big(|\phi_{h}\rangle\big) \ .
\end{equation}

\bibliographystyle{JHEP}
\bibliography{DescendantLiftPaper.bib}

\providecommand{\href}[2]{#2}\begingroup\raggedright\begin{thebibliography}{10}

\bibitem{Strominger:1996sh}
A.~Strominger and C.~Vafa, \emph{{Microscopic origin of the Bekenstein-Hawking
  entropy}}, \href{https://doi.org/10.1016/0370-2693(96)00345-0}{\emph{Phys.
  Lett. B} {\bfseries 379} (1996) 99}
  [\href{https://arxiv.org/abs/hep-th/9601029}{{\ttfamily hep-th/9601029}}].

\bibitem{Maldacena:1999bp}
J.M.~Maldacena, G.W.~Moore and A.~Strominger, \emph{{Counting BPS black holes
  in toroidal Type II string theory}},
  \href{https://arxiv.org/abs/hep-th/9903163}{{\ttfamily hep-th/9903163}}.

\bibitem{Bena:2022ldq}
I.~Bena, E.J.~Martinec, S.D.~Mathur and N.P.~Warner, \emph{{Snowmass White
  Paper: Micro- and Macro-Structure of Black Holes}},
  \href{https://arxiv.org/abs/2203.04981}{{\ttfamily 2203.04981}}.

\bibitem{Bena:2022rna}
I.~Bena, E.J.~Martinec, S.D.~Mathur and N.P.~Warner, \emph{{Fuzzballs and
  Microstate Geometries: Black-Hole Structure in String Theory}},
  \href{https://arxiv.org/abs/2204.13113}{{\ttfamily 2204.13113}}.

\bibitem{Heidmann:2021cms}
P.~Heidmann, \emph{{Non-BPS floating branes and bubbling geometries}},
  \href{https://doi.org/10.1007/JHEP02(2022)162}{\emph{JHEP} {\bfseries 02}
  (2022) 162} [\href{https://arxiv.org/abs/2112.03279}{{\ttfamily
  2112.03279}}].

\bibitem{Bah:2022pdn}
I.~Bah and P.~Heidmann, \emph{{Non-BPS bubbling geometries in AdS$_{3}$}},
  \href{https://doi.org/10.1007/JHEP02(2023)133}{\emph{JHEP} {\bfseries 02}
  (2023) 133} [\href{https://arxiv.org/abs/2210.06483}{{\ttfamily
  2210.06483}}].

\bibitem{Heidmann:2022zyd}
P.~Heidmann and A.~Houppe, \emph{{Solitonic excitations in AdS$_{2}$}},
  \href{https://doi.org/10.1007/JHEP07(2023)186}{\emph{JHEP} {\bfseries 07}
  (2023) 186} [\href{https://arxiv.org/abs/2212.05065}{{\ttfamily
  2212.05065}}].

\bibitem{Bah:2022yji}
I.~Bah, P.~Heidmann and P.~Weck, \emph{{Schwarzschild-like topological
  solitons}}, \href{https://doi.org/10.1007/JHEP08(2022)269}{\emph{JHEP}
  {\bfseries 08} (2022) 269}
  [\href{https://arxiv.org/abs/2203.12625}{{\ttfamily 2203.12625}}].

\bibitem{Bah:2023ows}
I.~Bah and P.~Heidmann, \emph{{Geometric Resolution of Schwarzschild Horizon}},
   \href{https://arxiv.org/abs/2303.10186}{{\ttfamily 2303.10186}}.

\bibitem{Lunin:2001jy}
O.~Lunin and S.D.~Mathur, \emph{{AdS / CFT duality and the black hole
  information paradox}},
  \href{https://doi.org/10.1016/S0550-3213(01)00620-4}{\emph{Nucl. Phys. B}
  {\bfseries 623} (2002) 342}
  [\href{https://arxiv.org/abs/hep-th/0109154}{{\ttfamily hep-th/0109154}}].

\bibitem{Mathur:2005zp}
S.D.~Mathur, \emph{{The Fuzzball proposal for black holes: An Elementary
  review}}, \href{https://doi.org/10.1002/prop.200410203}{\emph{Fortsch. Phys.}
  {\bfseries 53} (2005) 793}
  [\href{https://arxiv.org/abs/hep-th/0502050}{{\ttfamily hep-th/0502050}}].

\bibitem{Kanitscheider:2007wq}
I.~Kanitscheider, K.~Skenderis and M.~Taylor, \emph{{Fuzzballs with internal
  excitations}},
  \href{https://doi.org/10.1088/1126-6708/2007/06/056}{\emph{JHEP} {\bfseries
  06} (2007) 056} [\href{https://arxiv.org/abs/0704.0690}{{\ttfamily
  0704.0690}}].

\bibitem{Bena:2007kg}
I.~Bena and N.P.~Warner, \emph{{Black holes, black rings and their
  microstates}}, \href{https://doi.org/10.1007/978-3-540-79523-0_1}{\emph{Lect.
  Notes Phys.} {\bfseries 755} (2008) 1}
  [\href{https://arxiv.org/abs/hep-th/0701216}{{\ttfamily hep-th/0701216}}].

\bibitem{Chowdhury:2010ct}
B.D.~Chowdhury and A.~Virmani, \emph{{Modave Lectures on Fuzzballs and Emission
  from the D1-D5 System}},  in \emph{{5th Modave Summer School in Mathematical
  Physics}}, 1, 2010 [\href{https://arxiv.org/abs/1001.1444}{{\ttfamily
  1001.1444}}].

\bibitem{Shigemori:2020yuo}
M.~Shigemori, \emph{{Superstrata}},
  \href{https://doi.org/10.1007/s10714-020-02698-8}{\emph{Gen. Rel. Grav.}
  {\bfseries 52} (2020) 51} [\href{https://arxiv.org/abs/2002.01592}{{\ttfamily
  2002.01592}}].

\bibitem{Vafa:1995bm}
C.~Vafa, \emph{{Instantons on D-branes}},
  \href{https://doi.org/10.1016/0550-3213(96)00075-2}{\emph{Nucl. Phys. B}
  {\bfseries 463} (1996) 435}
  [\href{https://arxiv.org/abs/hep-th/9512078}{{\ttfamily hep-th/9512078}}].

\bibitem{Dijkgraaf:1998gf}
R.~Dijkgraaf, \emph{{Instanton strings and hyperKahler geometry}},
  \href{https://doi.org/10.1016/S0550-3213(98)00869-4}{\emph{Nucl. Phys. B}
  {\bfseries 543} (1999) 545}
  [\href{https://arxiv.org/abs/hep-th/9810210}{{\ttfamily hep-th/9810210}}].

\bibitem{Larsen:1999uk}
F.~Larsen and E.J.~Martinec, \emph{{U(1) charges and moduli in the D1 - D5
  system}}, \href{https://doi.org/10.1088/1126-6708/1999/06/019}{\emph{JHEP}
  {\bfseries 06} (1999) 019}
  [\href{https://arxiv.org/abs/hep-th/9905064}{{\ttfamily hep-th/9905064}}].

\bibitem{Seiberg:1999xz}
N.~Seiberg and E.~Witten, \emph{{The D1 / D5 system and singular CFT}},
  \href{https://doi.org/10.1088/1126-6708/1999/04/017}{\emph{JHEP} {\bfseries
  04} (1999) 017} [\href{https://arxiv.org/abs/hep-th/9903224}{{\ttfamily
  hep-th/9903224}}].

\bibitem{Arutyunov:1997gi}
G.E.~Arutyunov and S.A.~Frolov, \emph{{Four graviton scattering amplitude from
  S**N R**8 supersymmetric orbifold sigma model}},
  \href{https://doi.org/10.1016/S0550-3213(98)00326-5}{\emph{Nucl. Phys. B}
  {\bfseries 524} (1998) 159}
  [\href{https://arxiv.org/abs/hep-th/9712061}{{\ttfamily hep-th/9712061}}].

\bibitem{Arutyunov:1997gt}
G.E.~Arutyunov and S.A.~Frolov, \emph{{Virasoro amplitude from the S**N R**24
  orbifold sigma model}},
  \href{https://doi.org/10.1007/BF02557107}{\emph{Theor. Math. Phys.}
  {\bfseries 114} (1998) 43}
  [\href{https://arxiv.org/abs/hep-th/9708129}{{\ttfamily hep-th/9708129}}].

\bibitem{Jevicki:1998bm}
A.~Jevicki, M.~Mihailescu and S.~Ramgoolam, \emph{{Gravity from CFT on S**N(X):
  Symmetries and interactions}},
  \href{https://doi.org/10.1016/S0550-3213(00)00147-4}{\emph{Nucl. Phys. B}
  {\bfseries 577} (2000) 47}
  [\href{https://arxiv.org/abs/hep-th/9907144}{{\ttfamily hep-th/9907144}}].

\bibitem{David:2002wn}
J.R.~David, G.~Mandal and S.R.~Wadia, \emph{{Microscopic formulation of black
  holes in string theory}},
  \href{https://doi.org/10.1016/S0370-1573(02)00271-5}{\emph{Phys. Rept.}
  {\bfseries 369} (2002) 549}
  [\href{https://arxiv.org/abs/hep-th/0203048}{{\ttfamily hep-th/0203048}}].

\bibitem{Maldacena:1997re}
J.M.~Maldacena, \emph{{The Large N limit of superconformal field theories and
  supergravity}}, \href{https://doi.org/10.1023/A:1026654312961}{\emph{Adv.
  Theor. Math. Phys.} {\bfseries 2} (1998) 231}
  [\href{https://arxiv.org/abs/hep-th/9711200}{{\ttfamily hep-th/9711200}}].

\bibitem{Gava:2002xb}
E.~Gava and K.S.~Narain, \emph{{Proving the PP wave / CFT(2) duality}},
  \href{https://doi.org/10.1088/1126-6708/2002/12/023}{\emph{JHEP} {\bfseries
  12} (2002) 023} [\href{https://arxiv.org/abs/hep-th/0208081}{{\ttfamily
  hep-th/0208081}}].

\bibitem{Chang:2022mjp}
C.-M.~Chang and Y.-H.~Lin, \emph{{Words to describe a black hole}},
  \href{https://doi.org/10.1007/JHEP02(2023)109}{\emph{JHEP} {\bfseries 02}
  (2023) 109} [\href{https://arxiv.org/abs/2209.06728}{{\ttfamily
  2209.06728}}].

\bibitem{Chang:2023zqk}
C.-M.~Chang, L.~Feng, Y.-H.~Lin and Y.-X.~Tao, \emph{{Decoding stringy
  near-supersymmetric black holes}},
  \href{https://arxiv.org/abs/2306.04673}{{\ttfamily 2306.04673}}.

\bibitem{Budzik:2023vtr}
K.~Budzik, H.~Murali and P.~Vieira, \emph{{Following Black Hole States}},
  \href{https://arxiv.org/abs/2306.04693}{{\ttfamily 2306.04693}}.

\bibitem{Choi:2023znd}
S.~Choi, S.~Kim, E.~Lee, S.~Lee and J.~Park, \emph{{Towards quantum black hole
  microstates}},  \href{https://arxiv.org/abs/2304.10155}{{\ttfamily
  2304.10155}}.

\bibitem{Lima:2020boh}
A.A.~Lima, G.M.~Sotkov and M.~Stanishkov, \emph{{Microstate Renormalization in
  Deformed D1-D5 SCFT}},
  \href{https://doi.org/10.1016/j.physletb.2020.135630}{\emph{Phys. Lett. B}
  {\bfseries 808} (2020) 135630}
  [\href{https://arxiv.org/abs/2005.06702}{{\ttfamily 2005.06702}}].

\bibitem{Lima:2020kek}
A.A.~Lima, G.M.~Sotkov and M.~Stanishkov, \emph{{Renormalization of twisted
  Ramond fields in D1-D5 SCFT$_{2}$}},
  \href{https://doi.org/10.1007/JHEP03(2021)202}{\emph{JHEP} {\bfseries 03}
  (2021) 202} [\href{https://arxiv.org/abs/2010.00172}{{\ttfamily
  2010.00172}}].

\bibitem{Lima:2020nnx}
A.A.~Lima, G.M.~Sotkov and M.~Stanishkov, \emph{{Correlation functions of
  composite Ramond fields in deformed D1-D5 orbifold SCFT$_2$}},
  \href{https://doi.org/10.1103/PhysRevD.102.106004}{\emph{Phys. Rev. D}
  {\bfseries 102} (2020) 106004}
  [\href{https://arxiv.org/abs/2006.16303}{{\ttfamily 2006.16303}}].

\bibitem{Lima:2020urq}
A.A.~Lima, G.M.~Sotkov and M.~Stanishkov, \emph{{Dynamics of R-neutral Ramond
  fields in the D1-D5 SCFT}},
  \href{https://arxiv.org/abs/2012.08021}{{\ttfamily 2012.08021}}.

\bibitem{Lima:2021wrz}
A.A.~Lima, G.M.~Sotkov and M.~Stanishkov, \emph{{On the Dynamics of Protected
  Ramond Ground States in the D1-D5 CFT}},
  \href{https://arxiv.org/abs/2103.04459}{{\ttfamily 2103.04459}}.

\bibitem{Lima:2021xqj}
A.A.~Lima, G.M.~Sotkov and M.~Stanishkov, \emph{{Ramond States of the D1-D5 CFT
  away from the free orbifold point}},  in \emph{{14th International Workshop
  on Lie Theory and Its Applications in Physics}}, 12, 2021
  [\href{https://arxiv.org/abs/2112.10832}{{\ttfamily 2112.10832}}].

\bibitem{Lima:2022cnq}
A.A.~Lima, G.M.~Sotkov and M.~Stanishkov, \emph{{Four-point functions with
  multi-cycle fields in symmetric orbifolds and the D1-D5 CFT}},
  \href{https://arxiv.org/abs/2202.12424}{{\ttfamily 2202.12424}}.

\bibitem{Benjamin:2022jin}
N.~Benjamin, S.~Bintanja, A.~Castro and J.~Hollander, \emph{{The stranger
  things of symmetric product orbifold CFTs}},
  \href{https://doi.org/10.1007/JHEP11(2022)054}{\emph{JHEP} {\bfseries 11}
  (2022) 054} [\href{https://arxiv.org/abs/2208.11141}{{\ttfamily
  2208.11141}}].

\bibitem{Guo:2022and}
B.~Guo and S.D.~Mathur, \emph{{Dynamical evolution in the D1D5 CFT}},
  \href{https://doi.org/10.1007/JHEP12(2022)107}{\emph{JHEP} {\bfseries 12}
  (2022) 107} [\href{https://arxiv.org/abs/2208.05992}{{\ttfamily
  2208.05992}}].

\bibitem{Guo:2022sos}
B.~Guo and S.D.~Hampton, \emph{{Bootstrapping the effect of the twist operator
  in symmetric orbifold CFTs}},
  \href{https://doi.org/10.1007/JHEP02(2023)184}{\emph{JHEP} {\bfseries 02}
  (2023) 184} [\href{https://arxiv.org/abs/2206.01623}{{\ttfamily
  2206.01623}}].

\bibitem{Guo:2023czj}
B.~Guo and S.D.~Hampton, \emph{{Bootstrapping multi-wound twist effects in
  symmetric orbifold CFTs}},
  \href{https://arxiv.org/abs/2307.14255}{{\ttfamily 2307.14255}}.

\bibitem{Burrington:2022dii}
B.A.~Burrington and A.W.~Peet, \emph{{Fractional conformal descendants and
  correlators in general 2D S$_{N}$ orbifold CFTs at large N}},
  \href{https://doi.org/10.1007/JHEP02(2023)091}{\emph{JHEP} {\bfseries 02}
  (2023) 091} [\href{https://arxiv.org/abs/2211.04633}{{\ttfamily
  2211.04633}}].

\bibitem{Burrington:2022rtr}
B.A.~Burrington and A.W.~Peet, \emph{{Larger twists and higher n-point
  functions with fractional conformal descendants in S$_{N}$ orbifold CFTs at
  large N}}, \href{https://doi.org/10.1007/JHEP02(2023)229}{\emph{JHEP}
  {\bfseries 02} (2023) 229}
  [\href{https://arxiv.org/abs/2212.03993}{{\ttfamily 2212.03993}}].

\bibitem{Gaberdiel:2015uca}
M.R.~Gaberdiel, C.~Peng and I.G.~Zadeh, \emph{{Higgsing the stringy higher spin
  symmetry}}, \href{https://doi.org/10.1007/JHEP10(2015)101}{\emph{JHEP}
  {\bfseries 10} (2015) 101}
  [\href{https://arxiv.org/abs/1506.02045}{{\ttfamily 1506.02045}}].

\bibitem{Hampton:2018ygz}
S.~Hampton, S.D.~Mathur and I.G.~Zadeh, \emph{{Lifting of D1-D5-P states}},
  \href{https://doi.org/10.1007/JHEP01(2019)075}{\emph{JHEP} {\bfseries 01}
  (2019) 075} [\href{https://arxiv.org/abs/1804.10097}{{\ttfamily
  1804.10097}}].

\bibitem{Guo:2019ady}
B.~Guo and S.D.~Mathur, \emph{{Lifting of level-1 states in the D1D5 CFT}},
  \href{https://doi.org/10.1007/JHEP03(2020)028}{\emph{JHEP} {\bfseries 03}
  (2020) 028} [\href{https://arxiv.org/abs/1912.05567}{{\ttfamily
  1912.05567}}].

\bibitem{Guo:2020gxm}
B.~Guo and S.D.~Mathur, \emph{{Lifting at higher levels in the D1D5 CFT}},
  \href{https://doi.org/10.1007/JHEP11(2020)145}{\emph{JHEP} {\bfseries 11}
  (2020) 145} [\href{https://arxiv.org/abs/2008.01274}{{\ttfamily
  2008.01274}}].

\bibitem{Benjamin:2021zkn}
N.~Benjamin, C.A.~Keller and I.G.~Zadeh, \emph{{Lifting 1/4-BPS states in
  $AdS_{3}\times S^{3}\times T^{4}$}},
  \href{https://doi.org/10.1007/JHEP10(2021)089}{\emph{JHEP} {\bfseries 10}
  (2021) 089} [\href{https://arxiv.org/abs/2107.00655}{{\ttfamily
  2107.00655}}].

\bibitem{Guo:2022ifr}
B.~Guo, M.R.R.~Hughes, S.D.~Mathur and M.~Mehta, \emph{{Universal lifting in
  the D1-D5 CFT}}, \href{https://doi.org/10.1007/JHEP10(2022)148}{\emph{JHEP}
  {\bfseries 10} (2022) 148}
  [\href{https://arxiv.org/abs/2208.07409}{{\ttfamily 2208.07409}}].

\bibitem{Hughes:2023apl}
M.R.R.~Hughes, S.D.~Mathur and M.~Mehta, \emph{{Lifting of two-mode states in
  the D1-D5 CFT}},  \href{https://arxiv.org/abs/2309.03321}{{\ttfamily
  2309.03321}}.

\bibitem{Gaberdiel:2018rqv}
M.R.~Gaberdiel and R.~Gopakumar, \emph{{Tensionless string spectra on
  AdS$_{3}$}}, \href{https://doi.org/10.1007/JHEP05(2018)085}{\emph{JHEP}
  {\bfseries 05} (2018) 085}
  [\href{https://arxiv.org/abs/1803.04423}{{\ttfamily 1803.04423}}].

\bibitem{Eberhardt:2018ouy}
L.~Eberhardt, M.R.~Gaberdiel and R.~Gopakumar, \emph{{The Worldsheet Dual of
  the Symmetric Product CFT}},
  \href{https://doi.org/10.1007/JHEP04(2019)103}{\emph{JHEP} {\bfseries 04}
  (2019) 103} [\href{https://arxiv.org/abs/1812.01007}{{\ttfamily
  1812.01007}}].

\bibitem{Eberhardt:2019ywk}
L.~Eberhardt, M.R.~Gaberdiel and R.~Gopakumar, \emph{{Deriving the
  AdS$_{3}$/CFT$_{2}$ correspondence}},
  \href{https://doi.org/10.1007/JHEP02(2020)136}{\emph{JHEP} {\bfseries 02}
  (2020) 136} [\href{https://arxiv.org/abs/1911.00378}{{\ttfamily
  1911.00378}}].

\bibitem{Eberhardt:2020akk}
L.~Eberhardt, \emph{{AdS$_{3}$/CFT$_{2}$ at higher genus}},
  \href{https://doi.org/10.1007/JHEP05(2020)150}{\emph{JHEP} {\bfseries 05}
  (2020) 150} [\href{https://arxiv.org/abs/2002.11729}{{\ttfamily
  2002.11729}}].

\bibitem{Eberhardt:2019qcl}
L.~Eberhardt and M.R.~Gaberdiel, \emph{{String theory on AdS$_3$ and the
  symmetric orbifold of Liouville theory}},
  \href{https://doi.org/10.1016/j.nuclphysb.2019.114774}{\emph{Nucl. Phys. B}
  {\bfseries 948} (2019) 114774}
  [\href{https://arxiv.org/abs/1903.00421}{{\ttfamily 1903.00421}}].

\bibitem{Dei:2019osr}
A.~Dei, L.~Eberhardt and M.R.~Gaberdiel, \emph{{Three-point functions in
  AdS$_{3}$/CFT$_{2}$ holography}},
  \href{https://doi.org/10.1007/JHEP12(2019)012}{\emph{JHEP} {\bfseries 12}
  (2019) 012} [\href{https://arxiv.org/abs/1907.13144}{{\ttfamily
  1907.13144}}].

\bibitem{Schwimmer:1986mf}
A.~Schwimmer and N.~Seiberg, \emph{{Comments on the N=2, N=3, N=4
  Superconformal Algebras in Two-Dimensions}},
  \href{https://doi.org/10.1016/0370-2693(87)90566-1}{\emph{Phys. Lett. B}
  {\bfseries 184} (1987) 191}.

\bibitem{Sevrin:1988ew}
A.~Sevrin, W.~Troost and A.~Van~Proeyen, \emph{{Superconformal Algebras in
  Two-Dimensions with N=4}},
  \href{https://doi.org/10.1016/0370-2693(88)90645-4}{\emph{Phys. Lett. B}
  {\bfseries 208} (1988) 447}.

\bibitem{Gaberdiel:2015mra}
M.R.~Gaberdiel and R.~Gopakumar, \emph{{Stringy Symmetries and the Higher Spin
  Square}}, \href{https://doi.org/10.1088/1751-8113/48/18/185402}{\emph{J.
  Phys. A} {\bfseries 48} (2015) 185402}
  [\href{https://arxiv.org/abs/1501.07236}{{\ttfamily 1501.07236}}].

\bibitem{Guo:2021uiu}
B.~Guo and S.~Hampton, \emph{{Partial Spectral Flow in the D1D5 CFT}},
  \href{https://arxiv.org/abs/2112.10573}{{\ttfamily 2112.10573}}.

\bibitem{Lunin:2001fv}
O.~Lunin and S.D.~Mathur, \emph{{Metric of the multiply wound rotating
  string}}, \href{https://doi.org/10.1016/S0550-3213(01)00321-2}{\emph{Nucl.
  Phys. B} {\bfseries 610} (2001) 49}
  [\href{https://arxiv.org/abs/hep-th/0105136}{{\ttfamily hep-th/0105136}}].

\bibitem{Lunin:2000yv}
O.~Lunin and S.D.~Mathur, \emph{{Correlation functions for M**N / S(N)
  orbifolds}}, \href{https://doi.org/10.1007/s002200100431}{\emph{Commun. Math.
  Phys.} {\bfseries 219} (2001) 399}
  [\href{https://arxiv.org/abs/hep-th/0006196}{{\ttfamily hep-th/0006196}}].

\end{thebibliography}\endgroup
\end{document}